\pdfoutput=1

\documentclass[10pt,journal,cspaper,compsoc]{IEEEtran}

\usepackage[utf8]{inputenc}
\usepackage[T1]{fontenc}

\usepackage{amsmath}
\usepackage{amssymb}
\usepackage{amsfonts}
\usepackage{graphicx}
\usepackage{xspace}
\usepackage{listings}
\usepackage{url}
\usepackage{algorithm}
\usepackage{algorithmicx}
\usepackage[noend]{algpseudocode}
\usepackage{subfigure}

\usepackage{multirow}
\usepackage{dblfloatfix}
\usepackage{booktabs}
\usepackage{tikz}
\usepackage{pifont}

\usepackage{microtype}
\usepackage{tabularx}
\usepackage{float}
\usepackage{placeins}
\usepackage[english]{babel}
\usepackage[autolanguage]{numprint}
\usepackage[binary-units]{siunitx}
\usepackage{textcomp}

\newcommand{\myneedspace}{}

\newcommand{\Bytes}{Bytes\xspace}

\newcommand{\fig}{Figure\xspace}
\newcommand{\figs}{Figures\xspace}
\newcommand{\tab}{Table\xspace}
\newcommand{\alg}{Algorithm\xspace}
\newcommand{\algs}{Algorithms\xspace}
\newcommand{\append}{Appendix\xspace}
\newcommand{\Sec}{Section\xspace}
\newcommand{\Equ}{Equation\xspace}
\newcommand{\eg}{e.g.\xspace}
\newcommand{\ie}{i.e.\xspace}
\newcommand{\etal}{et~al.\@\xspace}
\newcommand{\etcet}{etc.\@\xspace}
\newcommand{\cf}{cf.\@\xspace}
\newcommand{\expdesc}{Exp. details\xspace}

\newcommand{\mycomment}[1]{\textit{//\ #1}\xspace}

\newcommand{\optionyes}{\ding{51}}
\newcommand{\optionno}{\ding{55}}

\newcommand{\machone}{\emph{TUWien}\xspace}
\newcommand{\machtwo}{\emph{VSC-1}\xspace}
\newcommand{\machthree}{\emph{Edel~(G5k)}\xspace}
\newcommand{\machfour}{\emph{Cartesius~(SURFsara)}\xspace}
\newcommand{\machfive}{\emph{VSC-3}\xspace}

\newcommand{\mpirun}{\texttt{mpirun}\xspace}
\newcommand{\mpiruns}{\texttt{mpirun}s\xspace} 

\newcommand{\runtime}{run-time\xspace}
\newcommand{\Runtime}{Run-time\xspace}
\newcommand{\runtimes}{run-times\xspace}
\newcommand{\Runtimes}{Run-times\xspace}

\newcommand{\mpicroscope}{\texttt{mpicroscope}\xspace}
\newcommand{\skampi}{{SKaMPI}\xspace}
\newcommand{\netgauge}{{Netgauge}\xspace}
\newcommand{\nbcbench}{{NBCBench}\xspace}
\newcommand{\phloem}{{Phloem MPI Benchmarks}\xspace}
\newcommand{\OSU}{{OSU Micro-Benchmarks}\xspace}
\newcommand{\intelbench}{{Intel MPI Benchmarks}\xspace}
\newcommand{\syncjk}{{JK}\xspace}
\newcommand{\synchca}{{HCA}\xspace}
\newcommand{\synchcatwo}{{HCA2}\xspace}
\newcommand{\mpptest}{\texttt{mpptest}\xspace}
\newcommand{\rdtsc}{\texttt{RDTSC}\xspace}
\newcommand{\rdtscp}{\texttt{RDTSCP}\xspace}
\newcommand{\hrtcal}{\texttt{HRT\_CALIBRATE}\xspace}
\newcommand{\jonesk}{Jones and Koenig\xspace}
\newcommand{\infiniband}{IB\xspace}
\newcommand{\gcc}{gcc\xspace}
\newcommand{\kolmtest}{Kolmogorov-Smirnov\xspace}
\newcommand{\shaptest}{Shapiro-Wilk\xspace}

\newcommand{\mpibcast}{\texttt{MPI\_Bcast}\xspace}

\newcommand{\mpialltoall}{\texttt{MPI\_Alltoall}\xspace}
\newcommand{\mpiscan}{\texttt{MPI\_Scan}\xspace}
\newcommand{\mpiallreduce}{\texttt{MPI\_Allreduce}\xspace}
\newcommand{\mpibarrier}{\texttt{MPI\_Barrier}\xspace}

\newcommand{\mpiwtime}{\texttt{MPI\_Wtime}\xspace}

\algnewcommand{\And}{\textbf{and}\xspace}
\algnewcommand{\Not}{\textbf{not}\xspace}
\algnewcommand{\Or}{\textbf{or}\xspace}

\newcommand{\mvapichtwoone}{MVAPICH~2.1a\xspace}

\newcommand{\mvapichtwoa}{MVAPICH~2.0a-qlc\xspace}
\newcommand{\mvapichonenine}{MVAPICH~1.9\xspace}
\newcommand{\necmpitwoeleven}{NEC\,MPI~1.2.11\xspace}
\newcommand{\necmpitwoeight}{NEC\,MPI~1.2.8\xspace}
\newcommand{\NEC}{NEC MPI\xspace}

\newcommand{\MVAPICH}{MVAPICH\xspace}

\newcommand{\intelmpi}{Intel~MPI~4.1\xspace}
\newcommand{\intelmpifive}{Intel~MPI~5\xspace}

\newcommand{\testttest}{\textsc{t-test}\xspace}
\newcommand{\testwilcox}{\textsc{Wilcoxon test}\xspace}
\newcommand{\testwelch}{\textsc{Welch's t-test}\xspace}

\newcommand{\designpartone}{Benchmark\xspace}
\newcommand{\designparttwo}{Scan\_Over\_MPI\_Functions\xspace}
\newcommand{\proctext}[1]{\textsc{#1}}

\newcommand{\groupone}{\textsc{group}~\num{1}\xspace}
\newcommand{\grouptwo}{\textsc{group}~\num{2}\xspace}

\newcommand{\msize}{\textit{msize}\xspace}
\newcommand{\myrank}{\textit{my\_rank}\xspace}
\newcommand{\np}{\textit{p}\xspace}
\newcommand{\nrep}{\textit{nrep}\xspace}
\newcommand{\rootproc}{\textit{root}\xspace}
\newcommand{\exectime}{\textit{t}\xspace}
\newcommand{\func}{\textit{func}\xspace}
\newcommand{\nmpiruns}{\textit{n}\xspace}
\newcommand{\ntrial}{\textit{ntrial}\xspace}

\newcommand{\initialts}{\textit{initial\_time}\xspace}
\newcommand{\stime}{\textit{s\_time}\xspace}
\newcommand{\etime}{\textit{e\_time}\xspace}
\newcommand{\localexectime}{\textit{local\_time}\xspace}

\newcommand{\startwin}{\textit{start\_time}\xspace}
\newcommand{\nextwin}{\textit{next\_win}\xspace}
\newcommand{\winsize}{\textit{win\_size}\xspace}

\newcommand{\diff}{\textit{diff}\xspace}
\newcommand{\rtt}{\textit{rtt}\xspace}
\newcommand{\rtts}{\textit{rtts}\xspace}

\newcommand{\tremote}{\textit{tremote}\xspace}

\newcommand{\recvbuf}{\textit{recvdiffs}\xspace}
\newcommand{\server}{\textit{server}\xspace}
\newcommand{\client}{\textit{client}\xspace}
\newcommand{\pref}{\textit{p\_ref}}
\newcommand{\currentround}{\textit{round}\xspace}
\newcommand{\hierintercepts}{\textit{hierarchical\_intercepts}\xspace}
\newcommand{\tmpmodel}{\textit{tmp\_lm}\xspace}
\newcommand{\lm}{\textit{lm}\xspace}

\newcommand{\difflist}{\ensuremath{l^{\diff}}\xspace}
\newcommand{\rttlist}{\ensuremath{l^{\rtt}}\xspace}
\newcommand{\proclist}{\ensuremath{l^{\np}}\xspace}
\newcommand{\mlist}{\ensuremath{l^{\msize}}\xspace}
\newcommand{\funclist}{\ensuremath{l^{\func}}\xspace}
\newcommand{\explist}{\ensuremath{l^\textit{exp}}\xspace}
\newcommand{\modellist}{\ensuremath{l^\textit{model}}}
\newcommand{\mlistclient}{\ensuremath{l^\textit{model}_\textit{client}}}

\newcommand{\exectimes}{\ensuremath{l^{\exectime}}}
\newcommand{\stimes}{\ensuremath{l^{\stime}}}
\newcommand{\etimes}{\ensuremath{l^{\etime}}}
\newcommand{\errorlist}{\ensuremath{l^\textit{error}}}
\newcommand{\localerrorlist}{\ensuremath{l^\textit{error\_local}}}
\newcommand{\freqlist}{\ensuremath{l^\textit{freq}}}
\newcommand{\finalstimes}{\ensuremath{l^{\stime}_{\textit{final}}}}
\newcommand{\finaletimes}{\ensuremath{l^{\etime}_{\textit{final}}}}
\newcommand{\finalexectimes}{\ensuremath{l^{\exectime}_{\textit{final}}}}

\newcommand{\localexectimes}{\ensuremath{l^\textit{t\_local}}}
\newcommand{\localstimes}{\ensuremath{l^\textit{s\_time\_local}}}
\newcommand{\localetimes}{\ensuremath{l^\textit{e\_time\_local}}}
\newcommand{\remotetimes}{\ensuremath{l^\textit{remote}}}
\newcommand{\exectimesfinal}{\ensuremath{l^\textit{t\_final}}}
\newcommand{\reftimes}{\ensuremath{l^\textit{ref}}}
\newcommand{\globaltimes}{\ensuremath{l^\textit{global}}}
\newcommand{\maxtimes}{\ensuremath{l^\textit{max}}}
\newcommand{\maxglobaltimes}{\ensuremath{l^\textit{max\_gl}}}
\newcommand{\tmpdifflist}{\textit{tmp}\xspace}
\newcommand{\maxlocaltime}{\ensuremath{t^\textit{max}}}

\newcommand{\warmup}{\textsc{warmup\_rounds}\xspace}

\newcommand{\nfitpoints}{\textsc{n\_fitpts}\xspace}
\newcommand{\nexchanges}{\textsc{n\_exchanges}\xspace}
\newcommand{\npingpongs}{\textsc{n\_pingpongs}\xspace}

\newcommand{\mpidouble}{\textsc{mpi\_double}\xspace}
\newcommand{\mpiuint}{\textsc{mpi\_uint64\_t}\xspace}
\newcommand{\mpimax}{\textsc{mpi\_max}\xspace}
\newcommand{\mpimin}{\textsc{mpi\_min}\xspace}

\newcommand{\nullval}{\textsc{null}\xspace}
\newcommand{\modeltype}{\textsc{t\_pair\_double}\xspace}

\newcommand{\myalgnumfont}{\scriptsize\selectfont}
\algrenewcommand{\alglinenumber}[1]{\myalgnumfont #1:}

\newcommand{\Ifdef}[1]{\textbf{\#Ifdef} #1}
\newcommand{\Ifndef}[1]{\textbf{\#Ifndef} #1}

\newcommand{\EndIfdef}{\textbf{\#EndIfdef}}
\newcommand{\EndIfndef}{\textbf{\#EndIfndef}}

\renewcommand*\Call[2]{\textproc{#1}(#2)}

\makeatletter
\renewcommand{\ALG@beginalgorithmic}{\scriptsize}
\makeatother

\begin{document}

\title{MPI Benchmarking Revisited:\\Experimental Design and Reproducibility}

\author{Sascha Hunold and~Alexandra Carpen-Amarie
  \IEEEcompsocitemizethanks{\IEEEcompsocthanksitem
    are with the TU Wien, Faculty of Informatics, Institute of Information
    Systems, Research Group for Parallel Computing, 
    Favoritenstrasse 16/184-5,  1040 Vienna, Austria\protect\\
    E-mail: \{hunold,carpenamarie\}@par.tuwien.ac.at}}

\IEEEtitleabstractindextext{
\begin{abstract}  
  The Message Passing Interface (MPI) is the prevalent programming
  model used on today's supercomputers. Therefore, MPI library
  developers are looking for the best possible performance (shortest
  \runtime) of individual MPI functions across many different
  supercomputer architectures. Several MPI benchmark suites have been
  developed to assess the performance of MPI implementations.
  Unfortunately, the outcome of these benchmarks is often neither
  reproducible nor statistically sound.  To overcome these issues, we
  show which experimental factors have an impact on the \runtime of
  blocking collective MPI operations and how to control them.  We
  address the problem of process and clock synchronization in MPI
  benchmarks.  Finally, we present a new experimental method that
  allows us to obtain reproducible and statistically sound MPI
  measurements.
\end{abstract}

}

\maketitle

\section{Introduction}
\label{sec:intro}

Since the Message Passing Interface (MPI) was standardized in the
1990s, it has been the prevalent programming model on the majority of
supercomputers. As MPI is an essential building block of
high-performance applications, performance problems in the MPI library
have direct consequences on the overall \runtime of applications.

Library developers and algorithm designers have one question in
common: which algorithm works better (is faster) for a given
communication problem?  For example, which implementation of broadcast
is faster on $p=128$ processors using a payload of $m= \SI{64}{\Bytes}$? As
today's parallel systems can hardly be modeled analytically, empirical
evaluations using \runtime tests of MPI functions are required to
compare different MPI implementations. It is therefore 
important to measure the \runtime of MPI functions correctly.

MPI library developers rely on benchmark suites to test their
implementations. The problem is that the results of these benchmarks
may vary significantly, where \tab~\ref{tab:moti_intel_mpi} shows one
example. The table compares the minimum and maximum \runtime of an
\mpibcast on \num{16} nodes that were reported by \num{30} different calls
(\mpiruns) to the \intelbench.
\begin{table}[t]
  \caption{\label{tab:moti_intel_mpi}Minimum and maximum \runtime of \mpibcast 
    obtained from \num{30} different runs of the Intel MPI Benchmarks on \num{16x1} processes with \necmpitwoeight on \machone.}
  \centering
\begin{scriptsize}
\begin{tabular}{rrrr}
  \toprule
Msg. size & min(avg) & max(avg) & diff \\
$[\Bytes]$ & $[\mu s]$ & $[\mu s]$ & $[\%]$ \\  
  \midrule
  0 & 0.04 & 0.04 & 0.00 \\ 
    1 & 2.93 & 3.12 & 6.09 \\ 
    2 & 2.83 & 3.20 & 11.56 \\ 
    4 & 2.82 & 3.06 & 7.84 \\ 
    8 & 2.86 & 3.13 & 8.63 \\ 
   16 & 2.84 & 3.14 & 9.55 \\ 
   32 & 3.13 & 3.44 & 9.01 \\ 
   64 & 3.15 & 3.51 & 10.26 \\ 
  128 & 3.62 & 4.03 & 10.17 \\ 
  256 & 4.34 & 4.90 & 11.43 \\ 
  512 & 5.41 & 5.91 & 8.46 \\ 
  1024 & 6.88 & 7.05 & 2.41 \\ 
  2048 & 9.52 & 9.71 & 1.96 \\ 
  4096 & 13.52 & 13.91 & 2.80 \\ 
  8192 & 19.30 & 19.66 & 1.83 \\ 
  16384 & 32.21 & 33.13 & 2.78 \\ 
  32768 & 53.57 & 54.47 & 1.65 \\ 
   \bottomrule
\end{tabular}
\end{scriptsize}
\end{table}
The third column lists the difference between the minimum and maximum
\runtime in percent. We can see that for payloads of up to
\SI{512}{\Bytes}, the \runtimes have an error of roughly
\SI{10}{\percent}.  One solution might be to change the default
parameters of the \intelbench. For example, one could force the
benchmark to perform more measurements. But the question then becomes:
how many runs are sufficient to obtain reproducible results?

It is a common practice---when comparing MPI implementations as part
of a scientific publication---to choose one of the available
benchmarks and compare the results. Many MPI benchmarks report either
the mean, the median, or the minimum \runtime. The problem is that
without using a dispersion metric and a rigorous statistical analysis,
we can hardly determine 
whether an observation is repeatable or the result of chance.

\begin{table*}[t]
\centering
\caption{\label{tab:mpibench}Comparison of measures for several MPI benchmark suites.}
\begin{scriptsize}
\begin{tabular}{l|c|c|c|c|c|c}
  \toprule
  benchmark name & ref. & version & mean & min & max & measure of dispersion \\ 
  \midrule
  \intelbench & \cite{intel_mpi_bench} & 4.0.0  & \optionyes & \optionyes & \optionyes &  \optionno \\ 
  MPIBench & \cite{GroveC05} & 1.0beta & \optionyes & \optionyes & &  sub-sampled data \\ 
  MPIBlib & \cite{LastovetskyRO08a} & 1.2.0 & \optionyes & & & conf. interv. of the mean (default 95\%) \\ 
  \mpicroscope & \cite{Traff12} & 1.0 & \optionyes  & \optionyes & \optionyes  &    \optionno \\ 
  \mpptest & \cite{GroppL99} & 1.5 &  \multicolumn{2}{|c|}{min of means} & & \optionno \\ 
  \nbcbench & \cite{Hoefler:2007fo} & 1.1 & \optionyes (+median) & \optionyes & \optionyes & \optionno \\ 
  \OSU & \cite{osu_benchmarks} & 4.4.1 & \optionyes & \optionyes & \optionyes &   \optionno \\ 
  \phloem & \cite{phloem_mpi} & 1.0.0 & \optionyes & \optionyes & \optionyes &  \optionno \\ 
  \skampi  & \cite{ReussnerST02} & 5.0.4 & \optionyes &  &  & std. error \\ 
  \bottomrule
\end{tabular}
\end{scriptsize}
\end{table*}
In this article, we make the following contributions to the problem of
\emph{accurately benchmarking blocking collective MPI operations}:
\begin{enumerate}
\item We show that a precise synchronization of clocks is key to
  measure MPI functions accurately.
\item We present a novel clock synchronization algorithm that has two
  advantages: (1)~it is accurate as it accounts for the clock drift
  between processes, and (2)~provides a good trade-off between time
  and accuracy.
\item We establish a list of (experimental) factors that, we show, do
  significantly influence the outcome of MPI performance measurements.
\item We propose a novel benchmarking method, including an
  experimental design and an appropriate data analysis strategy, that
  allows for a fair comparison of MPI libraries, but which is most
  importantly (1)~statistically sound and (2)~reproducible.
\end{enumerate}

We start by summarizing other MPI benchmarking approaches in
\Sec~\ref{sec:mpi_hist} and discuss their strengths and shortcomings.
\Sec~\ref{sec:exp_setup} introduces our general experimental framework
that we use for all experiments conducted as part of this article.  In
\Sec~\ref{sec:mpi_sync_exp}, we show the practical implications of
applying different process synchronization methods and also present
our novel clock synchronization algorithm.
\Sec~\ref{sec:characterization_mpi_benchmark} takes a closer look at
factors that may influence the experimental outcome (the \runtime
of an MPI function).  \Sec~\ref{sec:statistical_analysis} describes
our method for comparing the performance of MPI libraries in a
statistically sound manner. We summarize related work in
\Sec~\ref{sec:relwork} with an emphasis on statistically sound
experiments, before we conclude in \Sec~\ref{sec:conclusions}.

\section{A Brief History of MPI Benchmarking}
\label{sec:mpi_hist}

\newcolumntype{A}[1]{>{\centering\bfseries}m{#1}|}
\newcommand{\headcell}[2]{\multicolumn{1}{A{#1}}{\vspace{0.5\baselineskip}#2 \vspace{0.5\baselineskip}}}
\newcommand{\headfirstcell}[2]{\multicolumn{1}{|A{#1}}{\vspace{0.5\baselineskip}#2 \vspace{0.5\baselineskip}}}
\newcommand{\mytablerule}{\noalign{\hrule height \arrayrulewidth}}

\begin{table*}[t]
  \caption{\label{tab:timing_schemes}Measurement schemes (MS) for blocking MPI collectives found in different MPI benchmarks. In scheme (MS4), depending on the implementation, \texttt{Get\_Time} returns the local 
    time (measured with \mpiwtime or \rdtsc) or a logical global time.}
  \centering
\begin{scriptsize}
\begin{tabularx}{\textwidth}{@{}|p{.19\textwidth}|p{.2\textwidth}|p{.22\textwidth}|X|@{}}
\mytablerule
\headfirstcell{.19\textwidth}{(MS1) \skampi, \nbcbench, MPIBlib, MPIBench, \mpicroscope, \\ \OSU} &%
\headcell{.2\textwidth}{(MS2) Intel MPI Benchmarks, \mpptest} &%
\headcell{.22\textwidth}{(MS3) \phloem} &%
\headcell{.29\textwidth}{(MS4) \skampi, \nbcbench} \\
\mytablerule
  \begin{algorithmic}[1]
    \For{\textit{obs} in $1$ to \nrep}
    \State \mpibarrier
    \State \stimes[\textit{obs}] = \mpiwtime
    \State execute MPI function
    \State \etimes[\textit{obs}] = \mpiwtime
    \State \exectime += \etimes[\textit{obs}] - \stimes[\textit{obs}]
    \Statex \hspace*{.8em} \mycomment{\OSU}
    \EndFor
  \end{algorithmic}
&
  \begin{algorithmic}[1]
    \State \mpibarrier \mycomment{or omitted}
    \State \stime = \mpiwtime
    \For{\textit{obs} in $1$ to \nrep}    
    \State execute MPI function
    \EndFor
    \State \etime = \mpiwtime
    \State \exectime = $(\etime-\stime)/\nrep$
  \end{algorithmic}
&
  \begin{algorithmic}[1]
    \State \mpibarrier 
    \State \stime = \mpiwtime
    \For{\textit{obs} in $1$ to \nrep}    
    \State execute MPI function
    \State \mpibarrier \mycomment{or omitted}
    \EndFor
    \State \etime =  \mpiwtime
    \State \exectime = $(\etime-\stime)/\nrep$
  \end{algorithmic}
&
  \begin{algorithmic}[1]
    \State \Call{Sync Clocks}{}
    \State \textsc{Decide} on \startwin and \winsize
    \For{\textit{obs} in $1$ to \nrep}    
    \State \Call{Wait\_Until}{$\startwin + \textit{obs} \cdot \winsize$}
    \State \stimes[\textit{obs}] = \Call{Get\_Time}{}
    \State execute MPI function
    \State \etimes[\textit{obs}] = \Call{Get\_Time}{}
    \EndFor
  \end{algorithmic}%
\\
\mytablerule
\end{tabularx}
\end{scriptsize}
\end{table*}

\begin{table*}[t]
  \caption{\label{tab:processing_schemes}Commonly used data processing schemes (PS) for benchmarking blocking collective MPI operations.}
  \centering
\begin{scriptsize}
\begin{tabularx}{\textwidth}{@{}|p{.245\textwidth}|p{.245\textwidth}|p{.185\textwidth}|X|@{}}
\mytablerule
\headfirstcell{.245\textwidth}{(PS1) \skampi} &%
\headcell{.245\textwidth}{(PS2) MPIBlib, \mpicroscope} &%
\headcell{.185\textwidth}{(PS3) \OSU} &%
\headcell{.225\textwidth}{(PS4) MPIBlib  } \\
\mytablerule
  \begin{algorithmic}[1]
    \For{\textit{obs} in $1$ to \nrep}     
    \State \localexectimes[\textit{obs}] = \etimes[\textit{obs}] - \stimes[\textit{obs}] 
    \EndFor
    \State \textsc{Reduce} \nrep local \runtimes from each process on \rootproc:
    \Statex  \maxtimes[\textit{obs}] = \Call{max$_\np$}{$\localexectimes_\np[\textit{obs}]$} 
    \State \Call{Sort}{\maxtimes}
    \State \exectimes =   \maxtimes[ \nrep/4 : (\nrep\ - \nrep/4)]
    \State print \Call{mean$_{\nrep/2}$}{\exectimes}, \Call{stdev$_{\nrep/2}$}{\exectimes}
  \end{algorithmic} %
  &%
  \begin{algorithmic}[1]
    \For{\textit{obs} in $1$ to \nrep}     
    \State \localexectimes[\textit{obs}] = \etimes[\textit{obs}] - \stimes[\textit{obs}] 
    \EndFor
    \State \textsc{Reduce} \nrep local \runtimes from each process on \rootproc:
    \Statex  \maxtimes[\textit{obs}] = \Call{max$_\np$}{$\localexectimes_\np[\textit{obs}]$} 
    \State print \Call{mean$_{\nrep}$}{\maxtimes},  \Call{ci$_{\nrep}$}{\maxtimes} 
    \Statex \hfill \mycomment{MPIBlib}
    \Statex print \Call{mean$_{\nrep}$}{\maxtimes},  \Call{min$_{\nrep}$}{\maxtimes}, \Call{max$_{\nrep}$}{\maxtimes}
    \hfill \mycomment{\mpicroscope}
  \end{algorithmic} %
  &%
  \begin{algorithmic}[1]
    \State \localexectime = $\exectime / \nrep$
    \State \textsc{Reduce} \localexectime from each process to \rootproc:
    \Statex \textit{min\_lat} = \Call{min$_\np$}{$\localexectime$}
    \Statex \textit{max\_lat} = \Call{max$_\np$}{$\localexectime$}
    \Statex \textit{mean\_lat} = \Call{sum$_\np$}{$\localexectime$} 
    \Statex \hspace*{3.8em}  / \nrep
    \State print \textit{mean\_lat}, \textit{min\_lat}, \textit{max\_lat}
  \end{algorithmic}%
& %
  \begin{algorithmic}[1]
    \State \globaltimes = \Call{Normalize\_Times}{\etimes}
    \For{\textit{obs} in $1$ to \nrep}     
    \State \textsc{Reduce} \globaltimes[\textit{obs}] from each process to \rootproc:
    \Statex \hspace*{10pt} \maxglobaltimes[\textit{obs}] = \Call{max$_\np$}{$\globaltimes_\np[\textit{obs}]$}
    \State \exectimes[\textit{obs}] = \maxglobaltimes[\textit{obs}] - \stimes[\textit{obs}] 
    \EndFor
    \State print \Call{mean$_{\nrep}$}{\exectimes},  \Call{CI$_{\nrep}$}{\exectimes}
  \end{algorithmic} %
  \\
  \mytablerule
  \end{tabularx}\\[1ex]
\begin{tabularx}{\textwidth}{@{}|p{.245\textwidth}|p{.245\textwidth}|p{.185\textwidth}|X|@{}}
\mytablerule
\headfirstcell{.245\textwidth}{(PS5) MPIBench} &%
\headcell{.245\textwidth}{(PS6) \nbcbench } &%
\headcell{.185\textwidth}{(PS7) Intel MPI Benchmarks, \phloem} &%
\headcell{.225\textwidth}{(PS8) \mpptest} \\%
\mytablerule
  \begin{algorithmic}[1]
    \State \stimes = \Call{normalize\_times}{\stimes}
    \State \etimes = \Call{normalize\_times}{\etimes}
    \State \textsc{Gather} \etimes, \stimes~from each process on \rootproc into \finaletimes, \finalstimes
    \For{\textit{rank} in $1$ to \np}
    \For{\textit{obs} in $1$ to \nrep}
    \State \textit{i} = $(\textit{rank} - 1) \cdot \nrep  + \textit{obs} $
    \State \finalexectimes[\textit{i}] = \finaletimes[\textit{i}] - \finalstimes[\textit{i}] 
    \EndFor
    \EndFor
    \State  \exectimes =  \Call{remove\_outliers$_{\np \cdot \nrep}$}{\finalexectimes}
    \State print \Call{min}{\exectimes}, \Call{max}{\exectimes}, \Call{mean}{\exectimes}
  \end{algorithmic} %
  & %
  \begin{algorithmic}[1]
    \For{\textit{obs} in $1$ to \nrep}     
    \State \localexectimes[\textit{obs}] = \etimes[\textit{obs}] - \stimes[\textit{obs}] 
    \EndFor
    \State \exectime = $\Call{OP}{\localexectimes}$ 
    \Statex \mycomment{OP $\in \{$min, max, mean, median$\}$ }
    \State \textsc{Gather} \exectime from each process on \rootproc into \exectimes
    \State \maxlocaltime = \Call{max$_\np$}{$\exectimes$}
    \State print \maxlocaltime
  \end{algorithmic} 
& %
  \begin{algorithmic}[1]
    \State \textsc{Gather} average times \exectime\ on \rootproc process into \exectimes
    \State print \Call{min$_\np$}{\exectimes}, \Call{max$_\np$}{\exectimes}, \Call{mean$_\np$}{\exectimes}
  \end{algorithmic} %
     & %
  \begin{algorithmic}[1]
    \State \textsc{Broadcast} \exectime from the \rootproc process
    \State collect \Call{min$_{reps}$}{\exectime} over several repetitions of the measurement scheme
    \end{algorithmic} \\
\mytablerule
\end{tabularx}
\end{scriptsize}
\end{table*}

We now give a history of MPI benchmarking. Ever since the first MPI
standard was announced in 1995, several MPI benchmark suites have been
proposed. Some of the best-known MPI benchmark suites are summarized
in \tab~\ref{tab:mpibench}. The table includes information about the
measures (\eg, min, max) that each benchmark uses to present \runtimes
and which measure of dispersion is provided.  It is complemented with
the \runtime measurement approaches implemented by each benchmark,
which are separately summarized in the four pseudocode listings in
\tab~\ref{tab:timing_schemes}.  Furthermore,
\tab~\ref{tab:processing_schemes} details the methods selected by each
of the investigated benchmarks for computing and presenting the
measured \runtimes. The data in \tab~\ref{tab:mpibench} was gathered
to the best of our knowledge, since some benchmarks, like the Special
Karlsruher MPI-Benchmark (\skampi), have been released in many
incarnations and some other ones, like the MPIBench, are currently not
available for download\footnote{We obtained the source code of
  MPIBench 1.0beta through private communication.}. We therefore also
rely on the respective articles describing the benchmarks.

\mpptest was one of the first MPI benchmarks~\cite{GroppL99} and was a
part of the MPICH distribution. Gropp and Lusk carefully designed
\mpptest to allow reproducible measurements for realistic usage
scenarios. They pointed out common pitfalls when conducting MPI
performance measurements, such as ignoring cache effects. In
particular, to ensure cold caches when sending a message, \mpptest
uses a send and a receive buffer which are twice as big as the cache
level that should be ``cold''. Then, the starting address of a message
to be sent is always advanced in this larger buffer, trying to ensure
that the data accessed are not cached. If a starting address does not
leave enough space for the message to be sent, it is reset to the
beginning of the buffer.  At the time of designing \mpptest, most of
the hardware clocks were coarse-grained and therefore did not allow a
precise measurement of one call to a specific MPI function (as this
would have often resulted in obtaining a \num{0}).  To overcome this
problem and to improve the reproducibility of results, \mpptest
measures the time $\exectime$ of \nrep consecutive calls to an MPI
function and computes the mean $\bar{\exectime_i}=\exectime/\nrep$ of
these \nrep observations. This measurement is repeated $k$ times and
the minimum over these $k$ samples is reported, \ie,
$\min_{1 \le i \le k} \bar{\exectime_i}$.

The \skampi benchmark is a highly configurable MPI benchmark
suite~\cite{ReussnerST02} and features a domain-specific language for
describing individual MPI benchmark tests. \skampi also allows to
record MPI timings by using a window-based process synchronization
approach, in addition to the commonly used \mpibarrier (\cf
measurement schemes~(MS1) and (MS4) in
\tab~\ref{tab:timing_schemes}). \skampi reports the arithmetic mean
and the standard error of the \runtimes of MPI functions. It uses an
iterative measuring process for each test case, where a test is
repeated until the current relative standard error is smaller than a
predefined maximum.

MPIBlib by Lastovetsky~\etal \cite{LastovetskyRO08a} works similarly
to \skampi, as it computes a confidence interval of the mean based on
the current sample. It stops the measurements when the sample mean is
within a predefined range (\eg, a \SI{5}{\percent} difference) from
the end of a 95\% confidence interval. MPIBlib implements multiple
methods for computing the sample mean, as shown in schemes~(PS2) and
(PS4) in \tab~\ref{tab:processing_schemes}. It also provides an
additional scheme that measures the \runtime on the root process only,
but which we omitted for reasons of clarity.

\mpicroscope~\cite{Traff12} and \OSU~\cite{osu_benchmarks} perform
repeated measurements of one specific MPI function for a predefined
number of times. They report the minimum, the maximum, and the mean
\runtime of a sample. \mpicroscope attempts to reduce the number of
measurements using a linear (or optionally exponential) decay of
repetitions, \ie, if no new minimum execution time in a sample of
\nrep consecutive MPI function calls was found, the remaining number
of repetitions is decreased.

The Intel MPI Benchmarks~\cite{intel_mpi_bench} use a measurement
method similar to \mpptest, \ie, the time is taken before and after
executing \nrep consecutive calls to an MPI function. Then, the
benchmark computes the mean of the \runtimes over these \nrep
consecutive calls for each MPI rank. The final report includes the
minimum, maximum, and average of these means across all ranks.

The \phloem~\cite{phloem_mpi} for MPI collectives measure the total
time to execute \nrep consecutive MPI function calls and compute the
mean \runtime for each rank.  In addition, the \phloem can be
configured to interleave the evaluated MPI function calls with calls
to \mpibarrier in each iteration. Minimum, maximum, and average
\runtimes across MPI ranks are provided upon benchmark completion.

Grove and Coddington developed MPIBench~\cite{GroveC05}, which, in
addition to mean and minimum \runtimes, also plots a sub-sample of the
raw data to show the dispersion of measurements.  They discuss the
problem of outlier detection and removal. In their work, the \runtimes
that are bigger than some threshold time $t_{thresh}$ are treated as
outliers. To compute $t_{thresh}$, they determine the 99th percentile
of the sample, denoted as $t_{99}$, and then define
$t_{thresh} = t_{99} \cdot a$ for some constant $a \ge 1$ (default
$a=2$). Grove also shows the distribution of \runtimes obtained when
measuring \texttt{MPI\_Isend} with different message
sizes~\cite[p. 127]{Grove:2003to}. He highlights the fact that the
execution time of MPI functions is not normally distributed.

\nbcbench was initially introduced to assess the \runtime of
non-blocking collective implementations in comparison to their
blocking alternatives~\cite{Hoefler:2007fo}. Later, Hoefler~\etal
explained how blocking and non-blocking collective MPI operations
could be measured scalably and accurately~\cite{hoefler-pmeo08}.  The
authors show that calling MPI functions consecutively can lead to
pipelining effects, which could distort the results. To address these
problems, they implement a window-based synchronization scheme,
requiring $\mathcal{O}(\log{\np})$ rounds to complete, compared to the
$\mathcal{O}(\np)$ rounds needed by \skampi, where $\np$ denotes the
number of processes.

\section{Experimental Setup}
\label{sec:exp_setup}

As the results of this paper heavily rely on the empirical analysis of
hypotheses, we first introduce our measurement scheme for blocking,
collective MPI functions, the data processing methods applied, and the
parallel machines used for conducting our experiments.

\subsection{Notation}

The benchmarks \nbcbench and \netgauge are related.  For example,
Hoefler~\etal state the following: ``We used our new findings to
implement a new benchmark scheme in the benchmark suite Netgauge. The
implementation bases on NBCBench [..]''~\cite{Hoefler:2010vr}.  For
that reason, we use \nbcbench to refer to the MPI benchmark and
\netgauge to the algorithm that synchronizes clocks hierarchically.

We use the following notation in the remainder of the article, which
we borrowed from Kshemkalyani and
Singhal~\cite{Kshemkalyani:2008}. The \emph{clock offset} is the
difference between the time reported by two clocks. The \emph{skew of
  the clock} is the difference in the frequencies of two clocks and
the \emph{clock drift} is the difference between two clocks over a
period of time.

\subsection{Timing Procedure}

In the experiments presented in this article, we measure the time for
completing a single MPI function using the method shown in
\alg~\ref{alg:mpi_timing}. Before the start of a benchmark run, the
experimenter chooses the number of observations \nrep (sample size) to
be recorded for an individual test, where a test consists of an MPI
function, a message size, and a number of processes.
\begin{algorithm}[t]
  \begin{scriptsize}
  \caption{\label{alg:mpi_timing}MPI timing procedure.}
  \begin{algorithmic}[1]
  \Procedure{Time\_MPI\_function}{\func, \msize, \nrep} 
  \Statex \mycomment{\func\ - MPI function}
  \Statex \mycomment{\msize\ - message size}
  \Statex \mycomment{\nrep\ - nb. of observations}
  \State initialize time array \exectimes\ with \nrep elements
  \For{\textit{obs} in $1$ to \nrep}
  \State \Call{Sync\_Processes}{} \mycomment{either \textsc{MPI\_Barrier} or window-based sync.}
  \State \localstimes[\textit{obs}] = \Call{Get\_Time}{}
  \State execute \func(\msize)
  \State \localetimes[\textit{obs}] = \Call{Get\_Time}{}
  \EndFor
  
  \If{ \text{sync method} == \mpibarrier } \label{line:mpi_timing_start}
  \For{\textit{obs} in 1 to \nrep}
  \State $\localexectimes[\textit{obs}] = \localetimes[\textit{obs}] - \localstimes[\textit{obs}] $
  \EndFor
  \State \Call{MPI\_Reduce}{\localexectimes, \exectimes, \nrep, \mpimax, \rootproc}
  
  \Else
  \State normalize \localstimes, \localetimes to the global reference clock
  \State \Call{MPI\_Reduce}{\localstimes, \stimes, \nrep, \mpidouble, \mpimin, \rootproc}
  \State \Call{MPI\_Reduce}{\localetimes, \etimes, \nrep, \mpidouble, \mpimax, \rootproc}
  \For{\textit{obs} in $1$ to \nrep}
    \State $\exectimes[\textit{obs}] =  \etimes[\textit{obs}] - \stimes[\textit{obs}]$
  \EndFor \label{line:mpi_timing_end}
  \EndIf
  \If{ \myrank == \rootproc }
  \For{\textit{obs} in $1$ to \nrep}
  \State print \exectimes[\textit{obs}]
  \EndFor
  \EndIf
  \EndProcedure
  \end{algorithmic}
  \end{scriptsize}
\end{algorithm}
Before starting to measure the \runtime of an MPI function, all
processes need to be synchronized. We examine two kinds of
synchronization approaches: (1)~the use of
\mpibarrier and (2)~the window-based synchronization
scheme. Advantages and disadvantages of each synchronization method
will be discussed in more detail in \Sec~\ref{sec:mpi_sync_exp}.

Depending on the type of synchronization, we use different ways to
compute the time to complete a collective MPI operation, as detailed
in \alg~\ref{alg:mpi_timing} (lines
\ref{line:mpi_timing_start}--\ref{line:mpi_timing_end}).

\subsubsection{Completion Time based on Local Times}

In this case, each MPI process holds an array containing \nrep local
time measurements (\runtimes to complete a given MPI function).  We
apply a reduction operation ($\max$) on that array and collect
the results on the root process. Thus, the \runtime of an MPI function
\func using \np processes in iteration \textit{i},
$0 \le \textit{i} \le \nrep$, is given as
$\exectimes[i] = \max_{0 \le \textit{r} < \np}\{ \exectimes[i]
\}$.
In other words, the \runtime of an MPI function is defined as the
maximum local \runtime over all processes. This \runtime computation
procedure is typically applied for measurements where processes are
synchronized using \mpibarrier. 

\subsubsection{Completion Time based on Global Times}
\label{sec:completion_time_win}

When globally-synchronized clocks are available, we define the time to
complete an MPI operation as the difference between the maximum
finishing time and the minimum starting time among all 
processes. 
All \nrep starting and finishing timestamps from all processes are
gathered as vectors on the \rootproc node.  Then, the \rootproc node
computes the time of an MPI function \func using \np processes in
iteration $i$ like this
$\exectimes[i] = \max_{0 \le \textit{r} < \np}\{ \etimes[i] \}
- \min_{0 \le \textit{r} < \np}\{ \stimes[i] \}$.
We use this method to compute the completion time in all our
experiments in which we employ a clock synchronization method.

\subsection{Window-based Process Synchronization}
\label{sec:win_based_scheme_intro}

\skampi~\cite{skampi_collectives} was (to the best of our knowledge)
the first MPI benchmark suite that used a window-based synchronization
strategy to measure the \runtime of MPI functions.  Its window-based
synchronization method works as follows: (1)~The distributed clocks of
all participating MPI processes are synchronized relative to a
reference clock. To this end, each MPI process computes its clock
offset relative to a master process (\eg, process~\num{0}) to be able to
normalize its time to the master's reference clock.  (2)~The master
process selects a start time, which is a point in time that lies in
the future, and broadcasts this start time to all participating
processes. (3)~Since each process knows the time difference to the
master process, all processes are now able to wait for this start time
before executing the respective MPI function synchronously. When one
MPI function call has been completed, all processes will wait for
another future point in time before starting the next measurement. The
time period between these distinct points is called a ``window''.

This synchronization method shown in scheme~(MS4) of
\tab~\ref{tab:timing_schemes} is used for all benchmarking
experiments that rely on window-based process synchronization in this
paper.

\subsection{High-Resolution Time Measurements}
\label{sec:rdtsc_measure}

Hoefler~\etal discussed the problem that the resolution of \mpiwtime
is typically not high enough for measuring short time
intervals~\cite{Hoefler:2010vr}. They therefore use the CPU's clock
register to count the number of processor cycles since reset. More
specifically, \netgauge implements a time measurement mechanism based
on the atomic \rdtsc instruction, which provides access to the TSC
register and which is supported by the x86 and x86-64 instruction set
architectures (ISA).  However, several problems can arise when using
this mechanism. First, Hoefler~\etal point out that dynamic frequency
changes, which are automatically enabled in modern processors, can
modify the CPU clock rate and thus compromise the time
measurements. Second, in multi-processor systems, CPU clocks are not
necessarily synchronized, requiring the processes to be pinned to
cores to guarantee valid cycle counter values.

This \rdtsc mechanism is vulnerable to out-of-order instruction
executions supported by most modern processors~\cite{amd-arch-manual,
  intel-arch-manual}.  To overcome this issue, we performed our
measurements using the equivalent \rdtscp call, which guarantees
instruction serialization, that is, it makes sure that all
instructions have been executed when the timestamp counter is read.
Unless otherwise specified, we fixed the frequency to the highest
available value and pinned each process to a specific core in all our
experiments involving \rdtscp-based time measurements.

As all our experimental platforms are Linux systems, we checked the
TSC-related flags provided by \texttt{/proc/cpuinfo}.  On all our
systems, the flags \texttt{constant\_tsc} and \texttt{nonstop\_tsc}
were set, indicating that the speed of updating the TSC register is
independent of the current core frequency. Nevertheless, we need to
make sure that processes are pinned to cores throughout the
measurements to avoid accidentally reading the TSC register of another
core.

\subsection{Data Processing}
\label{sec:data_processing}

Most of the benchmarks listed in \tab~\ref{tab:mpibench} use some form
of implicit outlier removal (\eg, taking the minimum time recorded).
In addition, many benchmarks perform a number of warm-up rounds to
fill caches or to set up communication links. After the initial
warm-up phase has completed, the measurements taken are used to
compute the final statistics. One problem is that the operating system
noise can lead to relatively large variations of the measured \runtime
at any moment within the benchmark execution. A second problem is that
it is hard to estimate how many warm-up rounds are sufficient to reach
a ``steady state''. To make our benchmark method robust against these
two problems, we use Tukey's outlier filter to remove outliers after
all measurements have been
recorded~\cite[p.~126]{hogg2006probability}. When applying this
filter, we remove all measurements from the sample that are either
smaller than $Q_1 - 1.5 \cdot IQR$ or larger than
$Q_3 + 1.5 \cdot IQR$. $IQR$ denotes the interquartile range between
quartiles $Q_1$ and $Q_3$.

\begin{table*}[t]
  \centering
  \caption{Overview of parallel machines used in the experiments.}
  \label{tab:machines}
  \begin{footnotesize}
  \begin{tabular}{lllll}
    \toprule
    name & nodes & interconnect & MPI libraries  & compilers \\
    \midrule 
    \machone & \num{36} Dual Opteron 6134 @ \SI{2.3}{\giga\hertz}  & \infiniband QDR MT26428 & NEC MPI/LX 1.2.11 & \gcc~4.4.7 \\
             &   & &  \necmpitwoeight & \\
             &   & &  \mvapichonenine &\\
             &   & &  \mvapichtwoone & \\
    \machtwo & \num{436} Dual Xeon 5550 @ \SI{2.66}{\giga\hertz} & \infiniband QLogic 12200 & \intelmpi  &  \gcc~4.4.7 \\
    \machfive &  \num{2000} Dual Xeon E5-2650V2 @ \SI{2.6}{\giga\hertz} &  \infiniband QDR-80 & \intelmpifive  & \gcc~4.4.7 \\
              &   & &  \mvapichtwoa \\
    \machthree & \num{72} Dual Xeon E5520 @ \SI{2.27}{\giga\hertz} & \infiniband QDR MT26428  & \mvapichonenine  & \gcc~4.7.2 \\
    \machfour & \num{64} Dual Xeon E5-2450V2 @ \SI{2.5}{\giga\hertz} & \infiniband Mellanox ConnectX-3 FDR & \intelmpi  & \gcc~4.4.7  \\
    \bottomrule
  \end{tabular}
  \end{footnotesize}
\end{table*}

\subsection{Parallel Machines}

The parallel machines used for conducting our experiments are
summarized in \tab~\ref{tab:machines}.  On the \machone system, we
have dedicated access to the entire cluster. The \machthree system
belongs to Grid'5000\footnote{\url{http://www.grid5000.fr}}, which
features the OAR job scheduler that allows us to gain exclusive access
to a set of nodes connected to the same InfiniBand (\infiniband)
switch. On \machtwo, \machfive, and \machfour, we also made sure that
our allocations include dedicated nodes only. However, we have no
dedicated access to the switches as in the case of the other two
machines.

\section{MPI Process Synchronization Revisited}
\label{sec:mpi_sync_exp}

Now, we turn our attention to the problem of synchronizing MPI
processes, and its implication on performance results. A commonly
employed synchronization method for MPI processes is the use of the
\mpibarrier call. The problem is that the completion of \mpibarrier
only ensures that all processes have called this function, but
processes can still leave the barrier skewed in time. Depending on the
amount of skew and the actual test case, the \runtime of subsequent
MPI calls can fluctuate considerably. For instance, a process skew of
\SI{5}{\micro\second} introduced by calling \mpibarrier has more
effect on an \mpibcast with a message size of \SI{8}{\Bytes} than it
has on an \mpibcast with \SI{8}{\kibi\Bytes}.  Hoefler~\etal also
point out that a call to \mpibarrier can influence the collective
operation being benchmarked when both operations use the same
network~\cite{hoefler-pmeo08}.

In order to prevent such problems with \mpibarrier, several MPI
benchmarks use a window-based process synchronization scheme to ensure
that all processes start calling a given MPI function at the same
time.

We will take a closer look at both synchronization methods and discuss
their advantages and disadvantages.
Then, we will investigate the synchronization quality of
\mpibarrier. Afterwards, we propose a novel clock synchronization
method, which combines features of several competitors, namely
(1)~learning and applying a model of the clock drift and 
(2)~optionally, synchronization in $\mathcal{O}(\log \np)$ steps.  Finally, we will examine 
whether the window-based synchronization methods are competitive 
with \mpibarrier.

\subsection{MPI Process Synchronization in the Wild}
\label{sec:mpi_proc_sync_rel_work}

The clock synchronization algorithm used in \skampi is similar to
Cristian's algorithm~\cite{Cristian:1989vn} and works as follows: one
of the \np MPI processes is selected as the master process with which
all other processes will synchronize. Then, a number of ping-pong
messages is exchanged between process pairs, and the processes' local
time is piggy-backed on a ping-pong message. Using this approach, the
master process can determine the clock differences between itself and
each of the other processes. The clock offset computed at the master
process is later broadcast to the others, which allows all processes
to compute a logical global time. The logical global clock is used to
synchronize processes, as shown in
\algs~\ref{alg:skampi_compute_offset} and~\ref{alg:skampi_time_sync}.

A major drawback of \skampi's approach is that it requires linear time
to synchronize distributed clocks. To speed up the clock
synchronization, Hoefler~\etal implemented a more scalable method,
which only requires a logarithmic number of steps to complete
(depending on the number of processes). It is used in both \nbcbench
and \netgauge~\cite{Hoefler:2010vr}.  Their method outlined in
\algs~\ref{alg:netgauge_time_sync}, \ref{alg:netgauge_compute_offset},
and \ref{alg:netgauge_sync} works as follows: the set of processes is
divided into two groups: one group (\groupone) contains all processes
up to the largest rank that is a power of two, and a second group
(\grouptwo) contains the remaining processes.  The clock
synchronization is done in two phases. In the first one, all processes
in \groupone synchronize their clocks in a tree-like fashion in
$\log \np$ rounds. In the second phase, each remaining process of
\grouptwo synchronizes its clock with one distinct partner process
from \groupone in one additional round.

\alg~\ref{alg:skampi_compute_offset} shows the pseudocode of
\skampi{}'s method to determine the clock offsets between two
processes.  \skampi synchronizes the clock of each process with the
clock of the \rootproc node
in $\mathcal{O}(p)$ steps (\cf procedure
\proctext{Compute\_And\_Set\_Clock\_Offsets} in
\alg~\ref{alg:skampi_time_sync}).  Hoefler~\etal showed how the time to
compute these clock offsets can be reduced by using a tree-like
synchronization process~\cite{Hoefler:2010vr} (see
\algs~\ref{alg:netgauge_time_sync} and~\ref{alg:netgauge_compute_offset}).

The benchmarking methods proposed in \skampi and \nbcbench rely on a
periodic re-adjustment of the window size to cope with \runtimes that
are too large to fit into the synchronization window. Further, they
implement a minimal re-synchronization mechanism that broadcasts a new
point in time used for synchronizing processes for each new experiment
(\eg, for a different message size to be benchmarked). However,
neither \skampi nor \nbcbench implement a re-synchronization of the
distributed clocks to counterbalance the clock drift, \cf
\algs~\ref{alg:skampi_bench} and~\ref{alg:nbc_bench}.

Jones and Koenig proposed a clock synchronization method that takes
the clock drift between distributed processes into
account~\cite{Jones:2012dn}. Their method is based on the assumption
that the clock drift is linear in time. Each process learns a linear
model of the clock drift by exchanging ping-pong messages with a
single reference process.  After computing the linear model of the
clock drift (using a linear regression), each process can determine a
logical global time by adjusting its local time relative to the time
of the reference process. In the ping-pong phase, local times are
exchanged between each process and the reference process. When
processes receive a local timestamp from the reference process, some
time has already passed, which is the time for transferring the
timestamp message.  To account for this delay, the received timestamps
are corrected by half of the mean round-trip time (RTT).
Jones and Koenig do not further detail how the RTT is
obtained, even though the estimation of the RTT could be a source of
error. 

\subsection{Sources of Error for Clock Synchronization}

In \Sec~\ref{sec:win_based_scheme_intro}, we have introduced the
window-based synchronization scheme to measure the \runtime of one MPI
function call.  As this scheme requires synchronized, distributed
clocks, we now show some pitfalls when applying this scheme.

In the context of MPI benchmarking, Hoefler~\etal have shown that two
processor clocks are linearly drifting over
time~\cite{Hoefler:2010vr}. We re-conducted their experiment to
examine the clock drift on our current machines, but using a finer
resolution than what was done in~\cite{Hoefler:2010vr} (we only
measure in the range of seconds instead of hours).
\begin{figure}[t]
  \centering
  \includegraphics[width=.8\linewidth]{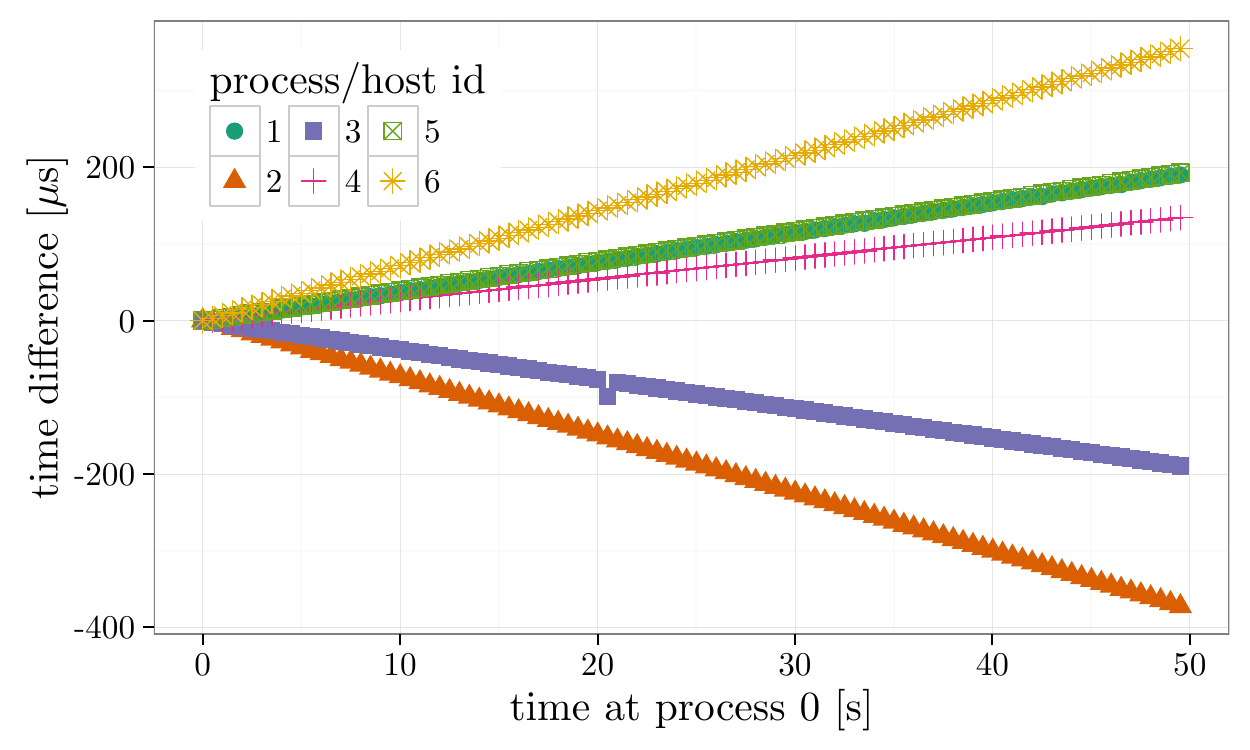}
  \caption{\label{fig:clock_skew_jupiter}Clock drift between a
    reference host and six other hosts on \machone (\expdesc:
    \append~\ref{sec:exp_clock_drift}).}
\end{figure}
\fig~\ref{fig:clock_skew_jupiter} shows that the maximum clock drift
between two hosts of our cluster is about \SI{700}{\micro\second}
($\lvert\SI{-400}{\micro\second}\rvert\SI[retain-explicit-plus]{+300}{\micro\second}$) 
after \SI{50}{\second}. Thus, not accounting for the clock drift will lead 
to highly inaccurate window-based measurements, with a drift error 
in the range of microseconds after only a few seconds of conducting 
measurements.
Hence, a window-based scheme must precisely deal with both clock
offset and clock drift. 

\subsubsection{The Error of the Frequency Estimation}

As mentioned in \Sec~\ref{sec:rdtsc_measure}, \netgauge supports
reading the TSC registers to obtain fine-grained timers. In order to
convert elapsed clock ticks into seconds, the update frequency of the
TSC registers is required. \netgauge applies the following method to
estimate this frequency: the estimation routine sleeps for a fixed
amount of time, in case of \netgauge for \SI{1}{\second}, and measures 
the number of ticks elapsed in this period. The process is repeated 
until a minimum number of ticks (given a search time threshold) has 
been recorded for the fixed-size sleeping period. In this way, 
\netgauge is able to estimate the update frequency of the TSC register.

\begin{figure}[t]
  \centering
  \includegraphics[width=.8\linewidth]{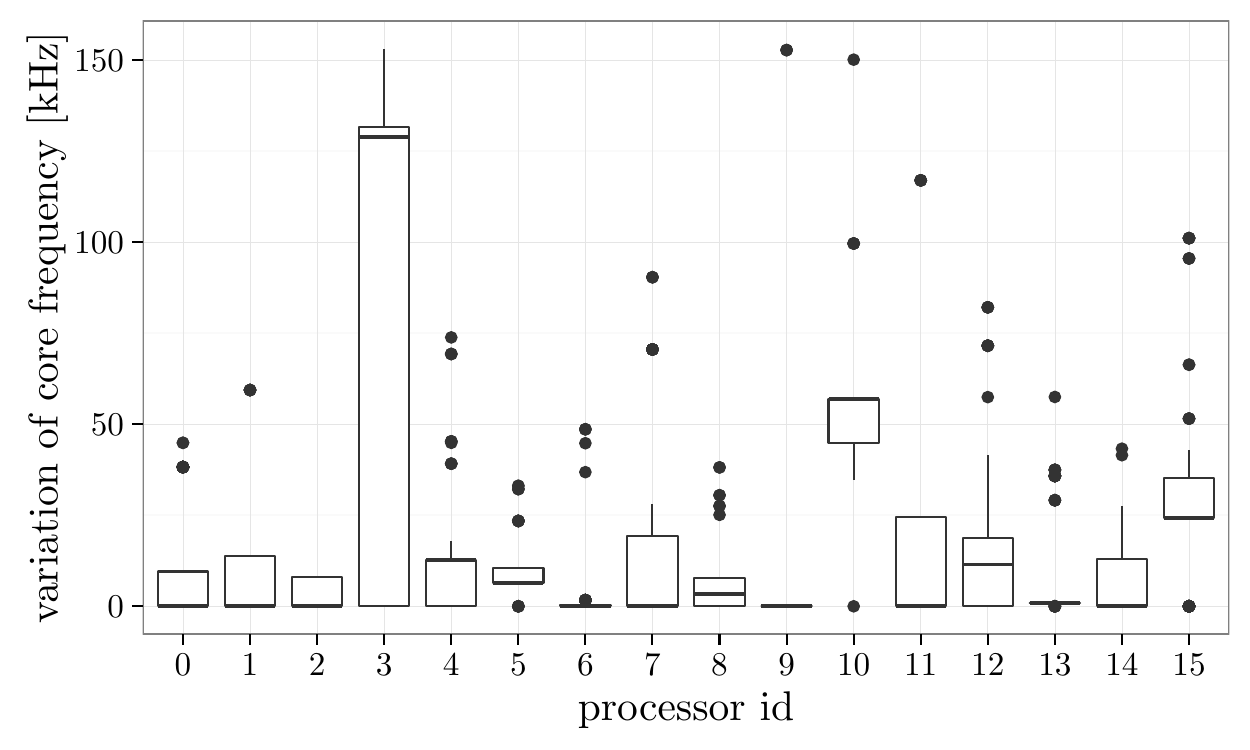}
  \caption{\label{fig:core_freq_diff}Differences of clock speeds
    obtained on \num{16}~nodes (processes) for \num{100}~calls to \hrtcal of
    \netgauge (\machone, \expdesc: \append~\ref{sec:exp_freq_calibration}).}
\end{figure}

We were interested in evaluating how accurate \netgauge{}'s method for
estimating the CPU frequency is, and whether it would be beneficial to
use it for our benchmarking purposes. We conducted an experiment (see
\alg~\ref{exp:freq_calibration}), in which we called \netgauge{}'s
frequency estimation macro (\hrtcal) \num{100}~times on \num{16} different nodes
(each clocked at \SI{2.3}{\giga\hertz}), and recorded the estimated frequency on
each core. \fig~\ref{fig:core_freq_diff} shows the frequency variation
obtained on each of the \num{16}~nodes. The results indicate that for most
nodes the estimated frequency variation lies in the range of
\SIrange{10}{20}{\kilo\hertz}, which at first glance suggests that such an estimation
method would be reproducible. However, if we analyze the error of this
frequency estimation, we obtain a variation of roughly \SI{10}{\kilo\hertz} for all
cores (which translates to \SI{d-5}{\giga\hertz}). That means, if we assume
that a processor runs at a fixed clock frequency of \SI{2.3}{\giga\hertz}, then the
error is $\frac{10^{-5}}{2.3} \approx 4.3 \cdot 10^{-6}$.
Consequently, applying this frequency estimation method in the context
of MPI benchmarking results in an inherent timing error of \SI{1}{\micro\second}
per second.

\begin{figure}[t]
  \centering
  \includegraphics[width=.8\linewidth]{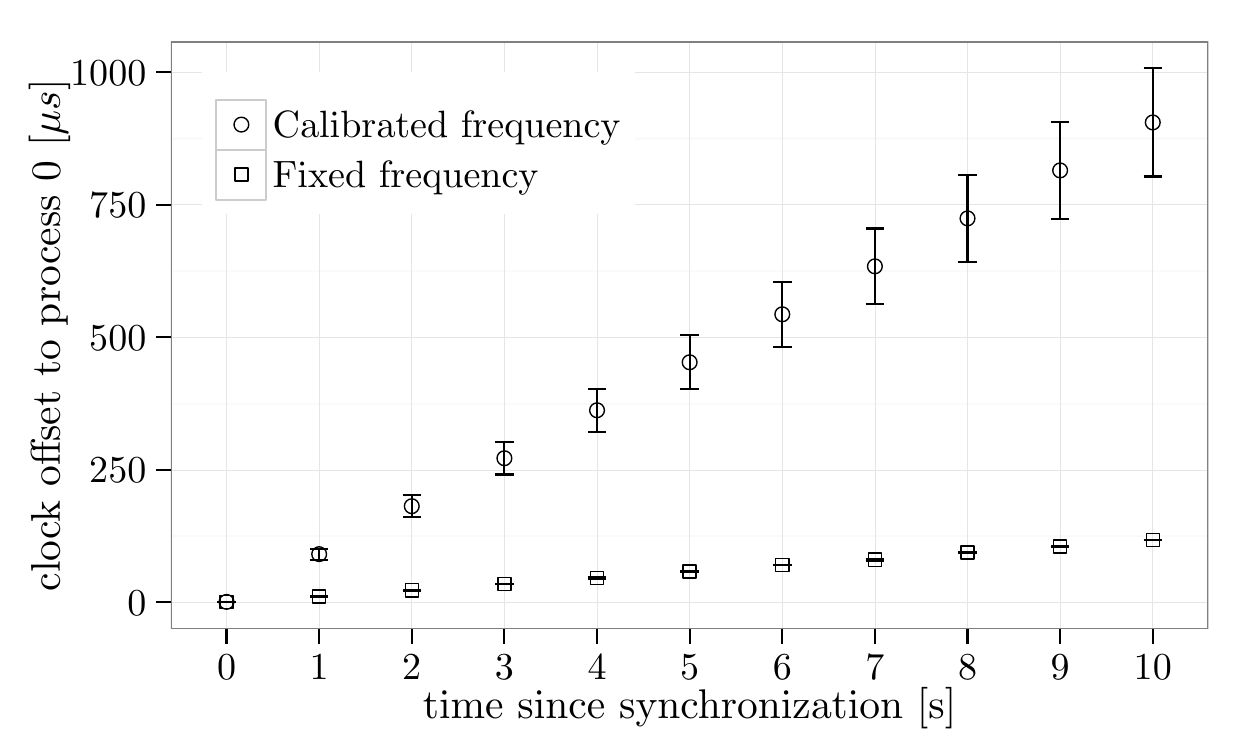}
  \caption{\label{fig:netgauge_drift}Mean clock drift and \SI{95}{\percent} confidence
    intervals for the \netgauge synchronization
    (\num{16 x 1}~processes,  \num{10}~calls to \mpirun, \mvapichtwoone, 
    \machone, \expdesc: \append~\ref{sec:exp:clock_drift_in_time}).}
\end{figure}
We also examined whether this error could be reduced when using a
fixed clock frequency to convert clock ticks obtained from
\rdtsc/\rdtscp into seconds. We therefore compared the clock drift of
\netgauge using its original frequency estimation to the clock drift
with a fixed frequency (here we fixed it to \SI{2.3}{\giga\hertz}). 
\fig~\ref{fig:netgauge_drift} compares the mean clock
drifts over \num{10}~\mpirun calls, which were measured for both 
frequency estimation methods (\cf \alg~\ref{exp:clock_drift_in_time}).
Surprisingly, the clock drifts are significantly different depending
on whether we estimate the clock frequency or not. According to these
results, the clock drift after \SI{10}{\second} using \netgauge{}'s frequency
estimation is almost \num{10}~times bigger than when we use a fixed
frequency.  As a consequence, we have decided to use a fixed clock
frequency when converting the clock ticks (obtained from \rdtsc) into
seconds for our window-based MPI benchmarking scheme.
\begin{figure}[t]
  \centering
  \includegraphics[width=.9\linewidth]{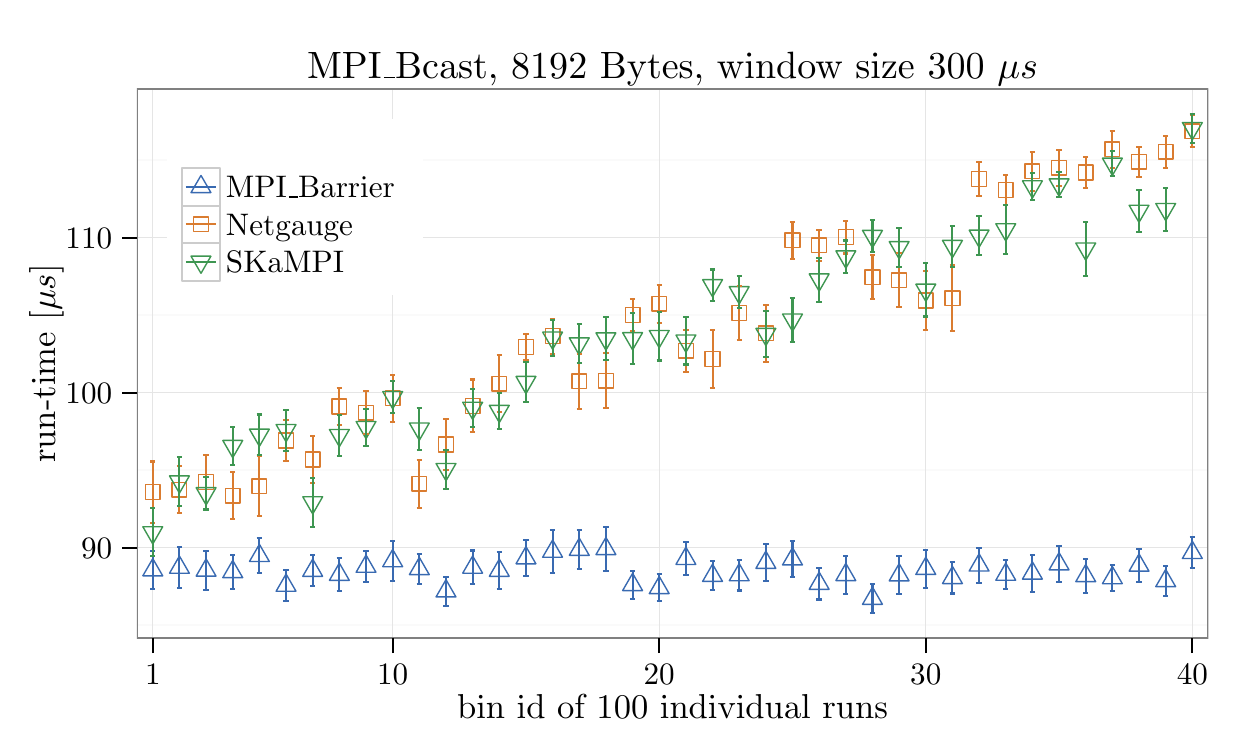}
  \caption{\label{fig:runtime_increase_sync}Drifting \runtimes of
    \mpibcast (mean~+~CI) using \netgauge and \skampi synchronization
    methods compared to the results obtained with \mpibarrier
    (\SI{8192}{\Bytes}, \num{32 x 16}~processes,
    \num{4000}~measurements, \num{1} call to \mpirun, 
    \mvapichtwoone, \machone, \expdesc:
    \append~\ref{sec:exp:clock_drift_ng_skampi}).}
\end{figure}
\subsubsection{The Error of Accounting for the Clock Offset only}

Now, we examine how accurately the clock synchronization schemes of
\netgauge and \skampi work in practice.  We designed an experiment
that measures the individual \runtimes of \num{4000}~consecutive calls to
\mpibcast using \num{512}~processes (distributed over
\num{32}~compute~nodes). The process synchronization between calls to
\mpibcast is either done by using an \mpibarrier call or by applying
the window-based schemes implemented in \netgauge or \skampi with a
fixed window size.  \fig~\ref{fig:runtime_increase_sync} shows the
development of the mean \runtime of \mpibcast over time. For
presentation purposes, we binned every group of \num{100} consecutive,
individual measurements and only plotted the bin means and their
confidence intervals. As expected, the \runtime of
\mpibcast stays relatively stable when synchronizing with \mpibarrier
(as this process synchronization method is independent of the
clock). However, the binned \runtimes increase over time when a
window-based scheme is applied. The underlying problem is that neither
\netgauge nor \skampi consider the clock drift when synchronizing
clocks. Instead, both benchmarks ``only'' determine the clock offset
between processes.

In conclusion, we contend that clock synchronization schemes need to
consider the clock drift when computing the logical global clock.

\subsection{Accounting for the Clock Drift and Offset}

\jonesk propose a method to synchronize distributed clocks while
considering the clock drift and the clock
offset~\cite{Jones:2012dn}. Their goal was to have accurately
synchronized clocks for implementing co-scheduling algorithms of
parallel applications and tracing functionality in the MPI domain.

\alg~\ref{alg:jk_clock_drift} presents the clock synchronization
method proposed by \jonesk. The idea is to learn a linear model of the
clock drift between a process and the reference process.  The
algorithm of \jonesk relies on two parameters. \nfitpoints specifies how 
many points (a \emph{fitpoint} is a tuple containing a reference clock 
timestamp and a clock offset) will be recorded as input for the linear 
regression analysis. 
\nexchanges denotes the number of ping-pong messages exchanged between
a pair of processes to obtain a single \emph{fitpoint}.  The entire method
works as follows: a process exchanges \nexchanges ping-pong messages
with the master process, where each message carries the local time of
a process. The remote process computes the time difference between its
local time and the local time of the reference process. Since these
measured clock offsets have a relatively large variance on many
networks, the process only keeps one of these (\nexchanges many) time
differences by selecting the median value. Then, this ping-pong
exchange phase between a process and the master (reference) process is
repeated \nfitpoints times. Finally, each remote process computes its
model of the clock drift relative to the master process by fitting a
linear regression model to the recorded clock offsets (time
differences).

In addition, since each of the measured clock offsets needs to be
corrected by $\rtt/2$ to account for the network latency, we present
our method for estimating the RTT between two processes
in~\alg~\ref{alg:hca_mean_rtt}.

Once the clock drift model is established, each measured local time
can be normalized to the reference time of the master process by
applying a drift correction, as shown in
\alg~\ref{alg:hca_normalize_time}.

\subsection{HCA Algorithm for Clock Synchronization}

\begin{algorithm}[t]
  \caption{\label{alg:hca_alg} \synchca clock synchronization.}
  \begin{scriptsize}
  \begin{algorithmic}[1] 
    \Statex \np\ - number of processes
    \Statex $r$ - current process rank ($0$ to $\np - 1$)
    \Statex \lm\ - linear model of the current process (defined by a \textit{slope} 
    \Statex \hspace*{3pt} and an \textit{intercept}) to adjust the local clock to the reference 
    \Statex \hspace*{3pt} time of \rootproc
    \Statex \modellist\ - array of \np linear models
    \Statex \modellist[0] = (0, 0) \ \mycomment{reference clock} 
    \Statex \hierintercepts\ -  if defined, compute intercepts 
    \Statex \hspace*{3pt} hierarchically (instead of directly between each $r$ and \rootproc)
    \Statex \startwin\ - next window start time, updated after each sync
    \Statex \initialts\ - local timestamp used to adjust the local clock to 
    \Statex \hspace*{3pt} the time $0$ of the synchronization start
    \Statex \texttt{maxpower} $= 2^{\lfloor log_2p \rfloor} $
    \Statex    
     \Procedure{Sync\_Clocks}{\nfitpoints, \nexchanges} 
        \State $\initialts = \Call{Get\_Time}{} $
        \Statex \mycomment{compute linear models of each clock's drift relative to \rootproc}
        \State \Call{Sync\_Clocks\_Pow2}{\nfitpoints, \nexchanges}
        \State \Call{Sync\_Clocks\_Remaining}{\nfitpoints, \nexchanges}                       
        \Statex \mycomment{send final linear models from \rootproc to each process}
        \State \Call{MPI\_Scatter}{\modellist, $1$, \modeltype,  \lm, $1$, \modeltype, \rootproc} 
        \Statex \Ifndef {\hierintercepts}
              \State \Call{Compute\_And\_Set\_All\_Intercepts}{\lm}
        \Statex \EndIfndef  
        \State \Call{MPI\_Barrier}{}

        \State $\startwin = \Call{Get\_Adjusted\_Time}{} + \winsize$
        \State  \Call{MPI\_Bcast}{\startwin, $1$, \mpidouble, \rootproc}
    \EndProcedure      
  \end{algorithmic}
\end{scriptsize}
\end{algorithm}

\begin{algorithm}[h!t]
  \caption{\label{alg:hca_clock_drift} Hierarchical linear models of the clock drift.}
  \begin{scriptsize}
  \begin{algorithmic}[1]     
   \Function{Get\_Adjusted\_Time}{} \label{func:get_adjusted_time}
     \State \textbf{return} \Call{Get\_Time}{} - \initialts   
   \EndFunction
   \Statex

   \Procedure{Sync\_Clocks\_Pow2}{\nfitpoints, \nexchanges}
       \Statex \mycomment {compute linear models of the clock drifts for processes with}
        \Statex \mycomment {indices between $0$ and ($\texttt{maxpower} - 1$)}  
       \State $\currentround = 1$
       
       \If {$r \ge \texttt{maxpower}$} \textbf{return}
       \EndIf
       \While {$2^{\currentround} \le \texttt{maxpower}$}
           \If {$(r \bmod 2^{\currentround}) == 0$} \hfill  \mycomment{process with reference clock}
               \State $\client = r + 2^{\currentround-1}$ 
               \State $\rtt =$ \Call{Compute\_Rtt}{$r$, \client}
               \State \Call{Learn\_Model\_HCA}{\nfitpoints, \nexchanges, \rtt, \textit{r}, \client}
	       \Statex \Ifdef {\hierintercepts}
               \State \Call{Compute\_And\_Set\_Intercept}{\nullval, \client, $r$}
	       \Statex \EndIfdef

               \Statex \mycomment {receive linear models collected by the client}
               \State \Call{MPI\_Recv}{\mlistclient, $2^{\currentround-1}$, \modeltype, \client}
		
	       \State $\modellist[\client] = \mlistclient[0]$ \hspace*{45pt}  \mycomment{save client model}

               \For{$i$ in $1$ to $(2^{\currentround-1} - 1)$} \hspace*{20pt}  \mycomment {compute resulting models}
 	           \State $\modellist[\client+i] =  $ \Call{Merge\_LMs}{$\modellist[\client]$, $\mlistclient[i]$}
 	       \EndFor
          \ElsIf {$(r \bmod 2^{\currentround}) == 2^{\currentround-1}$} \hfill  \mycomment{client}
              \State $\pref = r - 2^{\currentround-1}$  
              \State $\rtt =$ \Call{Compute\_Rtt}{\pref, $r$}      
              \State $\modellist[r] = $ \Call{Learn\_Model\_HCA}{\nfitpoints, \nexchanges, 
              \Statex  \hspace*{137pt} \rtt, $\pref, \textit{r}}$
	      \Statex \Ifdef {\textit{\hierintercepts}}
		      \State \Call{Compute\_And\_Set\_Intercept}{\modellist[$r$], $r$, \pref}
              \Statex \EndIfdef

              \Statex \mycomment {send all new linear models to the reference process}
              \State \Call{MPI\_Send}{$\modellist[r]$, $2^{\currentround-1}$, \modeltype, \pref}
          \EndIf
          \State $\currentround = \currentround + 1$
      \EndWhile
    \EndProcedure
    \Statex

    \Procedure{Sync\_Clocks\_Remaining}{\nfitpoints, \nexchanges}
        \Statex \mycomment{compute linear models of the clock drifts for processes with }
         \Statex \mycomment{indices between \texttt{maxpower} and ($\np - 1$)}      

       \If {$\texttt{maxpower} == \np$} \textbf{return}
       \EndIf

        \If {$r < \np - \texttt{maxpower}$} \hfill \mycomment{process with reference clock}
            \State $\client = r+\texttt{maxpower}$
            \State $\rtt =$ \Call{Compute\_Rtt}{$r$, \client}
            \State \Call{Learn\_Model\_HCA}{\nfitpoints, \nexchanges, \rtt, \textit{r}, \client}
	    \Statex \Ifdef {\hierintercepts}
	              \State \Call{Compute\_And\_Set\_Intercept}{\nullval, \client, $r$}
	     \Statex \EndIfdef
        \ElsIf  {$r \ge \texttt{maxpower}$} \hfill \mycomment{client}
            \State $\pref = r - \texttt{maxpower}$
            \State $\rtt =$ \Call{Compute\_Rtt}{\pref, $r$}
            \State $\lm = $ \Call{Learn\_Model\_HCA}{\nfitpoints, \nexchanges, 
            \Statex  \hspace*{111pt} \rtt, \pref, \textit{r}}
            \Statex \Ifdef {\textit{\hierintercepts}}
		      \State \Call{Compute\_And\_Set\_Intercept}{\lm, $r$, \pref}
            \Statex \EndIfdef
        \EndIf                
        
        \State $\textit{sub\_comm} =$ create communicator comprising process ranks
        \Statex \hspace*{55pt} $(0, \texttt{maxpower}, \texttt{maxpower}+1, \ldots, \np-1)$
        \If {$r == \rootproc$}
            \State \Call{MPI\_Gather}{$\lm$, $1$, \modeltype, 
            \State \hspace*{48pt}$\tmpmodel$, $1$, $\modeltype$, \rootproc, $\textit{sub\_comm}$}
            \For{$j$ in $0$ to ($\np - \texttt{maxpower} - 1)$}
                \State $q = \texttt{maxpower} + j$
                \State $\modellist[\textit{q}] =  $ \Call{Merge\_LMs}{$\modellist[j]$, 
                            $\tmpmodel[j+1]$}
            \EndFor
        \ElsIf {$r \ge \texttt{maxpower}$}
            \State \Call{MPI\_Gather}{$\lm$, $1$, \modeltype, 
            \State \hspace*{48pt}$\tmpmodel$, $1$, $\modeltype$, \rootproc, $\textit{sub\_comm}$}
        \EndIf        
    \EndProcedure     

    \Statex
    \Procedure{Compute\_And\_Set\_All\_Intercepts}{\textit{lm}} \label{func:compute_all_intercepts}
        \Statex \mycomment{compute intercepts for the model \textit{lm} of the current process $r$}
          \If {$r \ne \rootproc$}
             \State \Call{Compute\_And\_Set\_Intercept}{\textit{lm}, $r$, \rootproc}
         \Else
         \For{$i$ in $0$ to ($\np - 1$)  \textbf{s.t.}\ $i \ne \rootproc$ }
              \State \Call{Compute\_And\_Set\_Intercept}{\textit{lm}, $i$, \rootproc}
         \EndFor
        \EndIf  
    \EndProcedure    
  \end{algorithmic}
\end{scriptsize}
\end{algorithm}
\begin{algorithm}[t]
  \caption{\label{alg:clock_drift_2procs} Clock drift model for a pair of processes.}
  \begin{scriptsize}
  \begin{algorithmic}[1] 
     \Function{Learn\_Model\_Hca}{}(\nfitpoints, \nexchanges, \rtt, \textit{p1}, \textit{p2})
        \State $\textit{slope} = 0, \textit{intercept} = 0$
        \If {$\myrank == \textit{p1}$} \hfill \mycomment{process with reference clock}
            \For{\textit{idx} in 0 to $\nfitpoints - 1$}
                    \For{$i$ in 0 to $\nexchanges - 1$}
                        \State  \Call{MPI\_Recv}{\textit{tdummy}, $1$, \mpidouble, \textit{p2}}    
                        \State $\tremote =$ \Call{Get\_Adjusted\_Time}{}
                        \State \Call{MPI\_Send}{\tremote, $1$, \mpidouble, \textit{p2}}
                \EndFor
            \EndFor      
        \ElsIf {$\myrank == \textit{p2} $} \hfill \mycomment{client process}
            \For{\textit{idx} in 0 to $\nfitpoints - 1$}
                \For{$i$ in 0 to $\nexchanges - 1$}
                    \State  \Call{MPI\_Send}{\textit{tdummy}, $1$, \mpidouble, \textit{p1}}
                    \State  \Call{MPI\_Recv}{\tremote, $1$, \mpidouble, \textit{p1}}
                    \State $\textit{local\_times}[i] =$ \Call{Get\_Adjusted\_Time}{}
	            \State $\difflist[i] = \textit{local\_times}[i] - \tremote - \rtt/2$
	        \EndFor
	        \State $\difflist = $ \Call{Sort}{\difflist}
	        \State $\textit{yfit}[\textit{idx}] = $ \Call{Compute\_Median}{\difflist}
	         \State $\textit{idx\_median} = i \ \text{s.t.}\  (0 \le i < \nexchanges \ \And\ $
	         \Statex \hspace*{97pt} $\difflist[i] == \textit{yfit}[\textit{idx}])$
	        \State $\textit{xfit}[\textit{idx}] = \textit{local\_times}[\textit{idx\_median}] $	        
           \EndFor
           \State $\textit{slope}, \textit{intercept} = $ \Call{Linear\_Fit}{\textit{xfit}, \textit{yfit}, \nfitpoints}
        \EndIf
    
    \State \textbf{return} \Call{New\_LM}{\textit{slope}, $\textit{intercept}$}
    \EndFunction    
    \Statex

    \Procedure{Compute\_And\_Set\_Intercept}{}($\textit{lm}, \client, \pref$)\label{func:compute_intercept}
        \Statex \mycomment {compute the intercept using the SKaMPI method}      
          \If {$r == \client$}
             \State $\diff = \Call{SKaMPI\_PingPong}{\client, \pref}$
             \State $\textit{diff\_timestamp} = \Call{Get\_Adjusted\_Time}{}$
            \State $\textit{lm.intercept} = \textit{lm.slope} \cdot (-\textit{diff\_timestamp}) + \diff$
         \ElsIf{$r == \pref$}
	     \State$\diff = \Call{SKaMPI\_PingPong}{\client, \pref}$
        \EndIf  
    \EndProcedure   
    
    \Statex
   \Function{Merge\_LMs}{\textit{lm1}, \textit{lm2}} \label{func:merge_models}
       \State $\textit{new\_lm.intercept} = \textit{lm1.intercept} + \textit{lm2.intercept} -\ \textit{lm2.intercept} \cdot \textit{lm1.slope}  $
       \State $\textit{new\_lm.slope} = \textit{lm1.slope} + \textit{lm2.slope} - \textit{lm1.slope} \cdot \textit{lm2.slope} $
       \State \textbf{return} \textit{new\_lm}
   \EndFunction

  \end{algorithmic}
\end{scriptsize}
\end{algorithm}

We want to combine the advantages of the synchronization algorithm
developed by \jonesk (\syncjk) with the synchronization scheme applied
by \netgauge. We propose a novel algorithm that synchronizes
distributed clocks in a hierarchical way, but also takes the clock
drift into account.
\jonesk already noted in their article that they had chosen a
$\mathcal{O}(\np)$ scheme for better accuracy, ``whereas a balanced
$\mathcal{O}(\log{\np})$ scheme may complete in milliseconds with higher
variance (owing to the multiple reference
stratums)''~\cite{Jones:2012dn}.  We still want to explore the
possibility of applying such a $\mathcal{O}(\log{\np})$ scheme to improve the
scalability of the original algorithm of \jonesk.

\alg~\ref{alg:hca_clock_drift} shows the pseudocode of our novel
algorithm called \synchca (for \textbf{H}unold and
\textbf{C}arpen-\textbf{A}marie).  The computational structure of
\synchca works similarly to the algorithm described by
Hoefler~\etal~\cite{Hoefler:2010vr}.  The difference, however, is that
instead of determining only the clock offset of each process relative
to the \rootproc, \synchca computes a linear model of the clock drift
of each process to obtain a global clock.

The algorithm synchronizes clocks in two steps.  In the first step,
the clock drifts of processes with ranks smaller than the largest
power of two ($0, \ldots, 2^{\lfloor \log_2{p} \rfloor}-1$) are
estimated in function \proctext{Sync\_Clocks\_Pow2} of
\alg~\ref{alg:hca_clock_drift}.  Then, in the second step, the
remaining processes with larger ranks
($\ge 2^{\lfloor \log_2{p} \rfloor}$) compute their linear models of the
clock drift with respect to the already synchronized processes in one
additional round (\cf \proctext{Sync\_Clocks\_Remaining} function).

The major difference to the synchronization method found in \netgauge
is the call to \proctext{Learn\_Model\_HCA}{}, which determines the
model of the clock drift between two processes
(\alg~\ref{alg:clock_drift_2procs}). The parameters \nfitpoints and
\nexchanges play the same role as in the algorithm of \jonesk.
Further, we use the same RTT estimation
function as we did for the \syncjk synchronization, which is presented
\alg~\ref{alg:hca_mean_rtt}.

In the case of \netgauge, intermediate clock offsets are summed up in
a tree-like fashion to compute the offset of each process relative to
the reference \rootproc node.  To similarly build linear models of the
clock drift, we needed to solve the problem of combining linear
regression models.  More formally, let us assume we have three
processes located on different hosts called $p_1$, $p_2$, and $p_3$,
such that each process has its own clock.  If the clocks of hosts
$p_1$ and $p_2$ have an offset of $\diff_{p_1, p_2}$, and the clocks
of $p_2$ and $p_3$ have an offset of $\diff_{p_2, p_3}$, the clock
offset between $p_1$ and $p_3$ can be computed as
$\diff_{p_1, p_3} = \diff_{p_1, p_2} + \diff_{p_2, p_3}$.

Therefore, we apply a similar transitive computation to combine linear
regression models to obtain the clock drift between different
processes. If the clock drifts are computed in one round for process
pairs $(p_1, p_2)$ and $(p_2, p_3)$, the question 
becomes: how should these two linear
models be combined such that $p_3$ can obtain its clock drift with
respect to $p_1$?

Let us denote the model of the clock drift of $p_2$ relative to $p_1$
as
$t^{2 \rightarrow 1}(t_1) = t_1 - t_2 = s^{2 \rightarrow 1} t_1 + i^{2
  \rightarrow 1}$.
Similarly, the clock drift of $p_3$ relative to $p_2$ is given as
$t^{3 \rightarrow 2}(t_2) = t_2 - t_3 = s^{3 \rightarrow 2} t_2 + i^{3
  \rightarrow 2}$.
The computation of the clock drift between $p_3$
and $p_1$ is shown in \Equ~\ref{eq:hca_comb_models} and implemented in
\proctext{Merge\_LMs} (\cf line~\ref{func:merge_models} of
\alg~\ref{alg:clock_drift_2procs}).

\begin{small}
\begin{equation}
\label{eq:hca_comb_models}
\begin{split}
t^{3 \rightarrow 1}(t_1)  = &\ s^{3 \rightarrow 1} t_1 + i^{3 \rightarrow 1} \\
   = &\  t_1 - t_3 \\
  = &\ s^{2 \rightarrow 1} t_1 + i^{2 \rightarrow 1} \\
  & + s^{3 \rightarrow 2} t_2 + i^{3 \rightarrow 2} \\
  = &\ s^{2 \rightarrow 1} t_1 + i^{2 \rightarrow 1} \\
  & + s^{3 \rightarrow 2} ( t_1 - s^{2 \rightarrow  1} t_1 - i^{2 \rightarrow 1} ) + i^{3 \rightarrow 2} \\
  = &\ t_1 (  s^{2 \rightarrow 1} + s^{3 \rightarrow 2} - s^{2 \rightarrow  1} s^{3 \rightarrow 2} ) + i^{2 \rightarrow 1} \\
  & - s^{3 \rightarrow 2} i^{2 \rightarrow 1} + i^{3 \rightarrow 2} \ .\\
\end{split}
\end{equation}
\end{small}

To estimate the error of the computed linear model of the clock offset
as a function of time, we conducted a statistical analysis for each
pair of processes. We performed an experiment in which, for 
\num{15}~pairs of processes running on different nodes, we measured
\num{1000}~fitpoints. We estimated the confidence intervals of both
the slope and the intercept of each process pair.  In the case of the
slope, the length of the confidence interval is at most \num{2d-8}, whereas
the intercept computation revealed much wider confidence intervals, in
the order of \SI{100}{\milli\second}. 
The consequence 
of these larger intervals is that the intercept computed with a linear
regression analysis will decrease the accuracy of the initial clock
offset with a high probability. Thus, the global clock error will 
increase over time.

To minimize the impact of the intercept error, we do not use
the intercepts computed with a linear regression analysis.
Instead, we have explored two approaches that appeared to be promising
for computing the intercepts of the clock drift.
Both approaches rely on \skampi's
method for determining the clock offset between two processes at a
given point in time (\proctext{SKaMPI\_PingPong}).  The intercepts can be obtained by 
measuring the clock offset between two processes and then using the
already computed slope to find the intercept of the linear clock model
(\proctext{Compute\_And\_Set\_Intercept} of \alg~\ref{alg:clock_drift_2procs}).
The reason why we have selected the \skampi method for computing the
offset is that it provided us with the lowest initial clock offset
values, as it will be shown in the first experiment in
\Sec~\ref{sec:sync_methods_exp}.

The \emph{first approach} is to compute the intercepts in $\mathcal{O}(\np)$
rounds after completing the hierarchical computation of the clock
models, which only requires $\mathcal{O}(\log{\np})$ rounds.   
We employ \skampi's clock synchronization to measure the clock offset 
between the \rootproc and each of the other $\np-1$ processes as shown in
function~\proctext{Compute\_And\_Set\_All\_Intercepts}.
The advantage is that the intercept is measured for each clock model
separately. Thus, the intercept error only depends on the accuracy of
a single \skampi synchronization and on the error of the slope, which
was found to be very small (\num{d-8}).

The \emph{second approach} is to compute the intercepts during the
hierarchical computation of the clock model in
$\mathcal{O}(\log{\np})$ rounds.  This algorithmic option is enabled
by defining the global variable \hierintercepts.  In this case, we
measure the clock offset and compute the new intercept by adjusting
the offset using the clock model. Then, the intercept obtained from
the linear regression is replaced with this new intercept.  Here, the
\skampi method is used to measure the clock offset for a pair of
processes in each round.  In order to compute the clock model between
each process and the \rootproc, the linear models are combined
hierarchically using \Equ~(\ref{eq:hca_comb_models}).  The advantage
of this method compared to the first approach is its better
scalability. The downside is that relying on a combined intercept for
the linear model increases the error of the logical global clock.

\begin{figure}[t]
  \centering    
    \includegraphics[width=\linewidth]{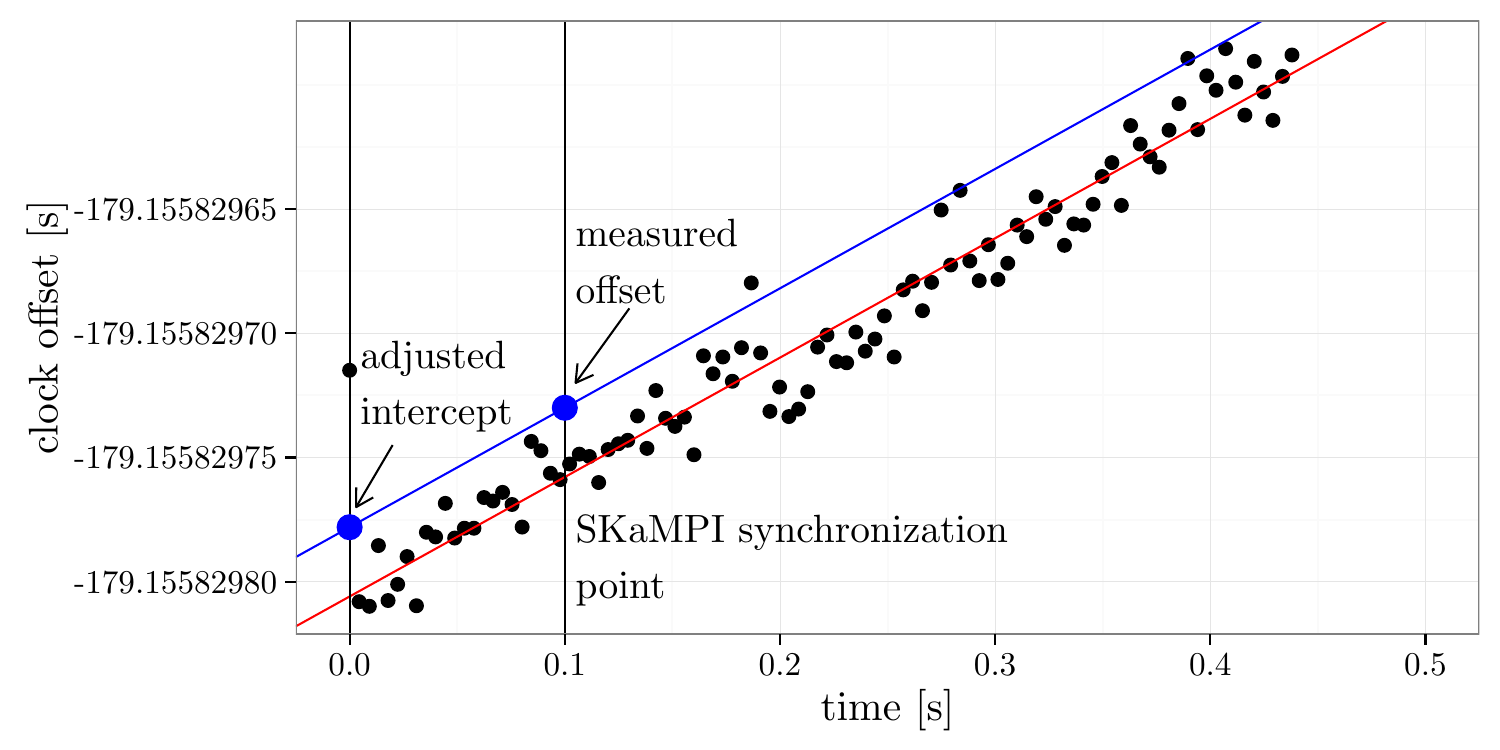}
    \caption{\label{fig:hca_skampi_example}Adjusting the intercept of
      the linear model using \skampi clock synchronization method
      after learning the slope of the clock drift.}
\end{figure}

The intercept used in the linear models represents the clock offset at
time zero. However, the clocks provided by the system (\rdtsc,
\mpiwtime) can start at arbitrary values.  For this reason, we use a
logical local clock that starts with value zero (by subtracting the
initially found timestamp, \cf line~\ref{func:get_adjusted_time} of
\alg~\ref{alg:hca_clock_drift}).  \fig~\ref{fig:hca_skampi_example}
depicts our method for obtaining accurate linear models of the
drift. First, the fitpoints are measured using timestamps that are
adjusted to the initial local time.  Next, the slope and the
(temporary) intercept are computed using these fitpoints. Then, we
re-measure the clock offset at a given point in time (called
synchronization point) by applying the \skampi approach. Finally, the
\emph{adjusted} (and therefore final) intercept is computed based on
the slope and the measured clock offset.

We would like to point out that \synchca should be considered as a
general framework to synchronize clocks. In the present paper, we have
used the method of \jonesk to compute the clock drift model and
\skampi's method to improve the accuracy of the model
intercept. However, the concrete implementations of (1) how to obtain
the linear model or (2) how to measure the clock offsets can be modified by
substituting the functions \proctext{Learn\_Model\_HCA} and
\proctext{Compute\_And\_Set\_Intercept}, respectively.

We now need to estimate the errors introduced by hierarchically
combining linear models using \synchca. Let
$\left[ \underline{s}^{2 \rightarrow 1}, \overline{s}^{2 \rightarrow
    1} \right]$
be the confidence interval of the slope (of the clock drift) of
process $p_2$ relative to $p_1$, and
$\left[ \underline{i}^{2 \rightarrow 1}, \overline{i}^{2 \rightarrow 1} \right]$
 the confidence interval of the corresponding intercept.
The impact of merging
two linear models (according to \Equ~\ref{eq:hca_comb_models}) on the
resulting slope and intercept can be computed as follows:
\begin{footnotesize}
\begin{equation}
\label{eq:hca_comb_model_errors}
\begin{split}
 \underline{s}^{3 \rightarrow 1}  =  &\ \underline{s}^{2 \rightarrow 1} + \underline{s}^{3 \rightarrow 2} \\ 
 		&\ - \max(\underline{s}^{2 \rightarrow 1} \underline{s}^{3 \rightarrow 2},
			\underline{s}^{2 \rightarrow 1}  \overline{s}^{3 \rightarrow 2},
			\overline{s}^{2 \rightarrow 1} \underline{s}^{3 \rightarrow 2}, 
			\overline{s}^{2 \rightarrow 1} \overline{s}^{3 \rightarrow 2}
		) \\
 \overline{s}^{3 \rightarrow 1}  =  &\ \overline{s}^{2 \rightarrow 1} + \overline{s}^{3 \rightarrow 2} \\ 
 		&\ - \min(\underline{s}^{2 \rightarrow 1} \underline{s}^{3 \rightarrow 2},
			\underline{s}^{2 \rightarrow 1} \overline{s}^{3 \rightarrow 2}, 
			\overline{s}^{2 \rightarrow 1} \underline{s}^{3 \rightarrow 2},
			\overline{s}^{2 \rightarrow 1} \overline{s}^{3 \rightarrow 2}
		) \ .\\
  \underline{i}^{3 \rightarrow 1}  =  &\ \underline{i}^{2 \rightarrow 1} + \underline{i}^{3 \rightarrow 2} \\ 
  		&\ - \max(\underline{i}^{2 \rightarrow 1} \underline{s}^{3 \rightarrow 2},
         		\underline{i}^{2 \rightarrow 1}  \overline{s}^{3 \rightarrow 2},
         		\overline{i}^{2 \rightarrow 1} \underline{s}^{3 \rightarrow 2}, 
         		\overline{i}^{2 \rightarrow 1} \overline{s}^{3 \rightarrow 2}
         	) \\
  \overline{i}^{3 \rightarrow 1}  =  &\ \overline{i}^{2 \rightarrow 1} + \overline{i}^{3 \rightarrow 2} \\ 
  		&\ - \min(\underline{i}^{2 \rightarrow 1} \underline{s}^{3 \rightarrow 2},
         		\underline{i}^{2 \rightarrow 1} \overline{s}^{3 \rightarrow 2}, 
         		\overline{i}^{2 \rightarrow 1} \underline{s}^{3 \rightarrow 2},
         		\overline{i}^{2 \rightarrow 1} \overline{s}^{3 \rightarrow 2}
         	)
\end{split}
\end{equation}
\end{footnotesize}

In the case of the \emph{first approach} of our algorithm,
we are only interested in the errors of merging slopes, as
the computed (linear regression) intercepts are disregarded 
in favor of clock offsets measured with \skampi in a linear fashion.
The confidence intervals (CIs) of the computed slopes in our
experiments on machine \machone were in the order of \num{d-8}. The CI
of the resulting slope only depends on the initial slopes and
increases linearly with the number of performed merge operations due
to the negligible resulting slope products. Thus, the error of the
slope grows logarithmically in the number of processes when using our
hierarchical way of combining linear regression models.  Consequently,
the merging error will only reach the order of microseconds when the
experiment is conducted on \num[exponent-base=2]{d100} processes.
The \emph{second approach} of the HCA method relies on hierarchically 
combining both intercepts and slopes. In our case, the values of the computed 
slopes are several orders of magnitude smaller than those of the intercepts, 
as the clocks do not steeply drift apart. Thus, when combining intercepts 
according to \Equ~\ref{eq:hca_comb_model_errors}, the sum of the initial 
intercept values will have a major impact on the resulting intercept. 
The confidence interval of the final intercept will consequently increase linearly 
with the number of merge operations applied.

\subsection{Evaluation of Clock Synchronization Methods} 
\label{sec:sync_methods_exp}

\begin{figure*}[t]
  \centering
  \subfigure[1 process per node]%
  {
    \label{fig:clock_drift_ppn1}
    \includegraphics[width=.45\linewidth]{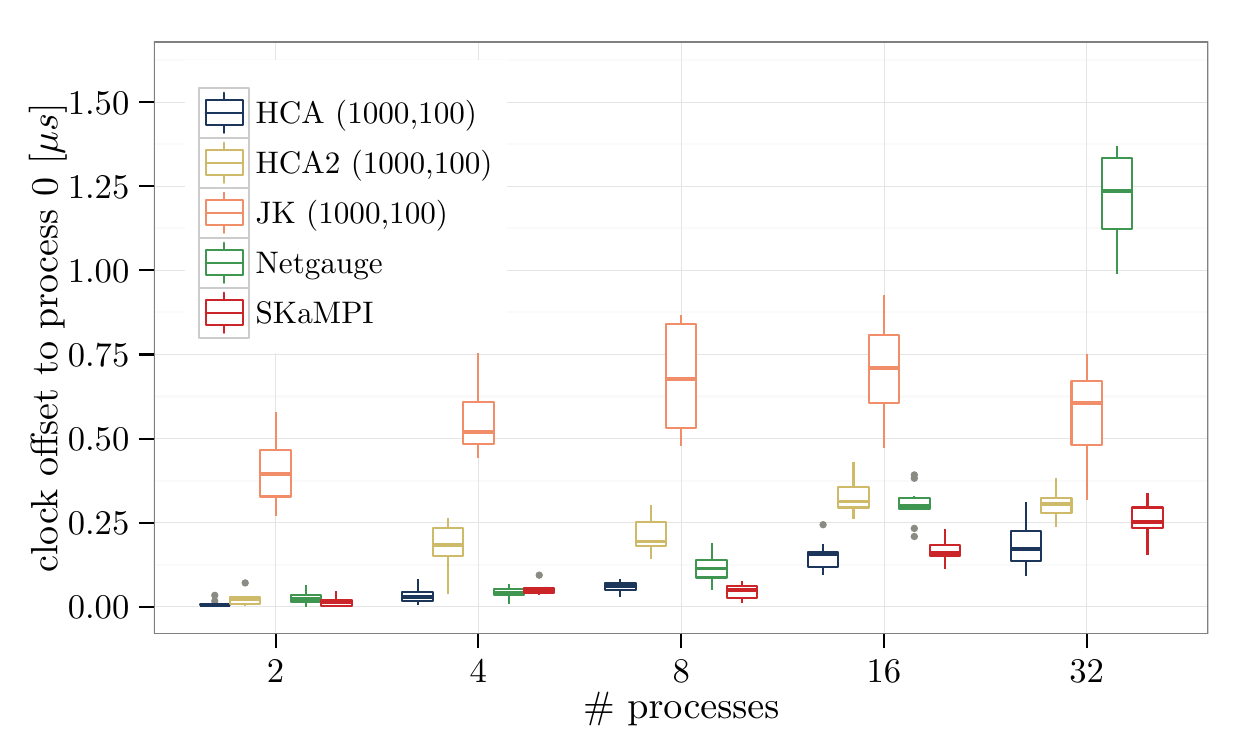}
  }
  \hfill
  \subfigure[16 processes per node]{%
    \label{fig:clock_drift_ppn16}
    \includegraphics[width=.45\linewidth]{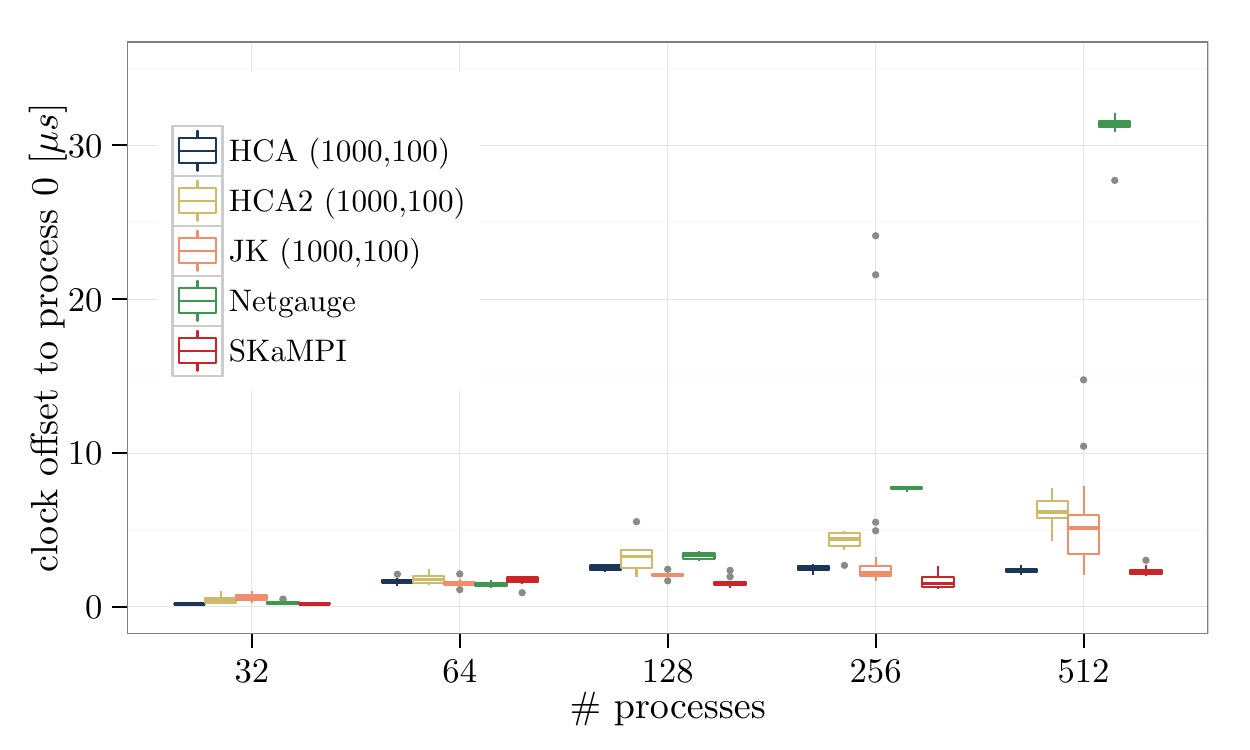}
  }
  \caption{ Clock offset directly after synchronizing the processes
    (\num{10}~calls to \mpirun, \mvapichtwoone, \machone, \expdesc:
    \append~\ref{sec:exp:clock_sync_offset}).}
\end{figure*}

To compare the synchronization schemes of \skampi and \netgauge
(\nbcbench) to the competitors, we have extracted the relevant clock
synchronization algorithms from their respective benchmarking
framework. In particular, it means that we use a fixed window size and
disable the dynamic adaptation implemented by \skampi and
\nbcbench. This allows for a fairer analysis of the clock drifts.
Furthermore, we rely on scheme (MS4) of
\tab~\ref{tab:timing_schemes} for measuring \runtimes. The \runtimes
are based on global times as described in
\Sec~\ref{sec:completion_time_win}.
The tuple (\nfitpoints, \nexchanges) needed for \synchca and \syncjk is 
specified in each figure.

In the following experiments, we show results obtained for \synchca
when applying each of the two approaches we previously described.
In the case of the \emph{first approach}, we use a hierarchical way of
computing the slopes and the linear way of obtaining the intercepts
($\mathcal{O}(\log{\np}) + \mathcal{O}(\np)$ rounds), while the \emph{second 
approach} computes both the slope and the intercept hierarchically in 
$\mathcal{O}(\log{\np})$ rounds.
The \synchca method applying the 
\emph{first approach} is simply denoted ``\synchca'' in the legend of all 
the experiments, whereas the label ``\synchcatwo'' is used for the 
\emph{second approach}.

In practice, the estimation of the drift slope using linear regression typically
requires many more ping-pong messages than the offset computation with
\skampi for a pair of processes.  Thus, $p-1$ \skampi rounds can be
much shorter than $\mathcal{O}(\log{\np})$ rounds of the hierarchical
slope computation.  In our experimental setting (\eg, number of
processes), a simple analytic model revealed that the \emph{first
  approach} of \synchca does not incur a significant \runtime overhead
compared to the \emph{second approach}, since the time for obtaining
the values for the linear regression is dominating.
As a consequence, we only show results for both implementations of 
the \synchca method in this section and only apply the \emph{first approach} 
in the remainder of the paper.

In our first experiment, we apply each of the previously 
described synchronization methods to obtain a global clock for 
every process.
Then, we measure the clock offset
between the \rootproc process and each of the other processes
\emph{directly after} the synchronization phase has been completed.
To that end, the \rootproc process exchanges a number of ping-pong
messages with all other processes and estimates its clock offset
relative to the global time computed on each process. 

\fig~\ref{fig:clock_drift_ppn1} presents the maximum clock offset
measured between any process and the reference process directly after
finishing the clock synchronization. We used one MPI process per
compute node in this experiment. Let $\diff_{r,\rootproc}^j$ be the
clock offset between process $r$ and process \rootproc in ping-pong
round~$j$ (in total $\textit{nrounds}=\num{10}$).  The maximum clock
offset is computed as
$\max_{0 \le r< \np}(\min_{0 \le j <
  \textit{nrounds}}(\diff_{r,\rootproc}^j))$
for each synchronization algorithm.  Each experiment was repeated
\num{10}~times (different calls to \mpirun).  The graph shows that the
clock offset measured directly after synchronizing the clocks with
\skampi or \netgauge is very small, \ie, we measured an offset of at
most \SI{0.2}{\micro\second} for up to \num{8} different compute
nodes.  However, the method of \netgauge leads to significantly larger
offsets when the number of processes (and therefore the number of
synchronization rounds) increases.  The \syncjk synchronization method
produces slightly larger clock offsets for a small number of processes
(\numrange[range-phrase = --]{2}{16}) compared to \skampi and
\netgauge, due to the inaccuracy of \syncjk{}'s approach for computing
linear models.

We also checked how the \synchca algorithm compares to the other clock synchronization
methods. \fig~\ref{fig:clock_drift_ppn1} shows that for up to 
\num{32}~processes (\num{1}~process per node), the maximum clock 
offsets obtained with the \emph{first approach} of \synchca are similar to the offsets yielded by the 
\skampi method.  In the particular case of \num{32}~processes, the 
maximum clock offset is clustered around \SI{0.25}{\micro\second}, 
which represents an improvement over all other methods. 
The \emph{second approach} of \synchca leads to slightly larger offset values, 
while it still outperforms the method of \jonesk, as well as the method of 
\netgauge for \num{32}~processes.
However, \synchca-based clock offsets show an increasing trend with 
the number of processes, which is a consequence of hierarchically 
combining linear models. 

The picture does not change for larger numbers of processes, as shown
in \fig~\ref{fig:clock_drift_ppn16}. Here, \skampi still synchronizes
the distributed clocks with the highest precision, but the relative
difference to \syncjk is smaller. \netgauge, in contrast, will lead to
the least synchronized clocks among its competitors for \num{256} or more
processes on our machine, due to its hierarchical way of combining the
computed offsets.  Both approaches of the \synchca method appear to be viable
alternatives to \skampi, as they result in clock offsets in the same
order of magnitude. 

However, it is important to recall that real clocks are drifting
apart, as shown in \fig~\ref{fig:clock_skew_jupiter}.  To evaluate
the synchronization methods in this scenario, we performed another 
experiment, in which we measured the clock offset over time (clock
drift). 
The \rootproc process waits in a loop for a given amount of
time (\eg, \SI{1}{\second}) and then measures its clock offset to all other
processes.  In this way, we can determine how much the logical global
time is drifting on each process.

\begin{figure}[t]
  \centering
  \includegraphics[width=\linewidth]{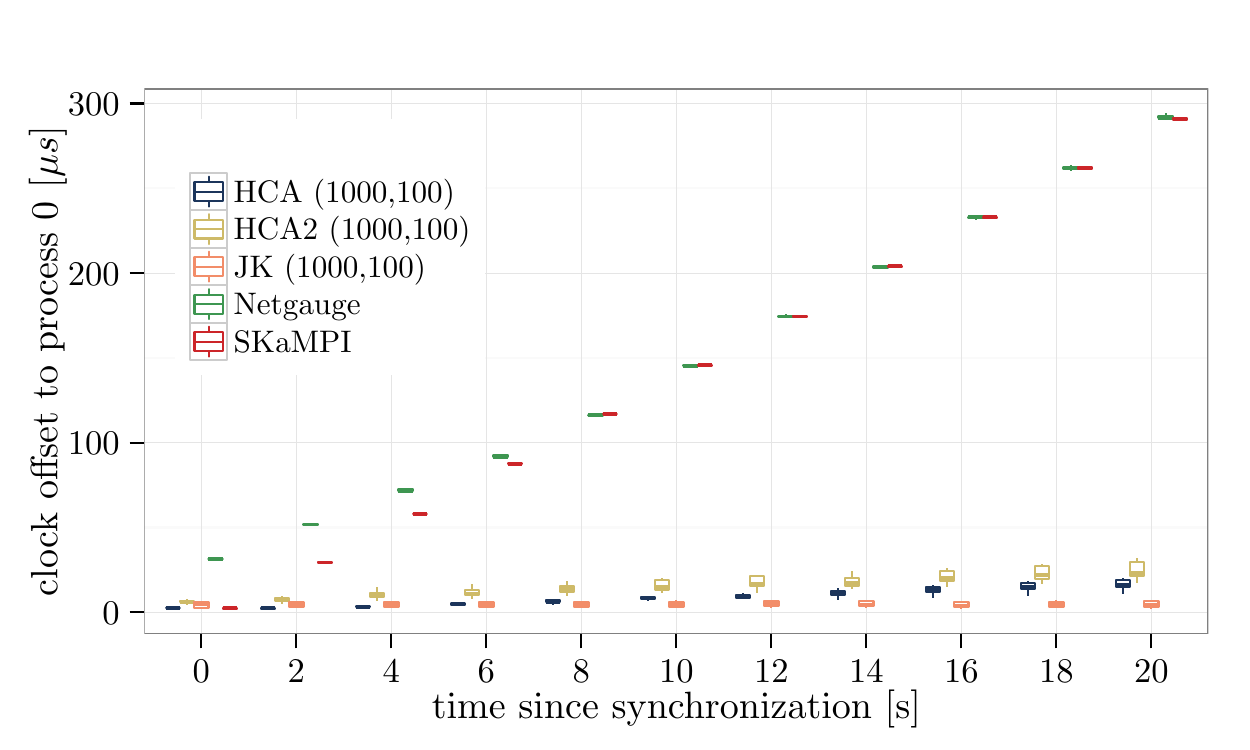}
  \caption{\label{fig:clock_skew_in_time_ppn16}Clock drift for 
  \num{512} (\num{32x16})~processes after \num{0},~\num{2},~\num{4},~$\ldots$,~\SI{20}{\second} 
  (distribution of maximum offsets over \num{10}~calls to \mpirun, 
  \mvapichtwoone, \machone, \expdesc: \append~\ref{sec:exp:clock_drift_all})}
\end{figure}

\fig~\ref{fig:clock_skew_in_time_ppn16} presents the clock drift
measured for 512 processes on 32~nodes of our 36~node cluster
(\machone).
We see that the clock synchronization methods that account for the
clock drift (\syncjk and both \synchca approaches) are superior to the 
ones that only compute the initial clock offset to a reference clock 
(\netgauge and \skampi). 
This experiment also highlights the disadvantage of hierarchically 
combining linear model intercepts in the case of \synchcatwo, which 
shows larger clock offsets than \synchca for \num{512}~processes.

While these results suggest that the method of \jonesk leads to the
most precise measurements for long execution times, its
synchronization mechanism is slow, as it serializes the computation of
linear models. We are therefore interested in understanding the
trade-off between the most accurate clock offset that is obtainable
and the time it takes to synchronize the processes.

\begin{figure}[t]
  \centering
  \includegraphics[width=\linewidth]{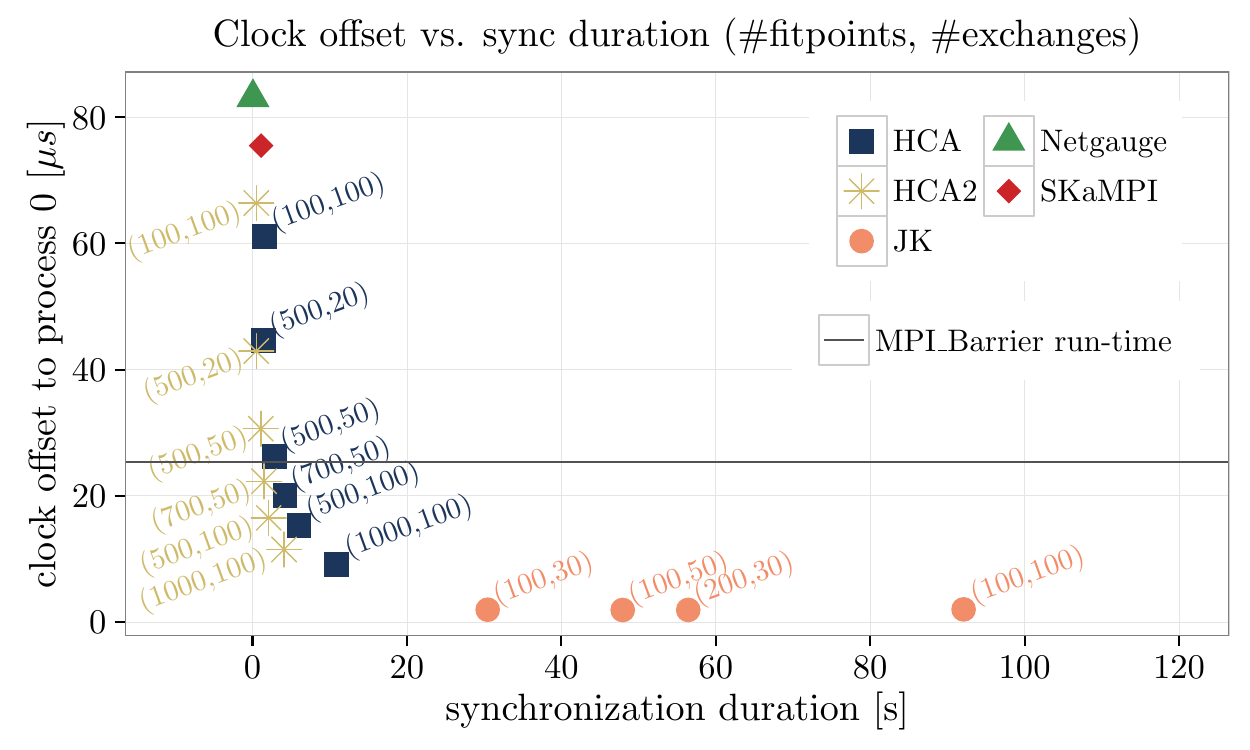}
  \caption{\label{fig:sync_efficiency_nodes512}Median clock offset
    (after \SI{5}{\second}) vs. synchronization phase duration for
    \num{512}~processes (\num{32x16}) (\num{10}~calls to \mpirun{}, 
    \mvapichtwoone, \machone, \expdesc: \append~\ref{sec:exp:sync_duration}).}
\end{figure}

\fig~\ref{fig:sync_efficiency_nodes512} shows the Pareto frontier of
the clock offset versus the synchronization time, which visualizes the
possible configuration choices.  We also added the mean time to
complete a call to \mpibarrier as a baseline. It provides an insight
on the magnitude of process imbalance when synchronizing measurements
through \mpibarrier calls and a limit to the clock offset that is
acceptable for window-based synchronization methods to prove useful in
benchmarking contexts.
The figure plots the clock offsets that were measured five seconds
after completing each clock synchronization method. We see that the
clock offsets obtained with \netgauge and \skampi are relatively large
($\approx{}\SI{80}{\micro\second}$), but both need less than one second 
to complete.
In contrast, the time to complete the clock
synchronization phases of \syncjk and \synchca depends on the number
of sent ping-pong messages needed to compute the regression models.
Thus, the parameters \emph{number of fitpoints} and \emph{number of
  exchanges} have a strong influence on the quality of the clock
synchronization. \fig~\ref{fig:sync_efficiency_nodes512} indicates 
that both implementations of \synchca are able to synchronize the
clocks with a higher precision than what \mpibarrier can provide,
while only requiring at most \SI{10}{\second} to finish the
synchronization process. The method of \jonesk, on the other hand,
produces even smaller clock offsets, but requires at least \SI{30}{\second}
(in the (\num{100}, \num{30}) case) to complete.

\subsection{A Closer Look at \mpibarrier}

So far, we have examined the accuracy of different clock
synchronization methods for the window-based benchmark scheme.  It
remains an open question how much the \mpibarrier synchronization
affects the results.

The use of \mpibarrier for process synchronization is portable but not
necessarily reproducible or fair. Every MPI implementation might use a
different algorithm to implement the barrier, with possibly different
synchronization characteristics. 
The advantage of synchronizing processes with \mpibarrier is that this
method is independent of a logical global clock, and thus,
subsequently measured \runtimes will not experience a drift. Further,
processes will typically require a shorter waiting time compared to a
window-based scheme, which makes the \mpibarrier-benchmarks usually
faster to complete a set of experiments.  However, we need to examine
how well the synchronization using \mpibarrier really works in
practice.

Typically, MPI benchmarks that use \mpibarrier to synchronize 
processes between measurements define the \runtime of an MPI function
as the maximum local \runtime measured on each process.  The problem
with this way of estimating the \runtime is that it is assumed that
all processes leave \mpibarrier and enter the MPI call to be
benchmarked almost synchronously. 

\begin{figure}[t]
  \centering
  \includegraphics[width=\linewidth]{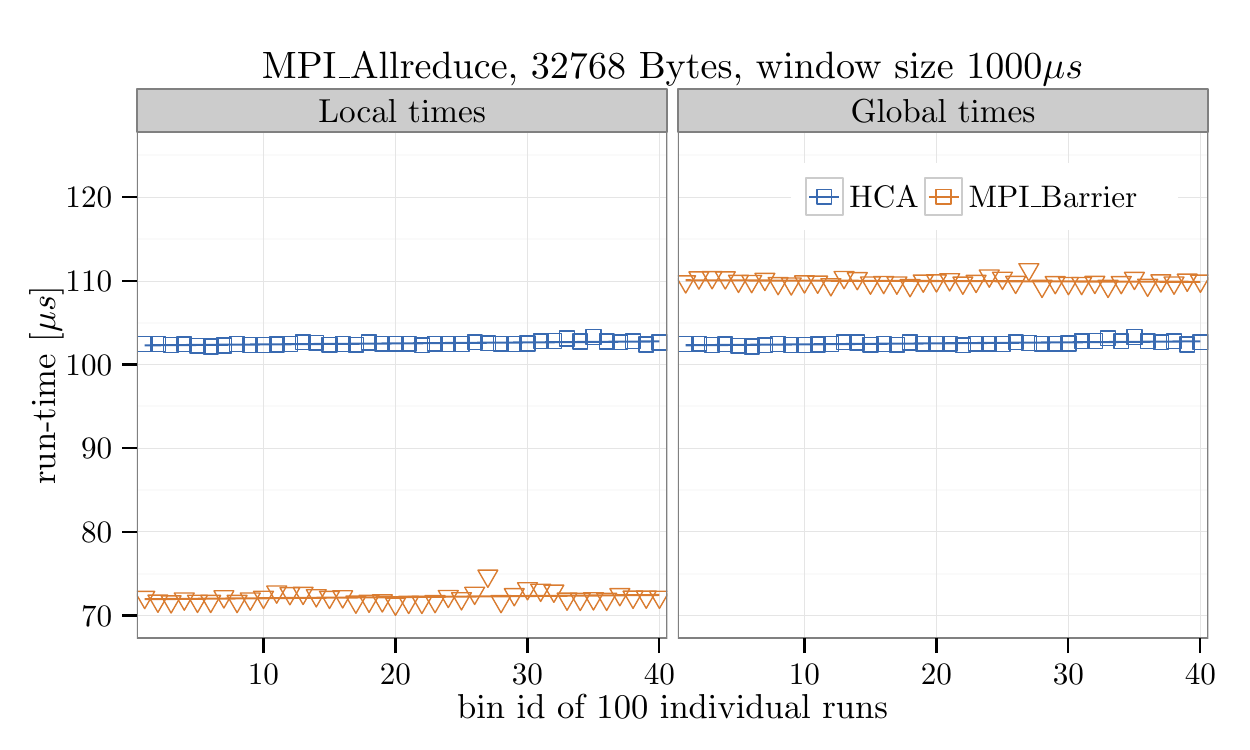}
  \caption{\label{fig:vsc3_barrier_vs_hca_Allreduce} \Runtime of
    \mpiallreduce obtained when using window-based synchronization and
    \mpibarrier{}-based synchronization and two different approaches
    for computing the \runtime (\SI{32}{\kibi\Bytes}, \num{16x1}~processes, 
    \num{4000}~runs, bin size: \num{100}, \mvapichtwoa, \machfive, \expdesc:
    \append~\ref{sec:exp:barrier_local_vs_global_times}).}
\end{figure}

When we compared measurements obtained with window-based and
\mpibarrier-based schemes, we encountered cases for which we initially
had no explanation. The graph on the left-hand side of
\fig~\ref{fig:vsc3_barrier_vs_hca_Allreduce} shows one of these
experiments, where we compare the \runtime of \mpiallreduce obtained
with a window-based scheme (in which clocks were synchronized using
\synchca) to the \runtime obtained when synchronizing with
\mpibarrier.  The mean \runtime of
\mpiallreduce when applying the \mpibarrier synchronization was about
\SI{70}{\micro\second} (computed by using local \runtimes), while the same call
took approximately \SI{100}{\micro\second} using the window-based scheme. 
As this difference seemed very large, it needed further investigation.

We repeated the experiment for \mpiallreduce, but this time, while we
still synchronized processes using \mpibarrier, we measured global
times on each process using our \synchca method to normalize local
times to the \rootproc's reference clock (\cf
\Sec~\ref{sec:completion_time_win}).
The chart on the right-hand side of \fig~\ref{fig:vsc3_barrier_vs_hca_Allreduce}
shows the resulting \runtimes of \mpiallreduce for both
synchronization methods (\mpibarrier and window-based with \synchca),
where all times were obtained using globally-synchronized clocks.
Now, the resulting \runtimes are much closer and their difference can
reasonably be explained by the way the two synchronization schemes work.

\begin{figure}[t]
  \centering
  \includegraphics[width=\linewidth]{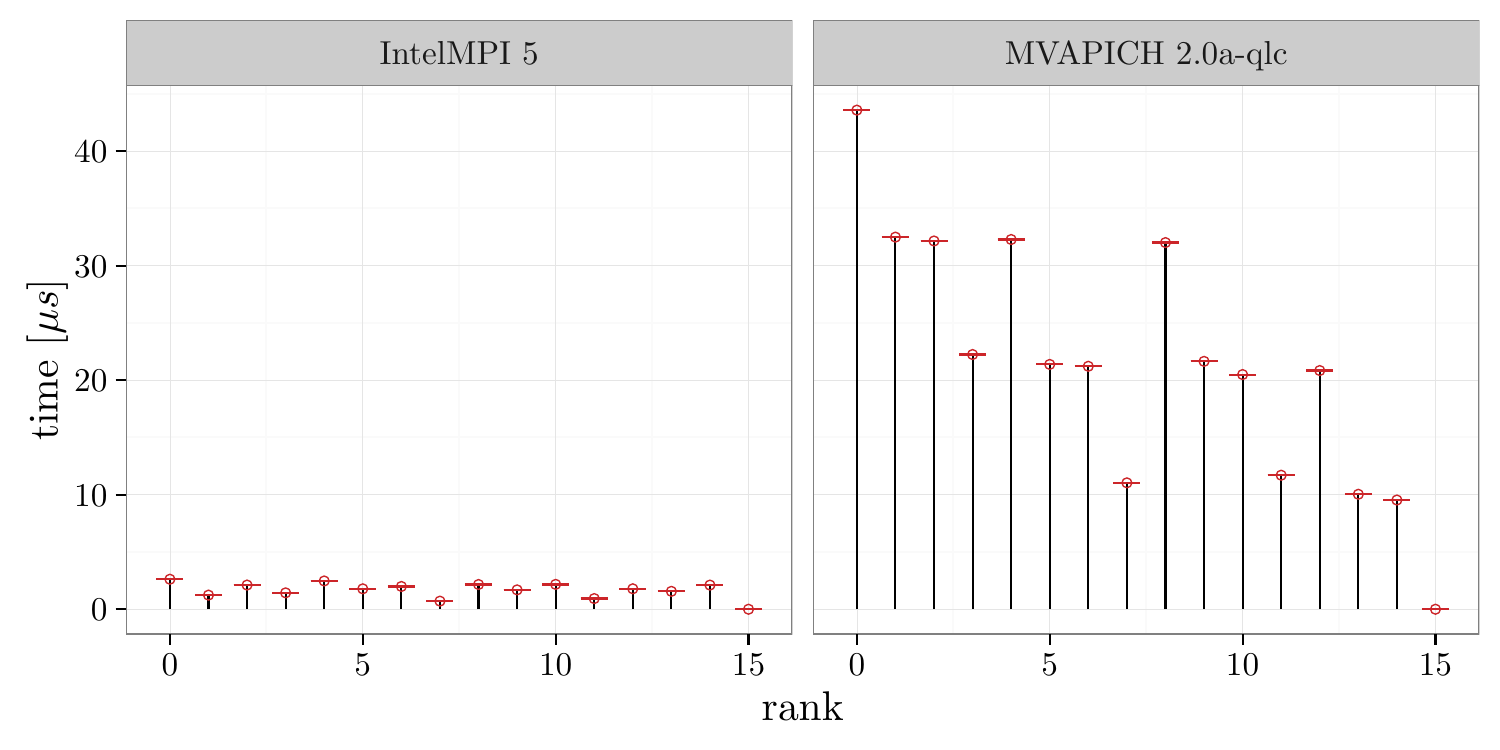}
  \caption{\label{fig:vsc3_barrier_processes} Synchronization
    imbalance of \mpibarrier implementations.  Exit times of each
    process relative to the first process that leaves the barrier
    (mean of \num{1000}~measurements and \SI{95}{\percent} confidence
    intervals, \num{16x1}~processes, one \mpirun, \synchca
    synchronization, window size: \SI{100}{\micro\second}, \machfive,
    \expdesc: \append~\ref{sec:exp:barrier_exit_times}). }
\end{figure}

Nevertheless, we still need to explain the gap between the observed
\runtimes when we switch from local to global times to determine the
overall \runtime. Ideally, both \runtime computation methods should
lead to similar results. Therefore, we investigated the skew of MPI
processes when they exit the \mpibarrier function. For this purpose,
we applied the \synchca method to synchronize clocks and recorded the
global timestamp of each process at the end of the \mpibarrier call.
The results of this experiment are shown in
\fig~\ref{fig:vsc3_barrier_processes}. The graphs compare the process
skew after completing \mpibarrier, measured with \intelmpifive (left)
and \mvapichtwoa (right). Surprisingly, a call to \mpibarrier using
\mvapichtwoa resulted in a large process skew. In particular, the mean
exit times between process~\num{0} and process~\num{15} differed by 
more than \SI{40}{\micro\second}.  This finding directly explains why the 
measurements in the previous experiment
(\cf~\fig~\ref{fig:vsc3_barrier_vs_hca_Allreduce}) showed such a large
difference in \runtime.

The experiments discussed here only indicate the potential impact of
using \mpibarrier to synchronize processes on MPI benchmarking
results. They are not meant to point out potential performance
problems of libraries such as \MVAPICH. Therefore, we show results
obtained with \mvapichtwoa, which is not the latest version of
\MVAPICH, but the one that was pre-installed on the system and for
which we experienced this significant process skew.

\begin{figure}[t]
  \centering
  \includegraphics[width=\linewidth]{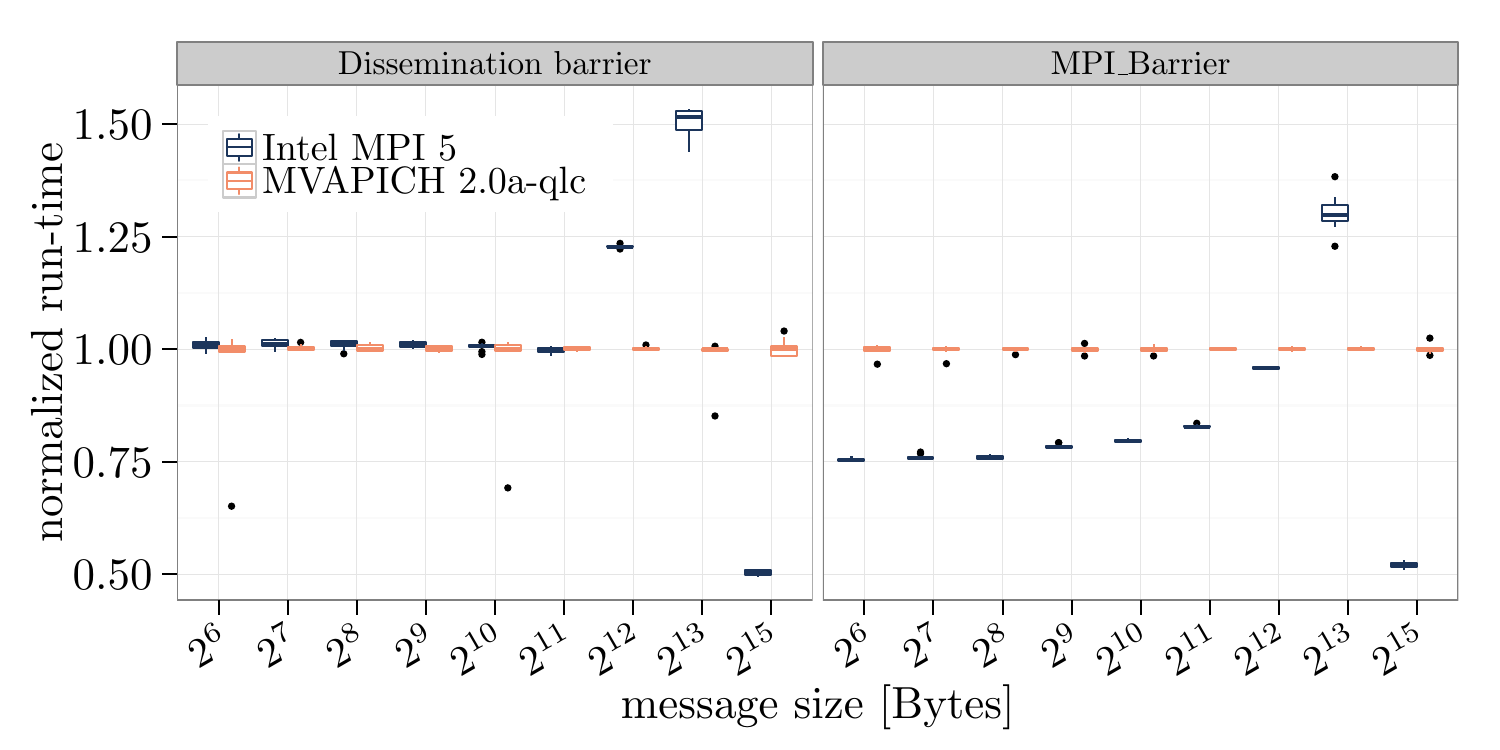}
  \caption{\label{fig:vsc3_intel_vs_mvapich_barriers} Distribution of
    normalized median \runtimes of \mpibcast for \intelmpifive and
    \mvapichtwoa, when measurements are synchronized using a call to
    \mpibarrier or an external 
    barrier implementation (\num{16x1}~processes, 
    \num{1000}~measurements, \num{10}~calls to \mpirun, \machfive, 
    \expdesc: \append~\ref{sec:exp:barrier_implem_impact}).  }
\end{figure}

Last, we would like to demonstrate how misleading the \runtime
measurements can be when the experimenter relies on \mpibarrier for
process synchronization. \fig~\ref{fig:vsc3_intel_vs_mvapich_barriers}
compares the normalized \runtimes of \mpibcast obtained with either an
external benchmark-provided dissemination barrier (\cf
\cite{Taubenfeld:2006}) or the barrier implementation provided by each
library. We have executed \num{10} distinct calls to \mpirun, in each of
which \num{1000}~measurements were recorded. We compute the median of each
sample and normalize the \runtimes for one message size to the median
\runtime of these \num{10}~medians of \mvapichtwoa. We observe, especially
for the smaller message sizes 
(\SIrange[exponent-base=2]{d6}{d11}{\Bytes}), that there is
no clear winner between \intelmpifive and \mvapichtwoa when our own
barrier implementation is used (left-hand side). However, when we
employ the library-provided \mpibarrier implementation for
synchronizing processes, we see a significant performance difference
between the libraries. In such cases, one could easily draw wrong
conclusions. 

\subsection{Summary of Distributed Clock Synchronization Methods}

We have shown that the choice of a clock synchronization method used
for MPI benchmarking has tremendous effects on the outcome. The clock
synchronization method implemented by \skampi can achieve very
accurate timings, but since the logical global clocks are drifting
quickly, only a small number of MPI operations can be measured
precisely, which of course depends on the length of each MPI function
call. In case the experimenter wants to measure for a longer period of
time (\eg, several milliseconds or even seconds), the approaches used
in \skampi and \netgauge simply lead to inaccurate measurements. To
overcome this problem, one could start re-synchronizing the clocks
after a given amount of time has passed or use a clock synchronization
algorithm that accounts for the clock drift.

The approach of \netgauge is more scalable than the clock
synchronization used in \skampi. However, it possesses an increased
synchronization error since it combines estimated clock offsets, which
themselves entail an error. Additionally, \netgauge uses a heuristic
to estimate the CPU frequency, which is a potential source of error.

The clock synchronization method of \jonesk accurately synchronizes a
set of distributed clocks as it considers the clock drift between
processes. Thus, it can be used for measuring MPI functions over a
longer time span. This approach could be used if very accurate
window-based measurements are required and if the relatively long time
for completing the clock synchronization can be tolerated.

Our clock synchronization method, called \synchca, can be seen as a
trade-off between achieving a higher accuracy for longer runs like
\syncjk and providing the speed of \netgauge.  It suffers from the
same problem as \netgauge, as it combines models with an inherent
experimental error. Nevertheless, in
our MPI benchmarking setup the \synchca algorithm emerged as the best option
for process synchronization compared to \mpibarrier, \skampi, or
\netgauge when measurements over longer periods of time (we have
tested for up to \SI{20}{\second}) for many processes are needed.

Last, we note that using a library-provided implementation of
\mpibarrier may lead to unforeseeable results, as processes can become
significantly skewed when they leave \mpibarrier. The decision whether
to rely on \mpibarrier should therefore be done after investigating
the behavior of the implementation on the given network. Nevertheless,
an MPI benchmark should provide its own \mpibarrier implementation
for fairly comparing two MPI libraries.

\section{Influencing Factors of MPI Benchmarks}
\label{sec:characterization_mpi_benchmark}

After examining the MPI benchmarking process, 
we now turn to characterizing and analyzing the performance data.
A good understanding of the performance data is essential for
selecting and applying the right statistical test for comparing MPI
alternatives.
But for a rigorous statistical analysis, we need a deeper insight into
our system and the factors that influence the \runtimes to be
measured.  Le\,Boudec points out that ``knowing all factors is a
tedious, but necessary task. In particular, all factors should be
incorporated, whether they interest you or
not''~\cite{boudec2010performance}.

Hence, we will first examine the shape of sampling distributions of
\runtime measurements. Then, we will analyze potential experimental
factors in the remainder of this section. However, we decided to
exclude ``obvious'' factors of MPI performance experiments, such as
the communication network, the number of processes, and the message
size.

\subsection{Sampling Distributions of MPI Timings}
\label{sec:sampling_distr}

To apply a statistical hypothesis test, we need to make sure that all
its assumptions are met. For example, many tests assume that the data
follow a specific probability distribution, \eg, the dataset is
normally distributed. We now examine the experimentally obtained
distributions of MPI function \runtimes.
\begin{figure}[t]
  \centering
  \includegraphics[width=\linewidth]{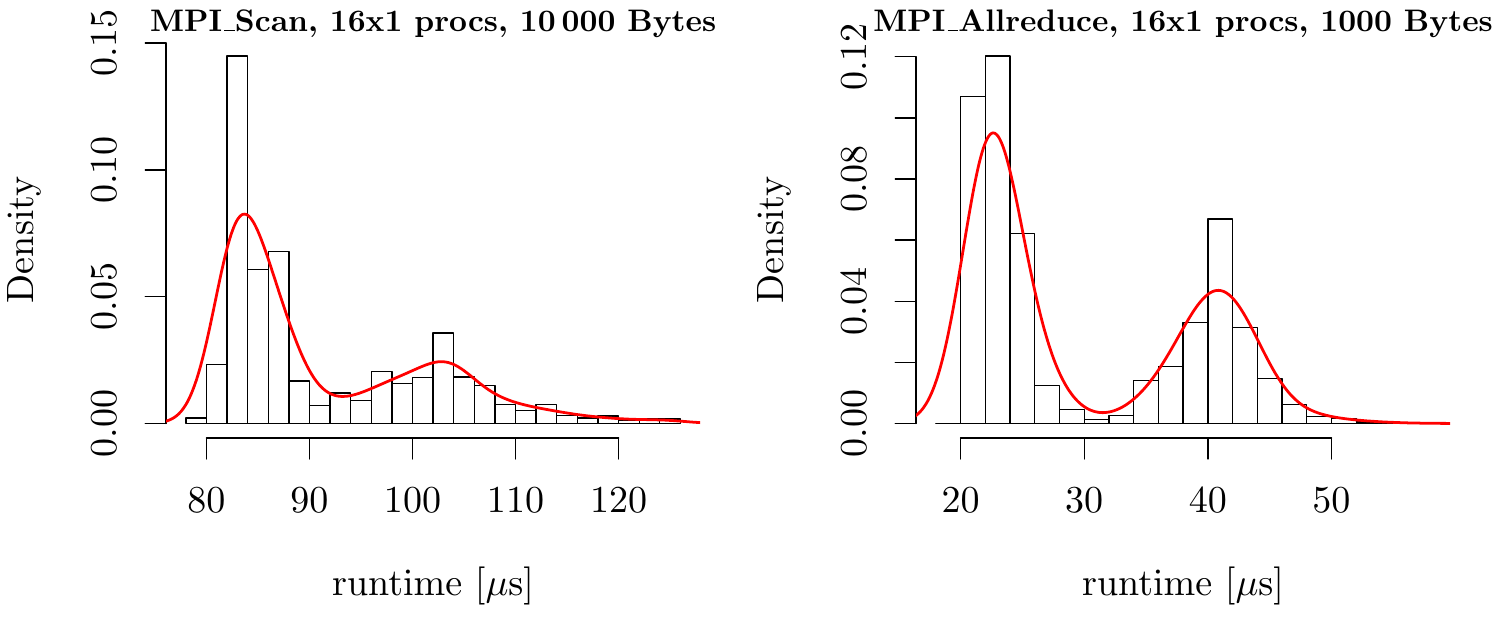}
  \caption{\label{fig:distribution_examples}Histogram of the time
    needed to complete a call to \mpiscan with \SI{10000}{\Bytes} (left)
    and to \mpiallreduce with \SI{1000}{\Bytes} (right) on \machone 
    (\mpibarrier synchronization, \necmpitwoeight,
    \expdesc: \append~\ref{sec:exp:runtime_hist}).}
\end{figure}

\begin{figure}[!ht]
  \centering
  \includegraphics[width=\linewidth]{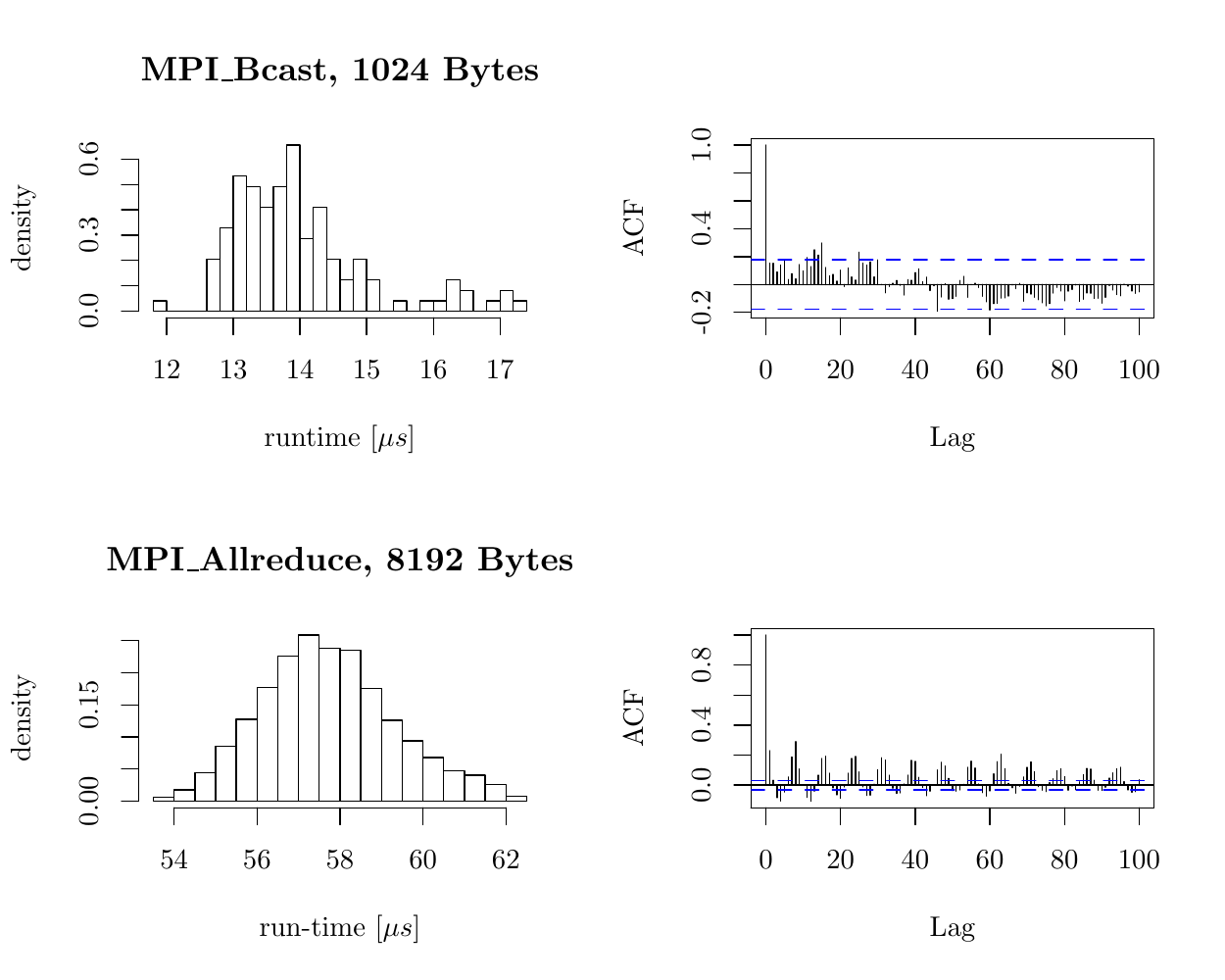}
  \caption{\label{fig:g5k_histograms_hca} Distribution of \runtimes
    and corresponding autocorrelation plots 
    (\num{16x8}~processes, \num{4000}~runs,
     \synchca synchronization, window size: \SI{500}{\micro\second},
    \mvapichonenine, \machthree, 
    \expdesc: \append~\ref{sec:exp:runtime_hist}).}
\end{figure}

\begin{figure}[!ht]
  \centering
  \includegraphics[width=\linewidth]{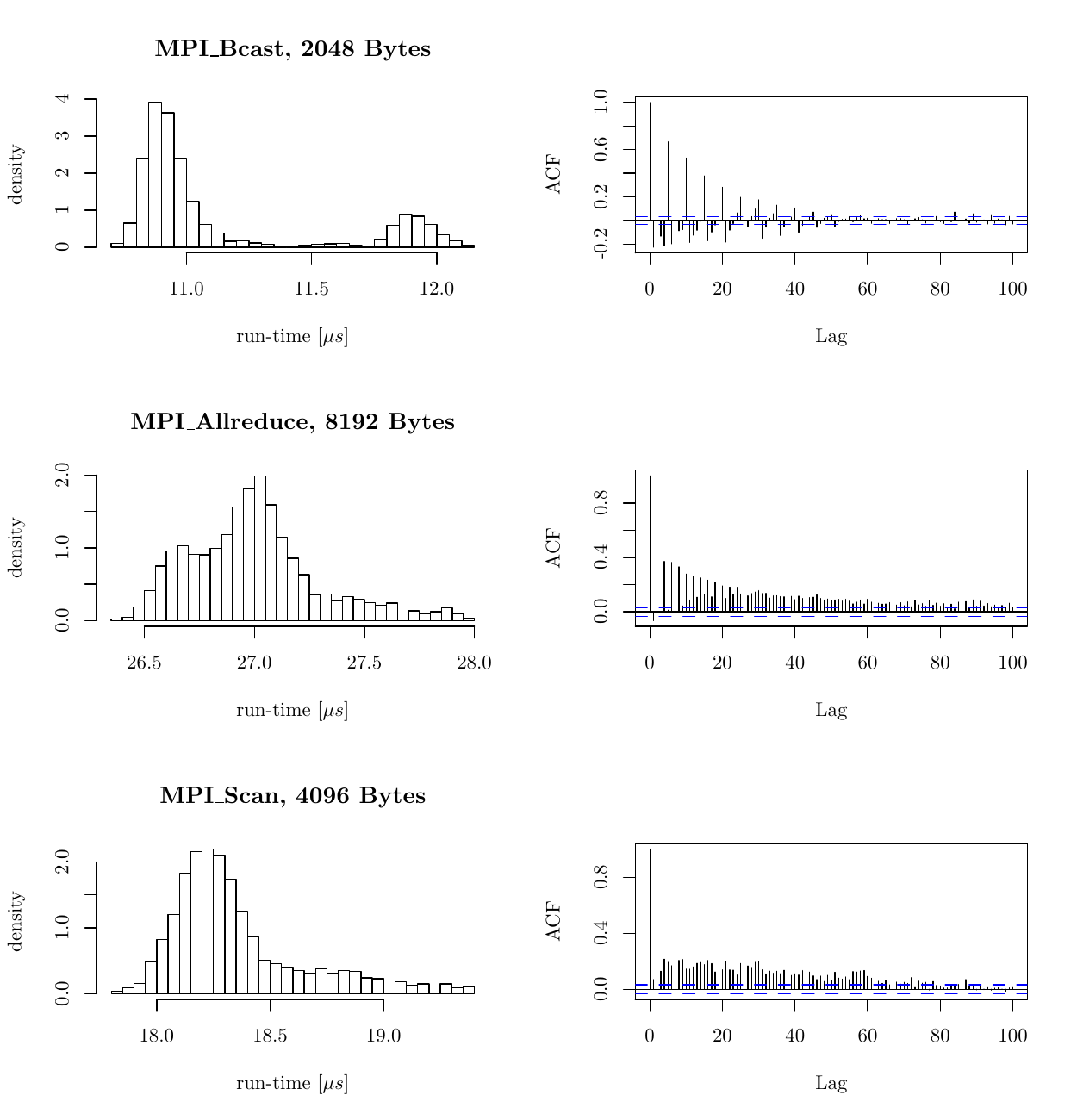}
  \caption{\label{fig:surfsara_histograms_hca} Distribution of
    \runtimes and corresponding autocorrelation plots 
    (\num{16x1}~processes, \num{4000}~runs, 
    \synchca synchronization, window size: \SI{500}{\micro\second},
    \intelmpi, \machfour, 
    \expdesc: \append~\ref{sec:exp:runtime_hist}).}
\end{figure}

\begin{figure}[!ht]
  \centering
  \includegraphics[width=\linewidth]{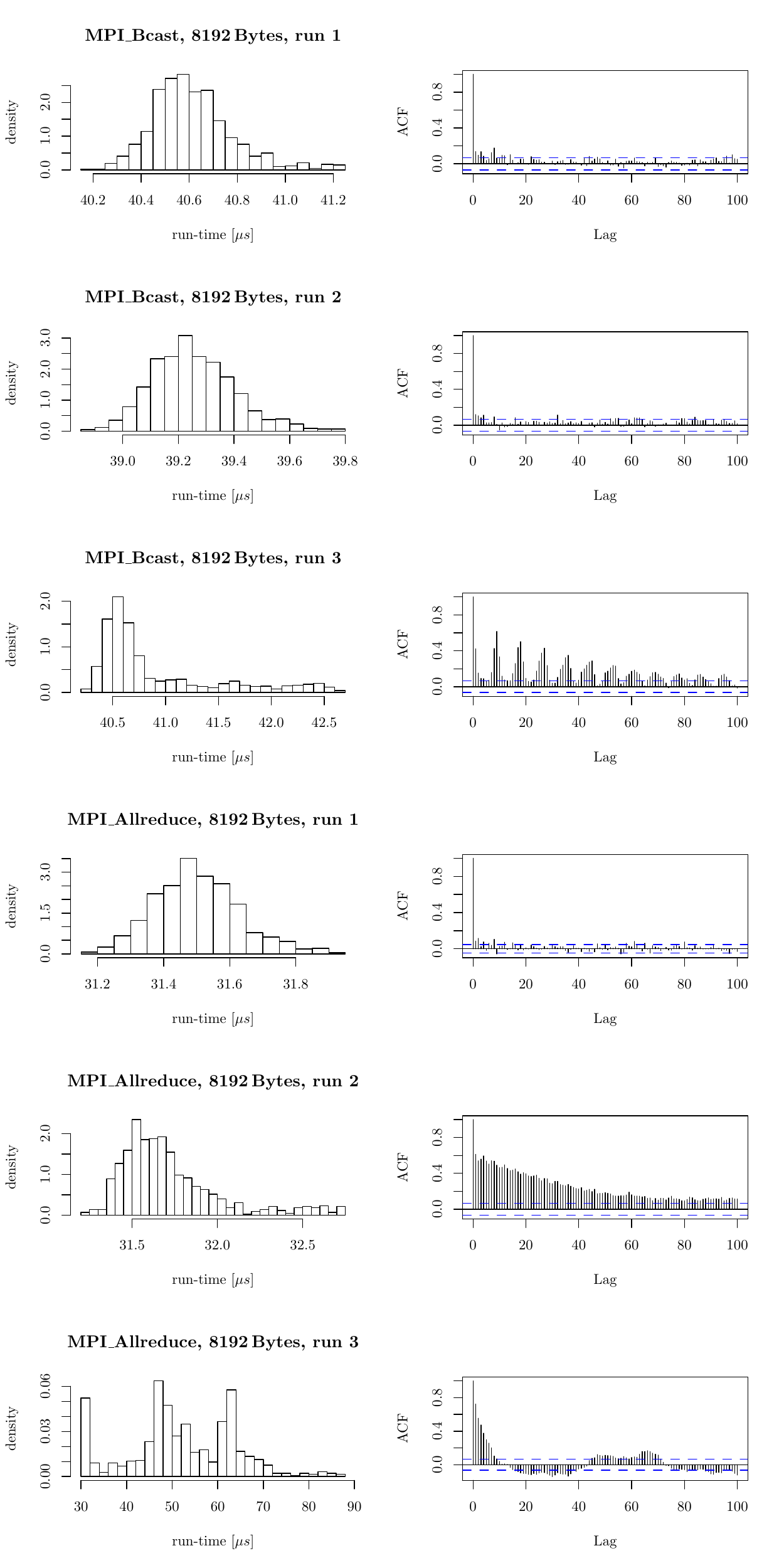}
  \caption{\label{fig:vsc3_histograms_hca} Various distributions of \runtimes
    and corresponding autocorrelation plots for several calls to \mpirun
    (\num{16x1}~processes,  \num{1000}~runs, 
     \synchca synchronization, window size: \SI{1}{\milli\second},
    \mvapichtwoa, \machfive, 
    \expdesc: \append~\ref{sec:exp:runtime_hist}).}
\end{figure}

We first ran a large number of MPI experiments to investigate various
sampling distribution of MPI timings. The experiments were conducted
for several MPI functions such as \mpibcast, \mpiallreduce,
\mpialltoall, or \mpiscan.  \fig~\ref{fig:distribution_examples} shows
the distribution of \runtimes for \num{10000}~calls to \mpiscan with a
payload of \SI{10000}{\Bytes} and to \mpiallreduce with a payload of
\SI{1000}{\Bytes}, both for \num{16}~processes 
(one process per node).  We used a kernel density estimator
(\texttt{density} in R) to obtain a visual representation of the
sampling distribution.  The figure indicates that the sampling
distributions are clearly not normal, and interestingly, in both
distributions we can see two distinct peaks. The peak on the right is
much smaller, but it appears in many of the histograms for 
small execution times (less than \SI{200}{\micro\second}). Similar distributions
were obtained for experiments with \mpialltoall and \mpibcast, as well
as on other parallel machines (see \figs~\ref{fig:g5k_histograms_hca},
\ref{fig:surfsara_histograms_hca}, and \ref{fig:vsc3_histograms_hca}
in \append~\ref{app:meas:distributions}).

Since the measured \runtimes do not follow a normal distribution, we
must be careful when computing statistics 
such as the confidence interval for the mean.  
The central limit theorem (CLT) states that sample means are normally
distributed if the sample size is large enough.  In practice, we most
often do not know in advance how large the sample size should be such
that the CLT holds. Many textbooks, like the books by
Lilja~\cite{liljameasuring} or Ross~\cite{ross2010introductory}, state
that a sample size of~\num{30} is large enough to obtain a normally
distributed mean. However, Chen
\etal~\cite{ChenCGTWH12} 
report in a recent study that samples need to include at least
\num{240}~observations, such that the sample means follow a normal
distribution.

\begin{figure}[t]
  \centering
  \includegraphics[width=\linewidth]{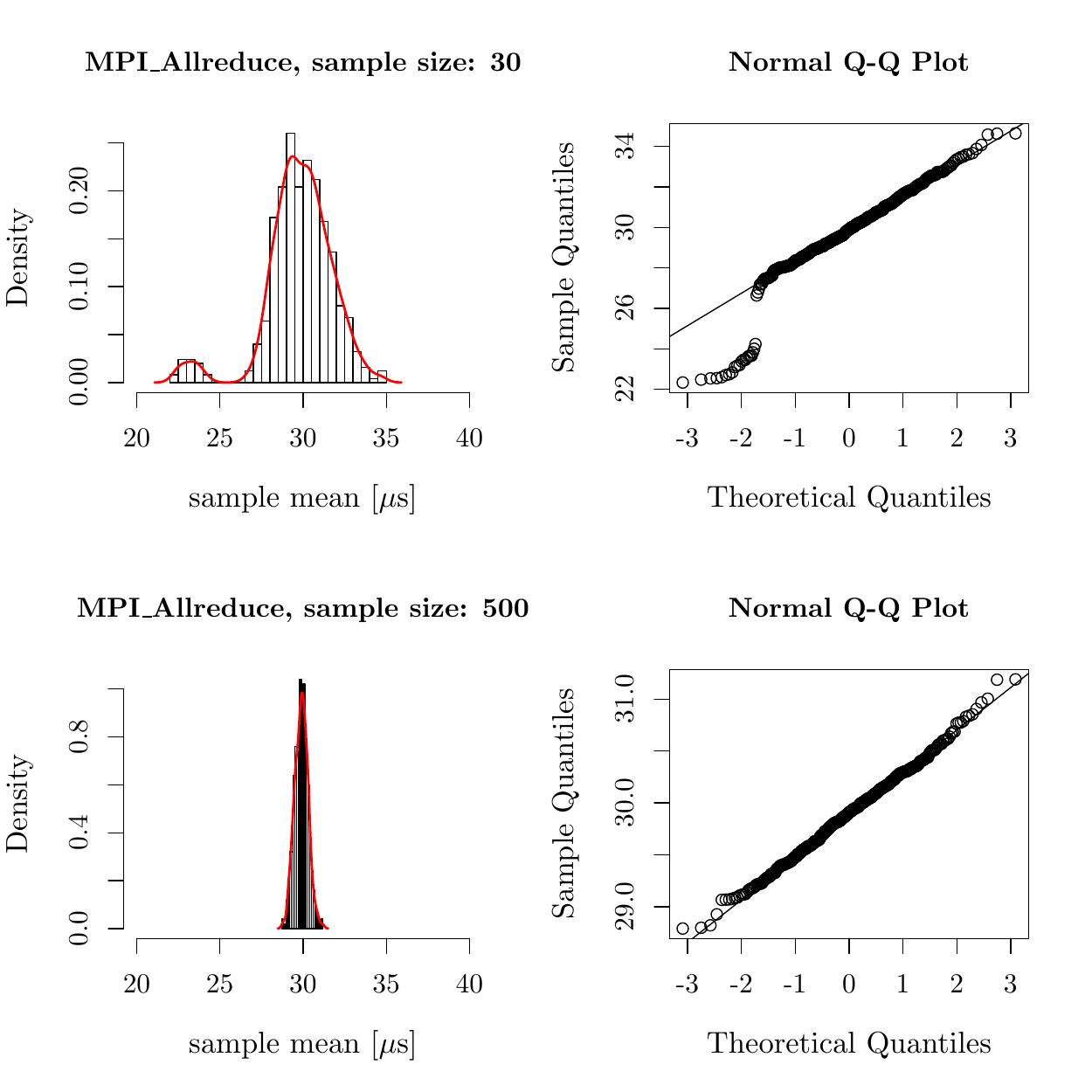}
  \caption{\label{fig:distribution_clt}Distribution of sample means
    and Q-Q plots when sampling using different sample sizes from the probability
    distribution for \mpiallreduce
    (\cf \fig~\ref{fig:distribution_examples}).
  }
\end{figure}

We are interested in how many repetitions of a single measurement are needed
within one call to \mpirun such that the CLT holds for the computed
sample mean. To answer this question, we analyzed distributions of
sample means by randomly sampling from the set of \num{10000}
previously measured MPI \runtimes of \mpiallreduce (\cf
\fig~\ref{fig:distribution_examples}). In particular, we drew
\num{500}~samples containing \num{10}, \num{20}, $\ldots$, \num{500}
observations each, computed the mean of each sample, and built a
histogram of the sample means for each sample
size. \fig~\ref{fig:distribution_clt} shows two histograms and their
corresponding Q-Q plots for a sample size of \num{30} and
\num{500}. The data provides evidence that a sample size of \num{30}
is not large enough to obtain a normally distributed sample mean.  In
our particular case, \num{500} observations were required so that the
distribution of sample means was normally shaped. We therefore advise
scientists to carefully verify the sample distributions in order to
compute meaningful confidence intervals of the sample mean when
benchmarking MPI functions. A similar suggestion has been made
recently by Hoefler and Belli~\cite{HoeflerB15}.

\subsection{Factor: The Influence of \mpirun}
\label{sec:influence_mpirun}

\begin{figure*}[!t]
\centering%
\subfigure[\machone, \necmpitwoeight]{%
\includegraphics[width=.32\linewidth]{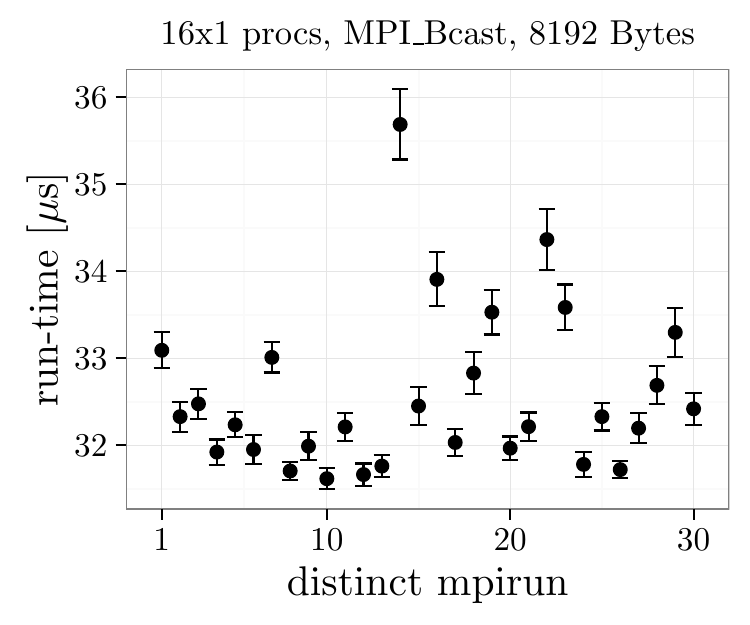}
}\hfill%
\subfigure[\machtwo, \intelmpi]{%
\includegraphics[width=.32\linewidth]{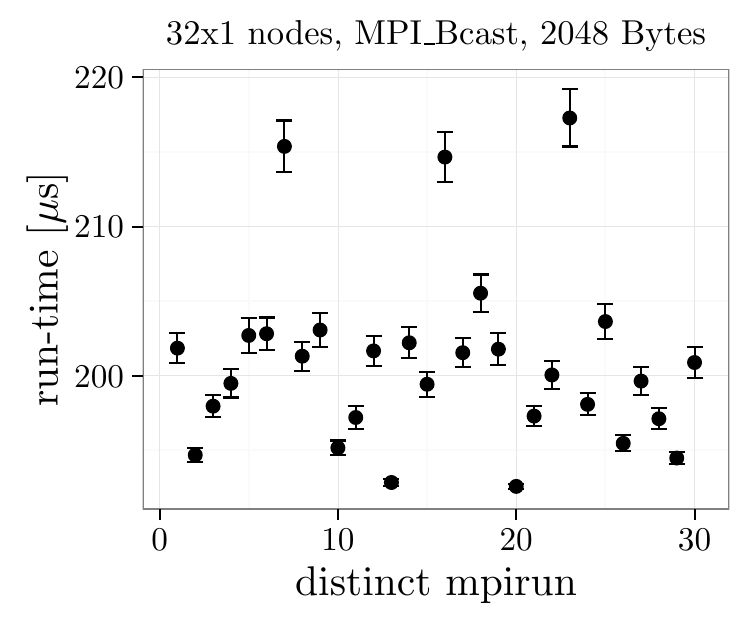}
}\hfill%
\subfigure[\machthree, \mvapichonenine]{%
\includegraphics[width=.32\linewidth]{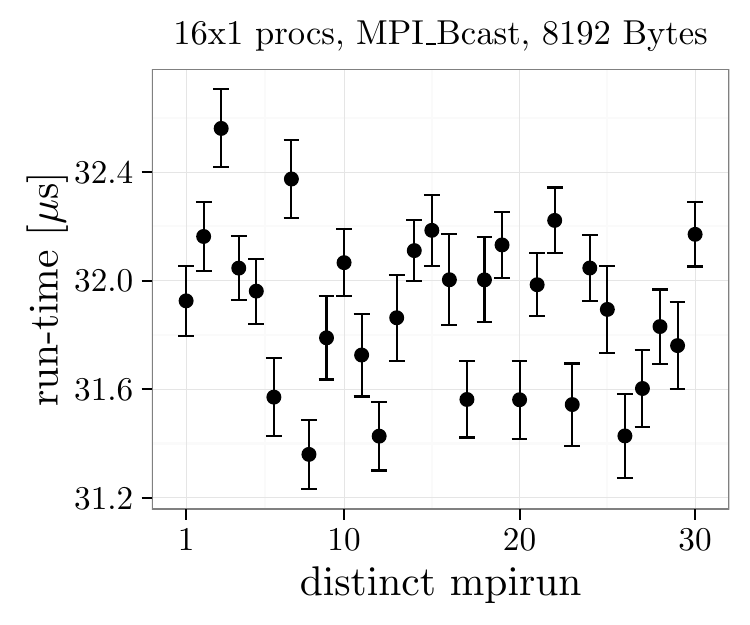}
}
\caption{\label{fig:mpirun_distinct}Mean and \SI{95}{\percent} confidence interval
  of the time to complete a call to \mpibcast (with a different size
  of the payload) for \num{30}~distinct calls to \mpirun, using \mpibarrier
  for synchronization (\expdesc: \append~\ref{sec:exp:infl_mpirun}).}
\end{figure*}

After taking a closer look at the results of the sampling experiment
shown in \Sec~\ref{sec:sampling_distr}, we noticed that the sample
means were slightly different between calls to \mpirun.  To
investigate the effect of \mpirun, we conducted a series of
experiments to determine whether distinct calls to \mpirun produce
different sample means (statistically significant).
The experimental setup was the following: We executed \num{30}~distinct
calls to \mpirun and within each \mpirun we measured each individual
\runtime of \num{1000}~calls to a given MPI function.
\fig~\ref{fig:mpirun_distinct} presents a subset of the gathered
experimental results. The graphs compare the means and their 
\SI{95}{\percent} confidence intervals for \num{30}~distinct calls 
to \mpirun, a given MPI function, and a message size.

\begin{figure}[t]
  \centering
  \includegraphics[width=\linewidth]{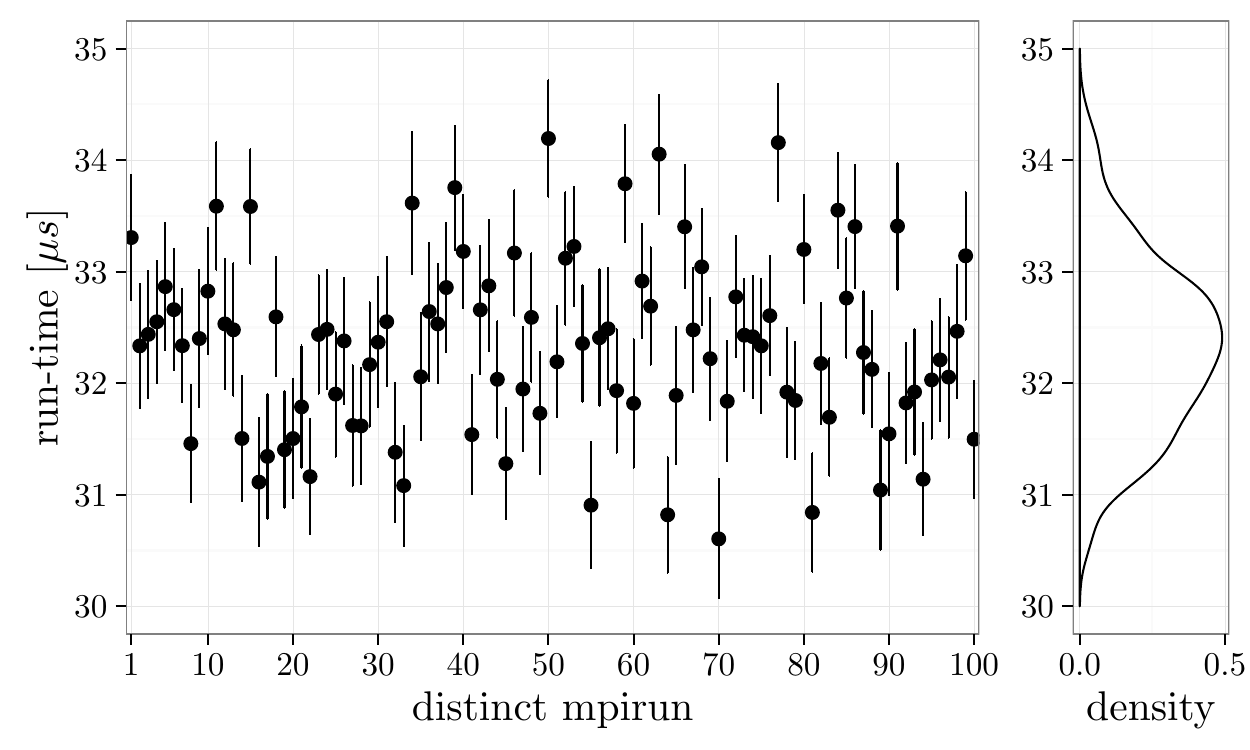}
  \caption{\label{fig:mpiruns_dist_allreduce}Distribution of mean
    \runtime values of \mpiallreduce and respective density estimation
    (\num{16x1}~processes, \num{1000}~measurements, \num{500}~calls to
    \mpirun, \syncjk synchronization, window size
    \SI{1}{\milli\second}, \mvapichtwoone, \machone, \expdesc:
    \append~\ref{sec:exp:runtime_dist_allreduce}).}
\end{figure}

The data yielded by this experiment provide convincing evidence that
the \runtime means obtained from distinct calls to \mpirun are
different. The differences between these means, however, are often not
very large (\SIrange[range-phrase = --]{3}{5}{\percent}), yet they are statistically significant.

Our finding that different calls to \mpirun have a significant effect
on the experimental outcome is very important for designing MPI
benchmarks.  As a consequence, it is insufficient for an MPI benchmark
to collect MPI \runtime measurements only from a single call to
\mpirun. Instead, several calls to \mpirun are required to correctly
account for the variance between different calls.  The problem of
finding out how many calls to \mpirun are needed is tightly connected
to the statistical hypothesis test to be applied. We discuss this
question in more detail in~\Sec~\ref{sec:statistical_analysis}.

\fig~\ref{fig:mpiruns_dist_allreduce} (right) shows the distribution
of \num{500}~sample means of the \runtime of \mpiallreduce obtained
from \num{500}~distinct calls to \mpirun.  For every call to \mpirun,
we recorded \num{1000}~\runtime measurements and computed their mean.
We can observe that the means are normally distributed. If the
distributions obtained from different \mpiruns are relatively similar,
the normality distribution of the means is a consequence of the
central limit theorem. However, we cannot formally assume that the
computed means are normally distributed, as each \mpirun could produce
a completely different distribution of \runtimes. In such cases, we
need to check for normality either using the \kolmtest or the
\shaptest test~\cite{Sheskin:2007}.

\subsection{Factor: Uncontrollable System Noise}

\begin{figure*}[t]
  \centering
\subfigure[all \runtimes]{%
\label{fig:system_noise2_all}%
\includegraphics[width=.49\linewidth]{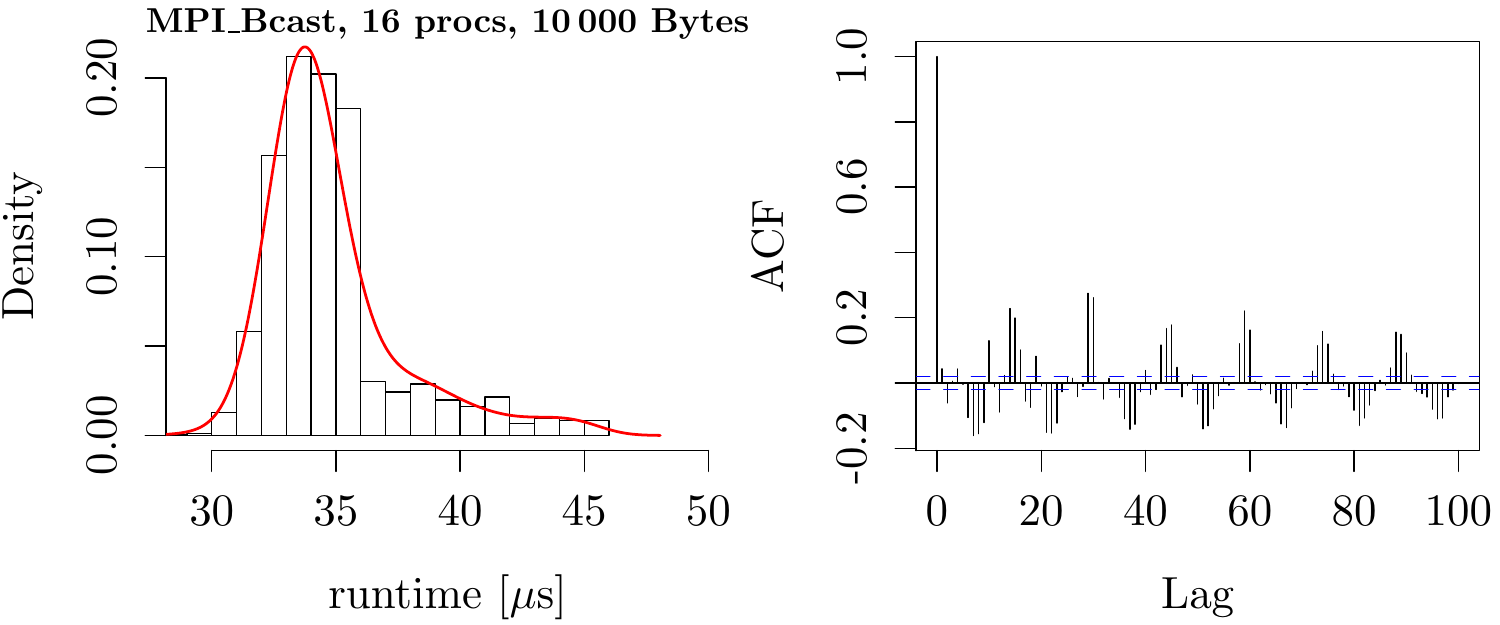}
}\hfill%
\subfigure[sub-sample of \runtimes]{%
\label{fig:system_noise2_subsample}%
\includegraphics[width=.49\linewidth]{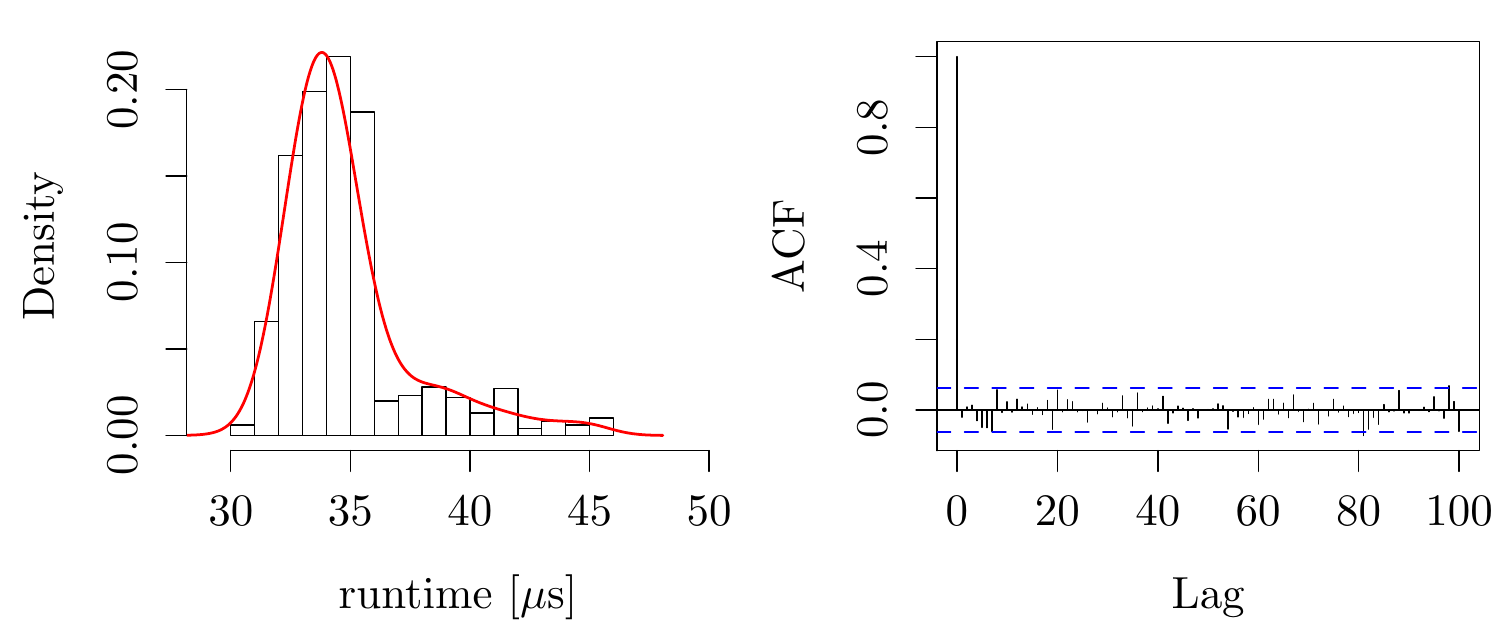}
}
\caption{\label{fig:system_noise2}Distribution of \runtimes for
  \mpibcast and the corresponding autocorrelation plot 
  (\SI{1000}{\Bytes}, \num{16x1}~processes, \num{1000}~runs, 
  \mpibarrier synchronization, \necmpitwoeight, \machone, \expdesc:
  \append~\ref{sec:exp:runtime_hist}).}
\end{figure*}

Several \runtime distributions shown so far exhibited a longer tail or
a second smaller peak on the right. Thus, it might be possible that
subsequent measurements of MPI functions are similar. Then the
question becomes whether different measurements, taken in sequence,
are independent from each other. This verification of the measurement
independence is essential, as virtually all statistical hypothesis
tests assume that random variables are independent and identically
distributed (iid). If this assumption is violated, statistical
measures could be misleading, \eg, the computed confidence interval of
the mean could be too small~\cite[p. 47]{boudec2010performance}.

Le\,Boudec suggests to evaluate the autocorrelation of the
experimental data~\cite{boudec2010performance}. Consequently, we
computed the autocorrelation 
of our experiments and show some of the results
in~\fig~\ref{fig:system_noise2}.  Autocorrelation is typically used in
time-series analysis, and it estimates the correlation between two
values of the same variable measured at different times as a function
of the time lag between them. In particular, the autocorrelation
coefficient\footnote{\url{http://www.itl.nist.gov/div898/handbook/eda/section3/autocopl.htm}}
at lag $h$ is computed as the ratio of the autocovariance $C_h$ to the
variance $C_0$. A significant correlation of measurements in the data
can be seen when a line in the lag plot at a specific lag value
exceeds the significance level. If all values were chosen randomly,
for example from a normal distribution, then individual measurements
are uncorrelated. \fig~\ref{fig:system_noise2_all} shows that the
experimental data exhibit a significant correlation between
measurements. An immediate consequence is that not all assumptions for
applying hypothesis tests hold true as measurements are correlated.

One way to remove the correlation is the use of data sub-sampling, as
stated by Le\,Boudec~\cite{boudec2010performance}. Indeed, when we
sub-sample \num{1000}~observations from the original sample of 
\num{10000}~observations, the \runtimes become uncorrelated as shown 
in \fig~\ref{fig:system_noise2_subsample}. When we compare both
histograms presented in~\fig~\ref{fig:system_noise2}, we can observe
that data sub-sampling has almost no effect on the sample mean. Hence,
we do not apply a sub-sampling strategy in the remainder of the
article, but we need to keep in mind that measurements are potentially
dependent.

\subsection{Factor: Synchronization Method}

After introducing and discussing several clock synchronization methods
in \Sec~\ref{sec:mpi_sync_exp}, we now want to evaluate their effect
on MPI benchmarking results.

\begin{figure}[t]
  \centering
  \includegraphics[width=\linewidth]{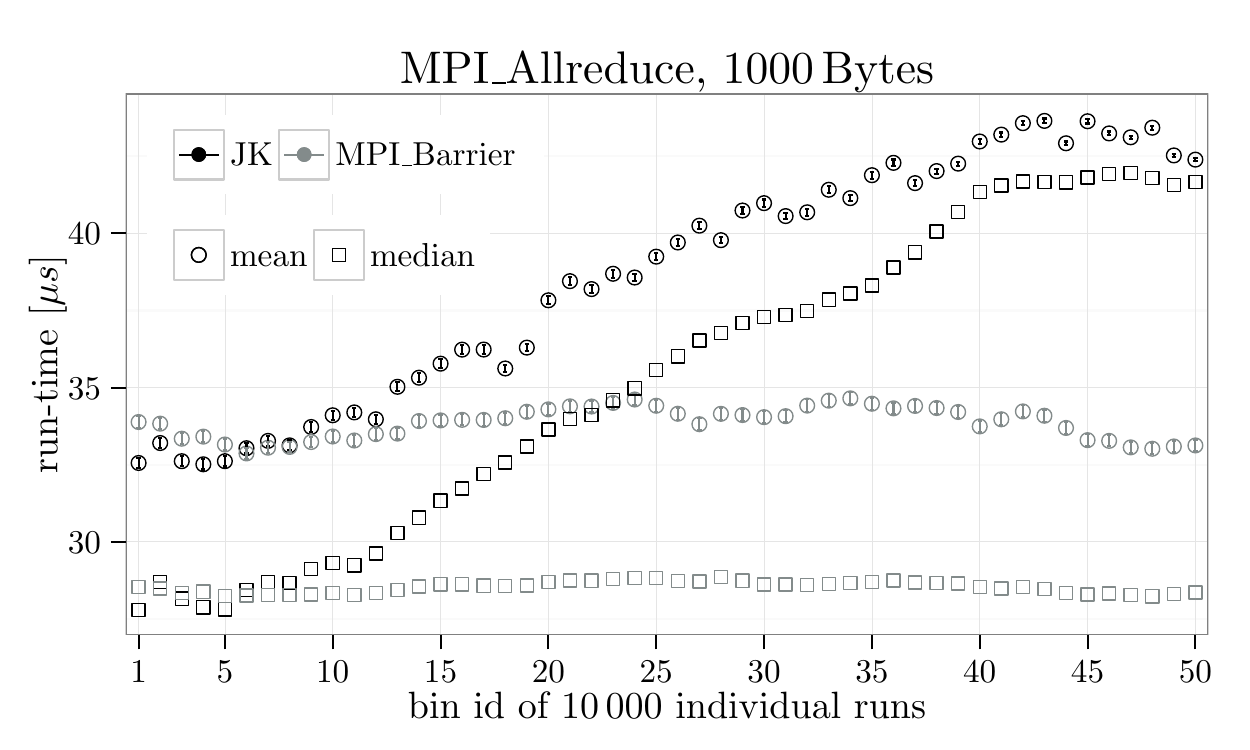}
  \caption{\label{fig:jones_skew_allreduce}Mean and median of
    \runtimes of \mpiallreduce when synchronizing either with the
    method of \jonesk (window size: \SI{1}{\milli\second}) or 
    \mpibarrier (\SI{1000}{\Bytes}, \num{16x1}~processes,
    \num{500000}~runs, bin size: \num{10000}, \mvapichtwoone,
    \machone, \expdesc: \append~\ref{sec:exp:runtime_drift_jk}).}
\end{figure}

We start by looking at the evolution of \runtime measurements over a
longer period of time in~\fig~\ref{fig:jones_skew_allreduce}.  This
graph compares the \runtimes of \mpiallreduce measured using the
synchronization method of \jonesk with the ones obtained using an
\mpibarrier.
The plot exposes two critical issues when measuring and analyzing MPI
performance data. First, we see a significant difference between the
mean and median \runtimes, which we computed for each bin of
\num{10000}~runs. The difference is also present when outliers are
removed (using Tukey's method, \cf~\Sec\ref{sec:data_processing}).
Second, the use of a window-based synchronization method might allow
the experimenter to obtain more accurate results. But even with a very
precise clock synchronization method, such as the algorithm of
\jonesk, the \runtimes will gradually drift apart over time.

\begin{figure}[t]
  \centering
  \includegraphics[width=\linewidth]{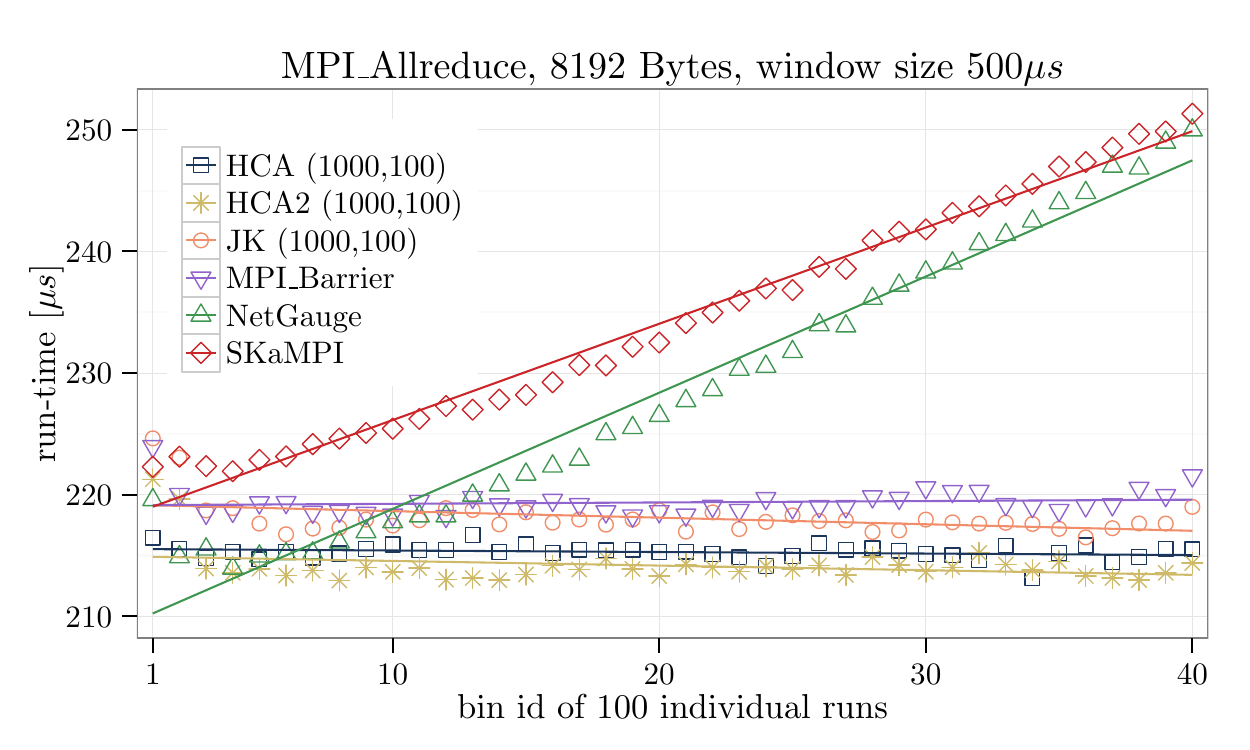}
  \caption{\label{fig:all_runtime_drift_allreduce}Drifting \runtimes
    of \mpiallreduce (median of \num{10}~experiments,
    \SI{8192}{\Bytes}, \num{32x16}~processes, \num{4000}~runs, bin
    size \num{100}, window size: \SI{500}{\micro\second}, 
    \mvapichtwoone, \machone, \expdesc:
    \append~\ref{sec:exp:runtime_drift_all}).}
\end{figure}

Now, we would like to know how the different clock synchronization
algorithms compare to each other in terms of clock
drift. \fig~\ref{fig:all_runtime_drift_allreduce} compares the
resulting \runtimes for \mpiallreduce with a message size of
\SI{8192}{\Bytes} obtained with different synchronization methods.
The \runtimes are computed as medians of bins over
\num{10}~experiments, where for each experiment we binned every
\num{100}~measurements and computed their means.
We can observe that a window-based approach might improve the accuracy
of the execution time of MPI functions. For example, in
\fig~\ref{fig:all_runtime_drift_allreduce}, the \runtimes measured
with \netgauge, \syncjk, or \synchca are smaller than the ones
measured with \mpibarrier. 
However, as explained before, the \runtimes obtained with \skampi and
\netgauge increase in time.

From the experiments shown above, we also see that the differences
between the \runtimes measured with either a window-based or an
\mpibarrier-based scheme are relatively small.  For that reason, the
practitioner may ask two questions: (1)~Is \mpibarrier good enough to
reasonably compare MPI measurements? and (2)~How large should the
window size be to get accurate measurements for window-based
synchronization schemes?

In our opinion, question~(1) cannot be answered generally as it
depends on the actual goal of an experiment and the implementation of
\mpibarrier. If the experimenter seeks to obtain the most accurate
timings for short-running MPI functions, the use of a window-based
scheme is recommended. For a fair comparison of MPI implementations,
relying on \mpibarrier may be completely sufficient if the same
\mpibarrier algorithm is used by all of them. 

\begin{figure}[t]
  \centering
  \includegraphics[width=\linewidth]{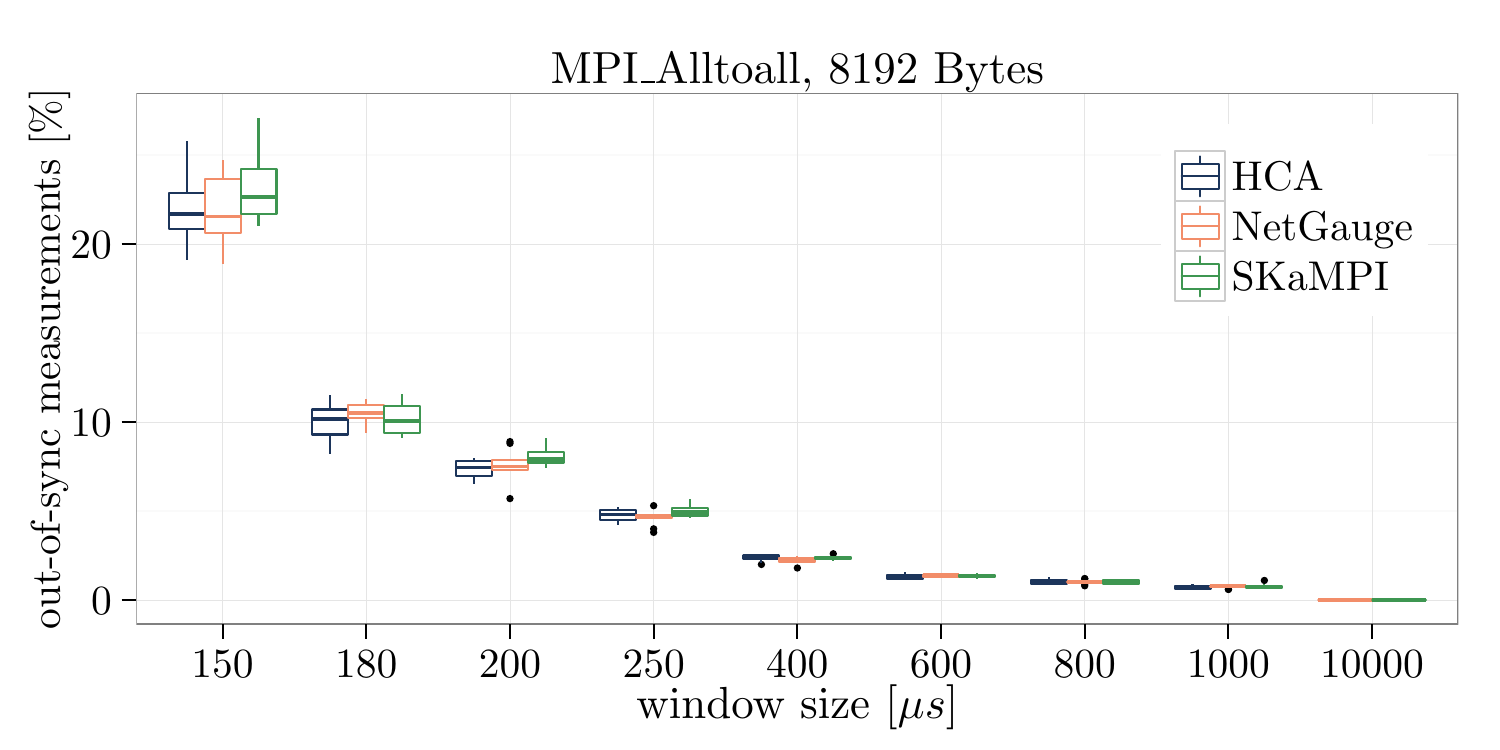}
  \caption{\label{fig:win_errors_alltoall}Percentage of incorrect
    measurements for \mpialltoall (\SI{8192}{\Bytes},
    \num{16x1}~processes, \num{1000}~measurements, \num{10}~calls to
    \mpirun,  \synchca synchronization, \mvapichtwoone, \machone, 
    \expdesc: \append~\ref{sec:exp:runtime_infl_winsize}).}
\end{figure}

To investigate how large the window size should be in order to achieve
a good trade-off between the number of correct measurements and the
duration of the entire experiment, we varied the window size and
recorded the number of out-of-sync measurements.  The implementations
of the window-based schemes found in \netgauge and \skampi increase
the window size when the number of incorrect measurements exceeds some
threshold. However, in the experiment shown in
\fig~\ref{fig:win_errors_alltoall}, we keep the window size constant
for comparison reasons.
We can see that the percentage of measurements that need to be
discarded is similar for all synchronization methods, when we measure
the \runtime of the \mpialltoall function \num{1000}~times.  It was
also expected that this percentage decreases when the window size is
increased as shown in the figure.

\begin{figure}[t]
  \centering
\includegraphics[width=\linewidth]{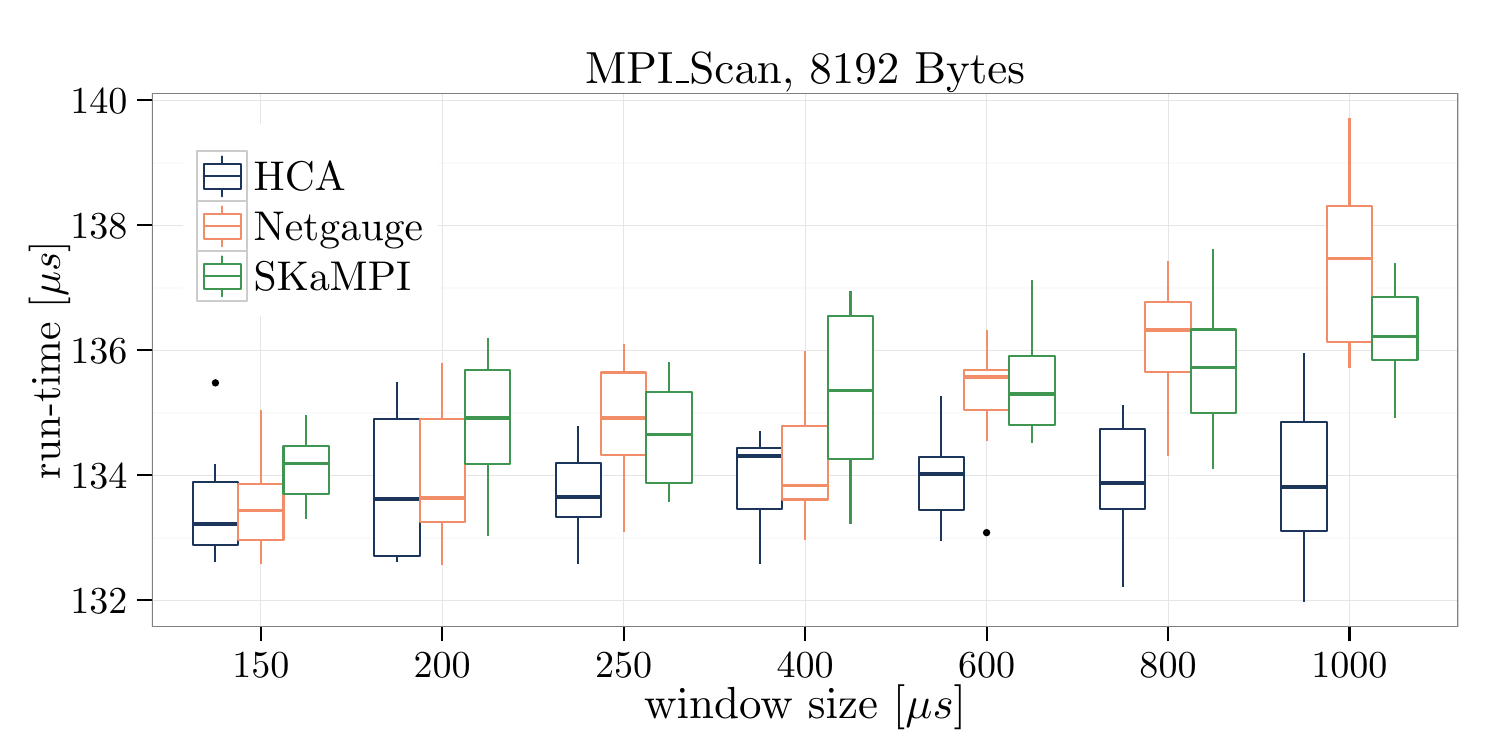}
  \caption{\label{fig:win_effect_scan}Impact of the window size on the
    \runtime of \mpiscan (median \runtime distribution of \num{10}~calls to \mpirun,
    \SI{8192}{\Bytes}, \num{16x1}~processes,
    \num{1000}~measurements, 
    \synchca synchronization,
    \mvapichtwoone, \machone, \expdesc:
    \append~\ref{sec:exp:runtime_infl_winsize}).}
\end{figure}

Now, one open question remains: How large should the window (size) be?
On the one hand, the larger the window size that we select, the more
time will elapse, which will result in a larger clock drift. On the
other hand, if the window size is too small, we will have to dismiss
many measurements. For this reason, \fig~\ref{fig:win_effect_scan}
compares the mean \runtimes measured with increasing window sizes for
different clock synchronization methods.
The figure shows that the \runtime of \mpiscan obtained using the
\synchca synchronization method stays relatively stable regardless of
the window size. This is not the case for the clock synchronization
methods used in \netgauge or \skampi; here, the \runtimes increase
when the window size grows. This behavior is again a consequence of
ignoring the clock drift in their clock synchronization methods.

\subsection{Factor: Pinning MPI Processes}
\begin{figure}[t]
  \centering
  \includegraphics[width=\linewidth]{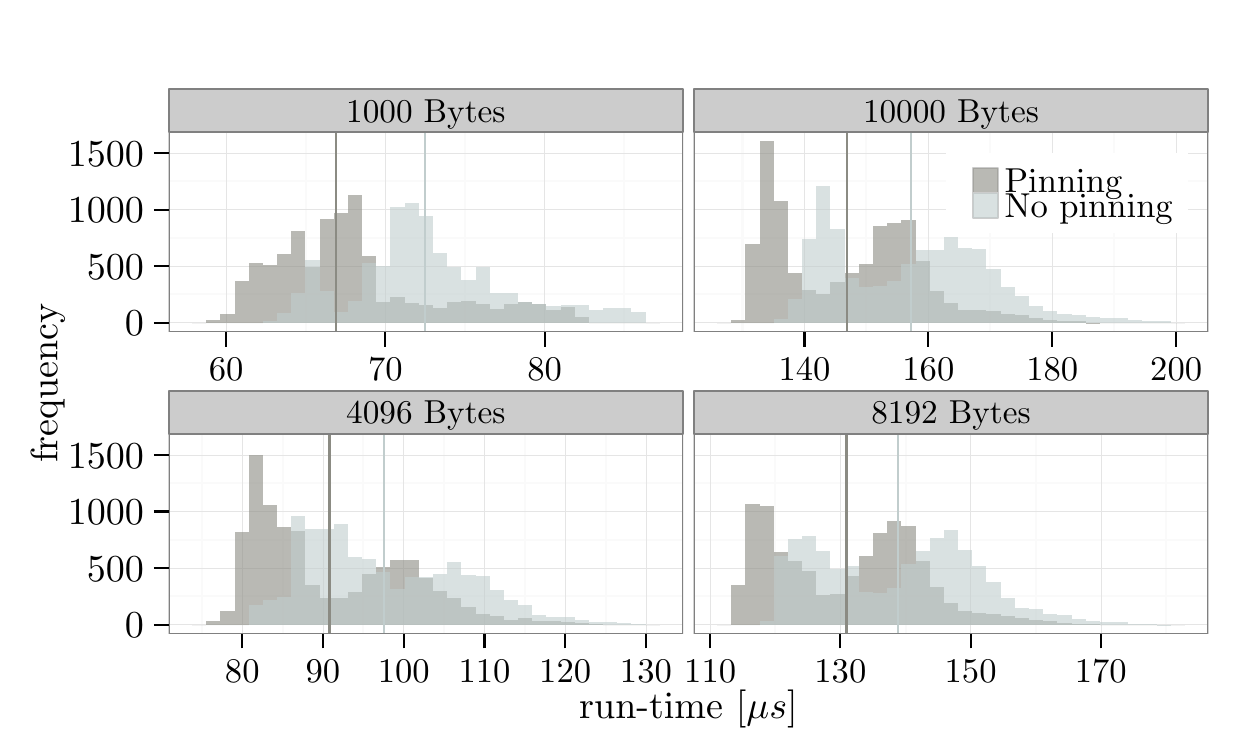}
  \caption{\label{fig:pinning_hist_nodes16_ppn16}Pinning effect on the
    \runtime of \mpiallreduce (\num{256}~processes (\num{16x16}),
    \num{1000}~measurements, \num{10}~calls to \mpirun, \mpiwtime,
    \synchca synchronization, window size: \SI{1}{\milli\second}, 
    \necmpitwoeleven,
    \machone (\expdesc: \append~\ref{sec:exp:pinning}).}
\end{figure}
It is well-known that the performance of MPI applications might be
sensitive to the way processes are pinned to CPUs, as pinning can
influence several performance-relevant properties, such as the cache
reuse or the applicability of intra-node communication.

In the context of MPI benchmarking, CPU pinning is certainly required
if we want to use the \rdtsc instruction to measure the \runtime,
since unpinned processes might result in erroneous results (\cf
\Sec~\ref{sec:rdtsc_measure}). Yet, the more general question is: Does
pinning affect the execution time of MPI functions?

\fig~\ref{fig:pinning_hist_nodes16_ppn16} shows the results of an
experiment in which we investigate whether the \runtime of an MPI
function changes if processes are pinned to CPUs or not (using
\mpiwtime for time measurements).  The figure
presents the histograms of \runtimes for \mpiallreduce and various
message sizes. Each histogram is generated by accumulating all
\runtime measurements from \num{10}~different calls to \mpirun.  We
can clearly see a significant difference in the shape of the
histograms and between the mean \runtimes, which are marked with a
vertical line.  Even though there could be cases where pinning has no
effect on measurements, we have shown that pinning is an experimental
factor to be considered in the context of MPI benchmarking.

\subsection{Factor: Compiler and Compiler Flags}

It seems self-evident to consider the compiler and the compiler flags
as being significant experimental factors of MPI benchmarking
applications. We still need to measure this effect to support our
conjecture.

\begin{figure}[t]
  \centering
  \includegraphics[width=\linewidth]{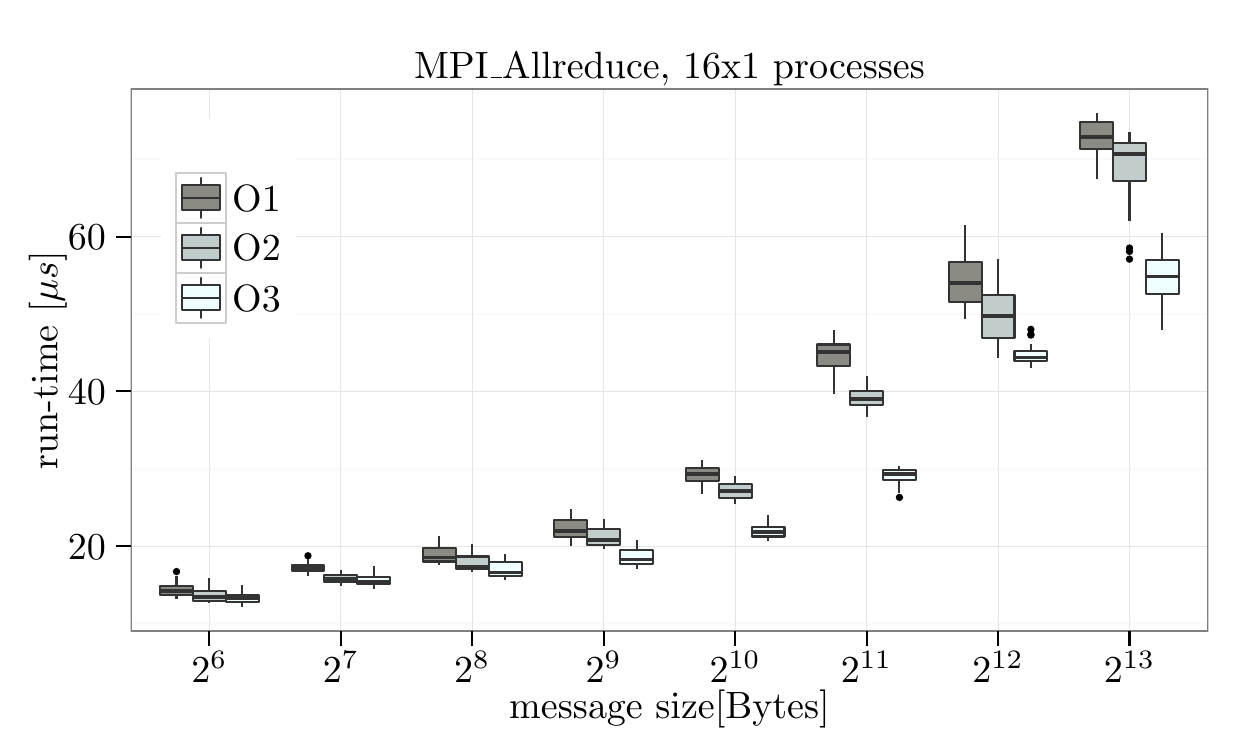}
  \caption{\label{fig:compiler_nodes16_ppn1}Compiler effect on the
    \runtime of \mpiallreduce 
    (median \runtime distribution of \num{30}~calls to \mpirun, 
    \num{16x1}~processes, \num{1000}~measurements, 
    \synchca synchronization, window size: \SI{1}{\milli\second}, 
    \mvapichtwoone, \machone, \expdesc:
    \append~\ref{sec:exp:compiler}).}
\end{figure}

We conducted an experiment in which we measured the \runtime of a call
to \mpiallreduce with the same version of \MVAPICH. We recompiled the
entire library (\mvapichtwoone) with \gcc~4.4.7, but for each
experimental run we changed the optimization flag to either
\texttt{-O1}, \texttt{-O2}, or \texttt{-O3}.
\fig~\ref{fig:compiler_nodes16_ppn1} 
clearly shows that compiling the library using \texttt{-O3}
outperforms the versions with other optimization flags.  Even though
it seems obvious, our message is this: if an MPI benchmarking
experiment does not clearly state the compiler and the compilation
flags used, the results will not be comparable or might not even be
trustworthy.

\subsection{Factor: DVFS}

The majority of today's processors offer dynamic voltage and frequency
scaling (DVFS) capabilities to reduce the energy consumption of the
chip. Changing the core frequency is therefore an obvious factor for
computationally-intensive workloads. In this work, we investigate
whether the choice of the DVFS level may alter the \runtimes of MPI
operations.

\begin{figure}[t]
  \centering
  \includegraphics[width=\linewidth]{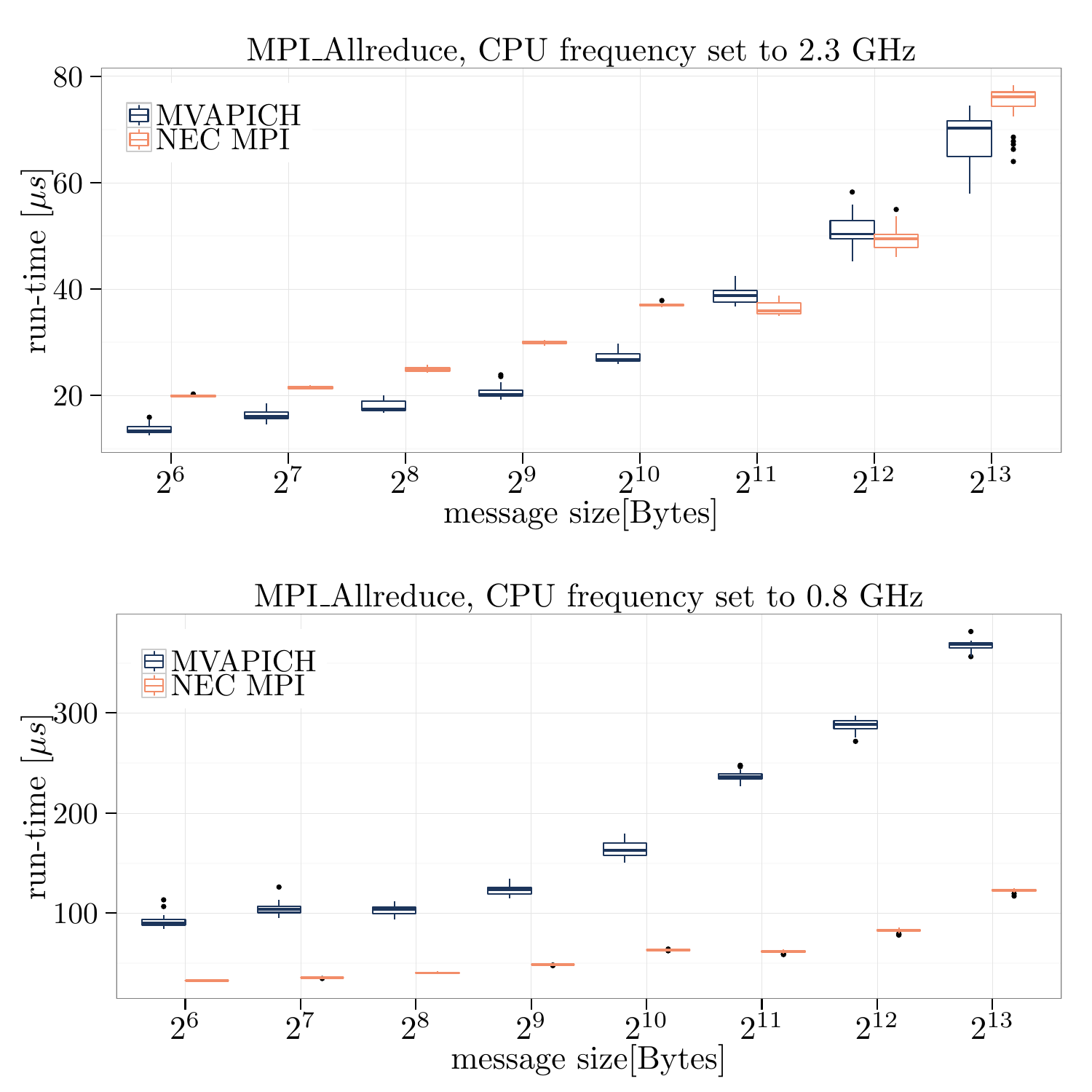}
  \caption{\label{fig:dvfs_nodes16_ppn1}DVFS configuration effect on
    \mpiallreduce \runtimes (median \runtime distribution of \num{30}~calls 
    to \mpirun, \num{16x1}~processes, \num{1000}~measurements,
    \synchca synchronization, window size: \SI{1}{\milli\second},
    \mvapichtwoone vs. \necmpitwoeleven, \machone, \expdesc:
    \append~\ref{sec:exp:dvfs}).}
\end{figure}

We conducted an experiment on \machone, in which we compared the
\runtimes of \mpiallreduce for two different MPI implementations,
\mvapichtwoone and \necmpitwoeleven, and for two different DVFS
levels, \SI{2.3}{\giga\hertz} and
\SI{0.8}{\giga\hertz}. \fig~\ref{fig:dvfs_nodes16_ppn1} presents the
results of this experiment. The upper graph shows that \MVAPICH
outperforms \NEC for message sizes of up to
\SI[exponent-base=2]{d10}{\Bytes} when all processors are running at a
fixed frequency of \SI{2.3}{\giga\hertz}. In contrast, when we change
the frequency to \SI{0.8}{\giga\hertz} for all the processors, \NEC
dominates \MVAPICH for all message sizes. Additionally, we see that
the individual \runtimes of \mpiallreduce increase significantly when
reducing the cores' frequencies.

The key observation is that the DVFS level needs to be carefully
stated. Two MPI implementations may compare and behave differently
depending on the chosen DVFS level.

\subsection{Factor: Warm vs. Cold Cache}

Gropp and Lusk~\cite{GroppL99} had already named the problem of
``ignor[ing] cache effects'' among the perils of performance
measurements. They pointed out that the time to complete a send or
receive operation depends on whether the send and receive buffers are
in the caches or not.  Therefore, \mpptest uses larger arrays for
sending and receiving messages, but the offset from where messages are
sent or received is changed in a block-cyclic way at every iteration,
to reduce the chance that data resides in cache.

The influence of caching was shown by Gropp and Lusk using \mpptest
for measuring the \runtime of point-to-point communication. In the
present work, we investigate how large the effect of caching is on
blocking collective MPI operations. Instead of using buffer-cycling,
we implemented another approach: we assume that the size of the last
level of data cache, which is private to each core, is known. On
current hardware this is often cache level~2. Let the size of this
data cache be $S^{\textit{LLC}}$ Bytes. We allocate an auxiliary
buffer $\textit{buf}_\textit{aux}$ containing
$S^{\textit{LLC}}$\,Bytes.  Now, we alter our MPI benchmark as
follows: we overwrite the entire buffer $\textit{buf}_\textit{aux}$
(using \texttt{memset}) after each iteration, \ie, when one
measurement of a collective MPI call has been completed. This way we
attempt (since we do not know the hardware details) to ensure that our
message buffer used for the MPI operation is not cached.

\begin{figure}[t]
  \centering
  \includegraphics[width=\linewidth]{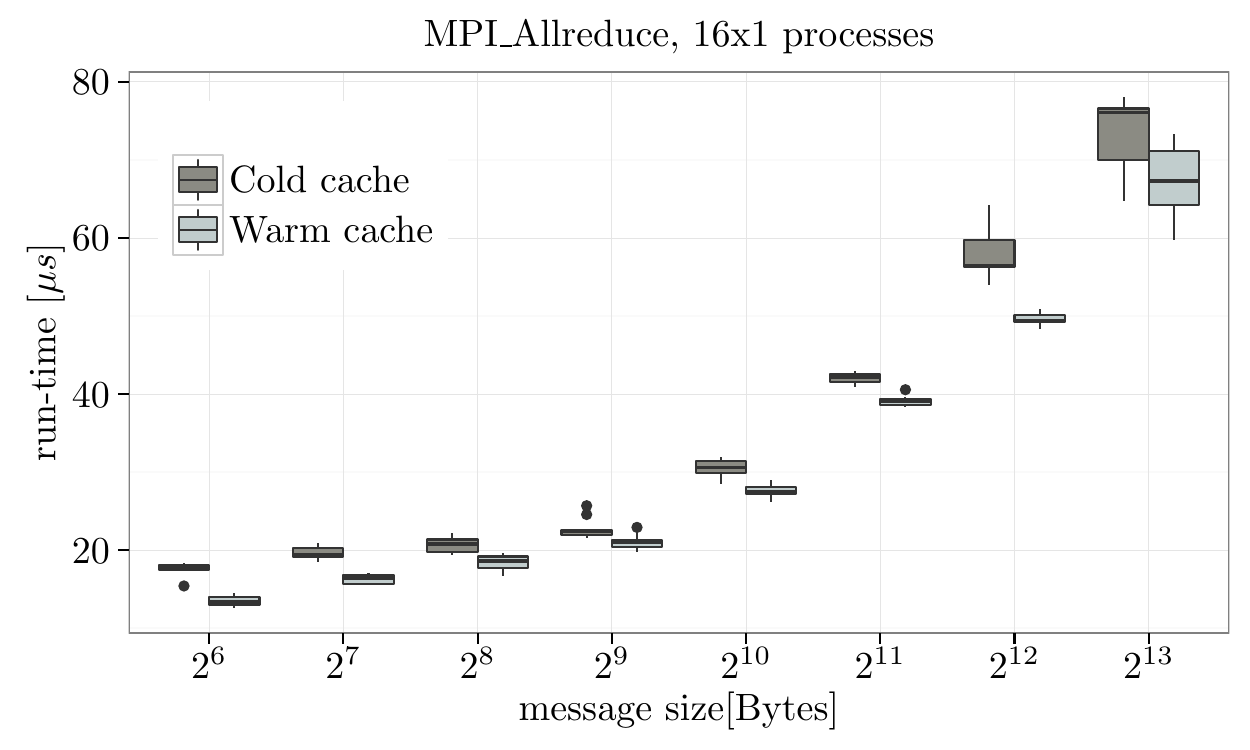}
  \caption{\label{fig:factor_caching_allreduce}Cold cache vs. warm
    cache: median \runtimes for \mpiallreduce 
    (median \runtime distribution of \num{10}~calls to \mpirun, 
    \num{16x1}~processes, \num{1000}~measurements, 
    \synchca synchronization,
    window size: \SI{1}{\milli\second}, \mvapichtwoone, \machone,
    \expdesc: \append~\ref{sec:exp:caching}).}
\end{figure}

The effect of caching is shown in
\fig~\ref{fig:factor_caching_allreduce}, in which we can see that the
reuse of message buffers between subsequent MPI calls, in this case
\mpiallreduce, has a significant impact on the \runtime.  As a result,
MPI benchmarks must clearly state whether and how the caching of
messages (buffers) is controlled.

\subsection{Summarizing Experimental Factors}

Our initial goal was to allow a fair and reproducible comparison of
the performance of MPI implementations. A well-defined experimental
design is one requirement to achieve that goal, and therefore, the
analysis of experimental factors is of major importance.  We have
analyzed factors, such as compiler flags or cache control, and
evaluated whether they have a significant effect on the experimental
outcome. The influence of some factors on the performance measurements
was not surprising, for example, we had expected that the DVFS level
would affect the \runtimes.

However, the experiments led to two main results: The first lesson we
learned was that the execution time of MPI benchmarks varies
significantly between calls to \mpirun. As a consequence, a
reproducible and fair comparison of \runtime measurements requires
that performance data are recorded from different calls to \mpirun.
The second lesson, that we found quite surprising, was that
determining which MPI implementation is better for a given case
depends on the configuration of the experimental factors. For example,
the \runtime of \mpibcast might be shorter with library~A using DVFS
level ``low'', but library~B will provide a faster \mpibcast
implementation in DVFS level ``high''.

Of course, our list of examined experimental factors is not
exhaustive, and we are aware that other factors could also impact the
experimental outcome.  One such example is the operating system. Since
many of such factors are often uncontrollable, we need to address them
statistically.

In conclusion, we advise MPI experimenters to carefully list the
settings of all experimental factors, besides the obvious factors such
as number of processes, message size, and parallel machine.
\tab~\ref{tab:exp_factors} is our proposal of a list of experimental
factors that, we believe, should be attached to all MPI benchmark
data.

\begin{table}[t]
  \centering
  \caption{Experimental factors in MPI benchmarking.}
  \label{tab:exp_factors}
  \begin{scriptsize}
    \begin{tabular}{ll}
      \toprule
      factor  & example \\
      \midrule
      MPI implementation & \mvapichtwoone \\
      network & \infiniband QDR MT4036 \\
      synchronization method & window-based scheme \\
              & clock synchronization: \synchca \\
              & window size: \SI{100}{\micro\second}\\
      \mpirun & \num{10}~distinct calls\\
      compiler / flags & \gcc~4.3 \texttt{-O3} \\
      DVFS level & \SI{2.3}{\giga\hertz}\\
      cache & no cache control \\
              & (messages might be cached and reused)\\
      pinning & \texttt{--bind-to-core}\\
      \bottomrule
    \end{tabular}
  \end{scriptsize}
\end{table}

\section{Statistically Rigorous and Reproducible MPI Benchmarking}
\label{sec:statistical_analysis}

After investigating the factors that may influence results of MPI
benchmarks, we now propose a method to compare MPI implementations by
using statistical hypothesis testing. Our goal is to establish an
experimental methodology that aims to reproduce the test outcome
between several experiments.

\begin{figure*}[t]
 \centering
 \includegraphics[width=.49\linewidth]{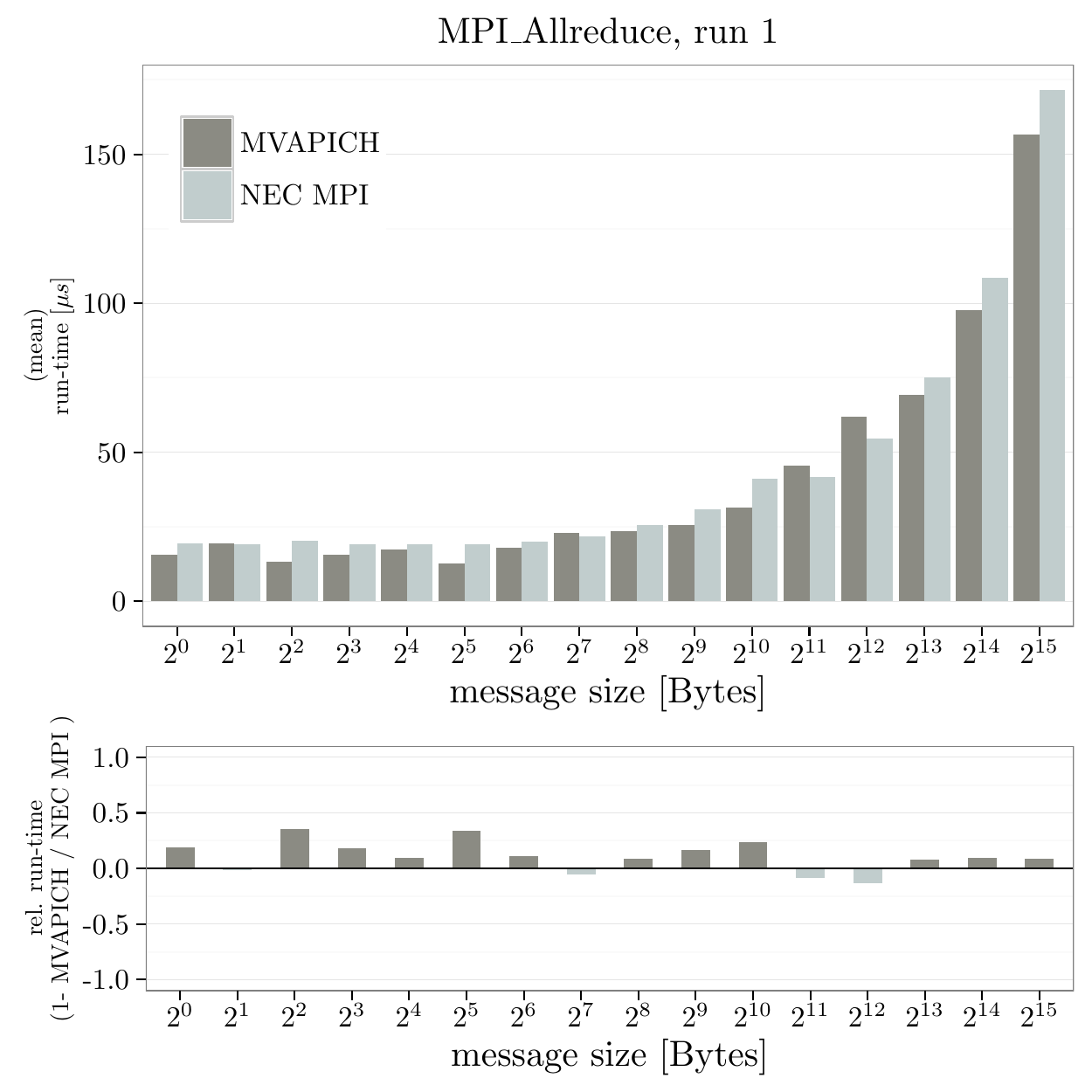}
 \hfill
 \includegraphics[width=.49\linewidth]{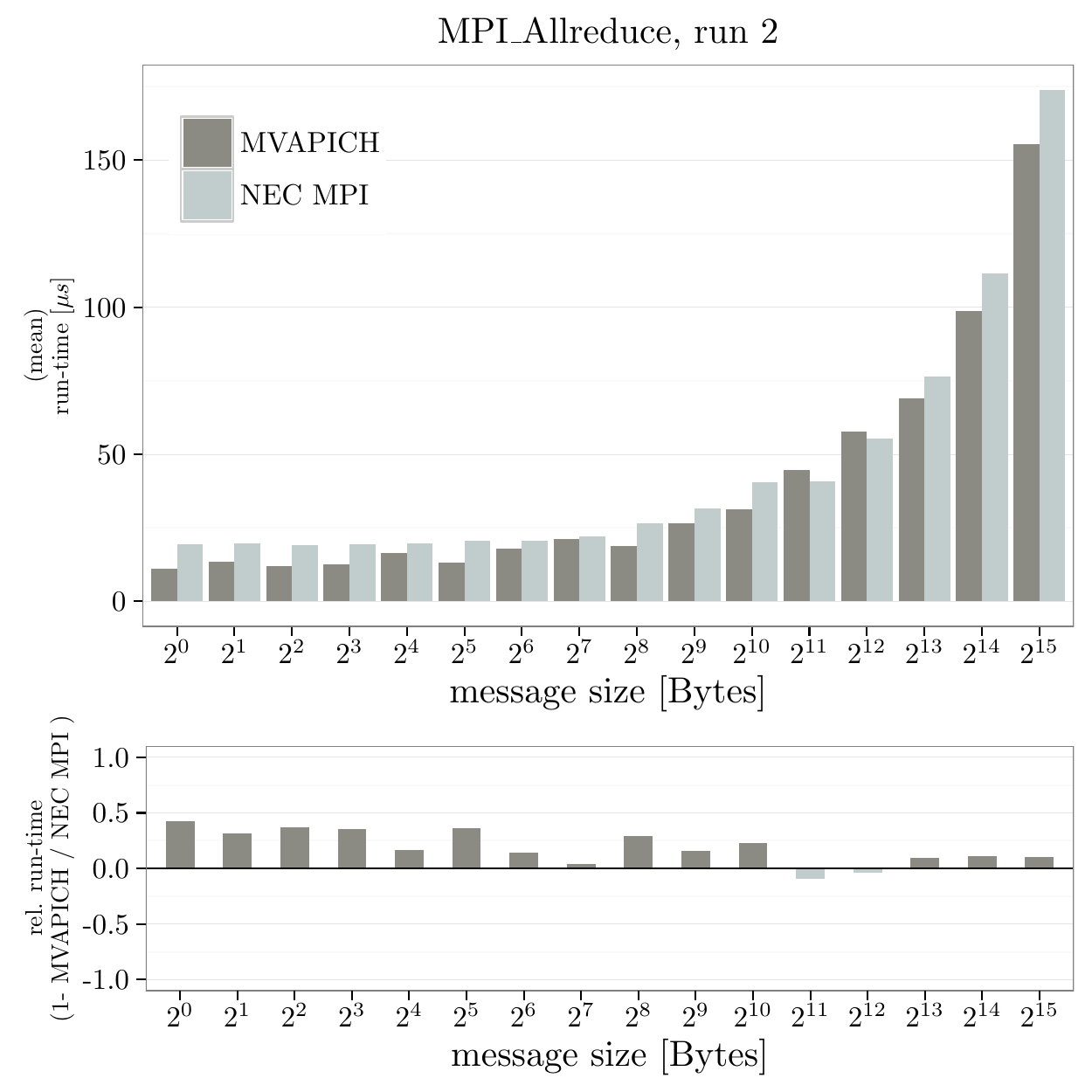}
 \caption{\label{fig:reg_stats_mean}Comparison of mean \runtimes of
   \mpiallreduce and different message sizes for two distinct calls to
   \mpirun (\num{16x1}~processes,
   \num{1000} measurements per message size, \synchca
   synchronization with window sizes adapted to each message size,
   \mvapichtwoone vs. \necmpitwoeleven, \machone, \expdesc:
   \append~\ref{sec:exp:stat_tests}).}
\end{figure*}

We motivate the need for a more robust evaluation method by showing
the results in \fig~\ref{fig:reg_stats_mean}.  On the left-hand side,
we see a comparison between the \runtime of \MVAPICH and \NEC when
executing \mpiallreduce with various message sizes. Each bar
represents the mean \runtime computed for \num{1000} individual
measurements of a single call to \mpirun. One might say that such a
comparison is fair (due to the large number of repetitions) and we
contend this is common practice when analyzing experimental results in
the context of MPI benchmarking. However, when we look at the results
shown on the right-hand side, the outcome changes significantly. For
example, the ratio of mean \runtimes for a message size of
$2^1$\,\Bytes has now changed. This observation matches the result of
our factor analysis, in which we have discovered that the call to
\mpirun is an experimental factor (\cf
\Sec~\ref{sec:influence_mpirun}). Therefore, we emphasize again that
an MPI benchmark needs to collect data from multiple \mpirun calls to
be fair and reproducible.

\subsection{Design of Reproducible Experiments}
\label{sec:data_analysis}

Our new experimental design for measuring MPI performance data is
shown in \alg~\ref{alg:repeat_procedure}. The procedure
\proctext{\designpartone} generates the experimental layout and has
five parameters, two of them being important for the statistical
analysis: (1)~\nmpiruns denotes the number of distinct calls to
\mpirun for each message size, and (2)~\nrep specifies the number of
measurements taken for each message size in each call to \mpirun. In
total, we measure the execution time of a specific MPI function for
every message size $\nmpiruns \cdot \nrep$ times. In the procedure
\proctext{\designpartone} in \alg~\ref{alg:repeat_procedure}, we issue
\nmpiruns calls to \mpirun, where the number of processes \np stays
fixed. To respect the principles of experimental design
(randomization, replication, blocking) as stated by
Montgomery~\cite{Montgomery:2006}, we randomize the experiment by
shuffling the order of experiments within a call to \mpirun. The
procedure \proctext{\designparttwo} has three parameters: the list of
message sizes (\mlist), the list of MPI functions to be tested
(\funclist), and the number of observations (\nrep) to be recorded for
each message size. The procedure creates a list (\explist) containing
the experiments covering all message sizes and MPI functions. The
order of elements in this list is shuffled before each item
(experiment) is executed.
\begin{algorithm}[t]
  \begin{scriptsize}
  \caption{\label{alg:repeat_procedure}Design of MPI experiment.}
  \begin{algorithmic}[1] 
  \Procedure{\designpartone}{\np, \nmpiruns, \mlist, \funclist, \nrep}
  \Statex \mycomment{\np\ - nb. of processes}
  \Statex \mycomment{\nmpiruns\ - nb. of \mpiruns} 
  \Statex \mycomment{\mlist\ - list of message sizes}
  \Statex \mycomment{\funclist\ - list of MPI functions}
  \Statex \mycomment{\nrep\ - nb. of measurements per run}
  \For{i in 1 to \nmpiruns}
   \State mpirun -np \np\ \Call{\designparttwo}{\mlist, \funclist, \nrep}
  \EndFor
  \EndProcedure
  \Statex
  \Procedure{\designparttwo}{\mlist, \funclist, \nrep}
  \State \explist $\leftarrow$ () 
  \ForAll{\msize in $l^m$} 
  \ForAll{\func in \funclist} 
     \State \explist.add( \textsc{Time\_MPI\_Function}(\func, \msize, \nrep) )
  \EndFor
  \EndFor
  \State shuffle(\explist) \mycomment{create random permutation of calls in place}
  \ForAll{$exp$ in \explist}
  \State call $exp$
  \EndFor
  \EndProcedure
  \end{algorithmic}
  \end{scriptsize}
\end{algorithm}

The procedure \proctext{\designpartone} of
\alg~\ref{alg:repeat_procedure} is executed for each MPI
implementation.  After the measurement results have been gathered, we
apply the data-analysis procedure shown in
\alg~\ref{alg:data_anal}. Here, we group \runtime measurements by the
message size, the type of MPI function, and the number of processes.
We remove statistical outliers from each of these measurement
groups. Last, we compute averages (the median and the mean) for each
group of measurements and store them in a table.
\begin{algorithm}[t]
  \begin{scriptsize}
    \caption{\label{alg:data_anal}Analysis of benchmark data.}
  \begin{algorithmic}[1] 
    \Procedure{Analyze\_Results}{\mlist, \proclist, \funclist, \nmpiruns} 
    \Statex \mycomment{\mlist\ - list of message
      sizes} 
    \Statex \mycomment{\proclist\ - list of processes}
    \Statex \mycomment{\funclist\ - list of MPI functions}
    \Statex \mycomment{\nmpiruns\ - nb. of \mpiruns}
    \ForAll{$\msize \in \mlist, \np \in \proclist, \func \in \funclist$} 
     \For{$i$ in 1 to \nmpiruns} 
     \State $\exectimes_i =
    \{\, \exectimes[\msize][\np][\func][i][j] \quad \text{for all} \, 1 \le
    j \le \nrep\}$ 
    \State $\exectimes_i = \text{remove\_outliers}(\exectimes_i)$ 
    \State
    $v[\msize][\np][\func][i] = ( \text{median}(\exectimes_i), \text{mean}(\exectimes_i) )$
    \EndFor
    \EndFor
    \State print $v$ \mycomment{table with results}
    \EndProcedure
  \end{algorithmic}
  \end{scriptsize}
\end{algorithm}
By applying this data-analysis method, we obtain a distribution of
averages (medians or means) over $\nmpiruns$ calls to \mpirun for each
message size, MPI function, and number of processes.

\subsection{Fair Performance Comparison Through Statistical Data
  Analysis}

\begin{figure*}[t]
 \centering
 \includegraphics[width=.49\linewidth]{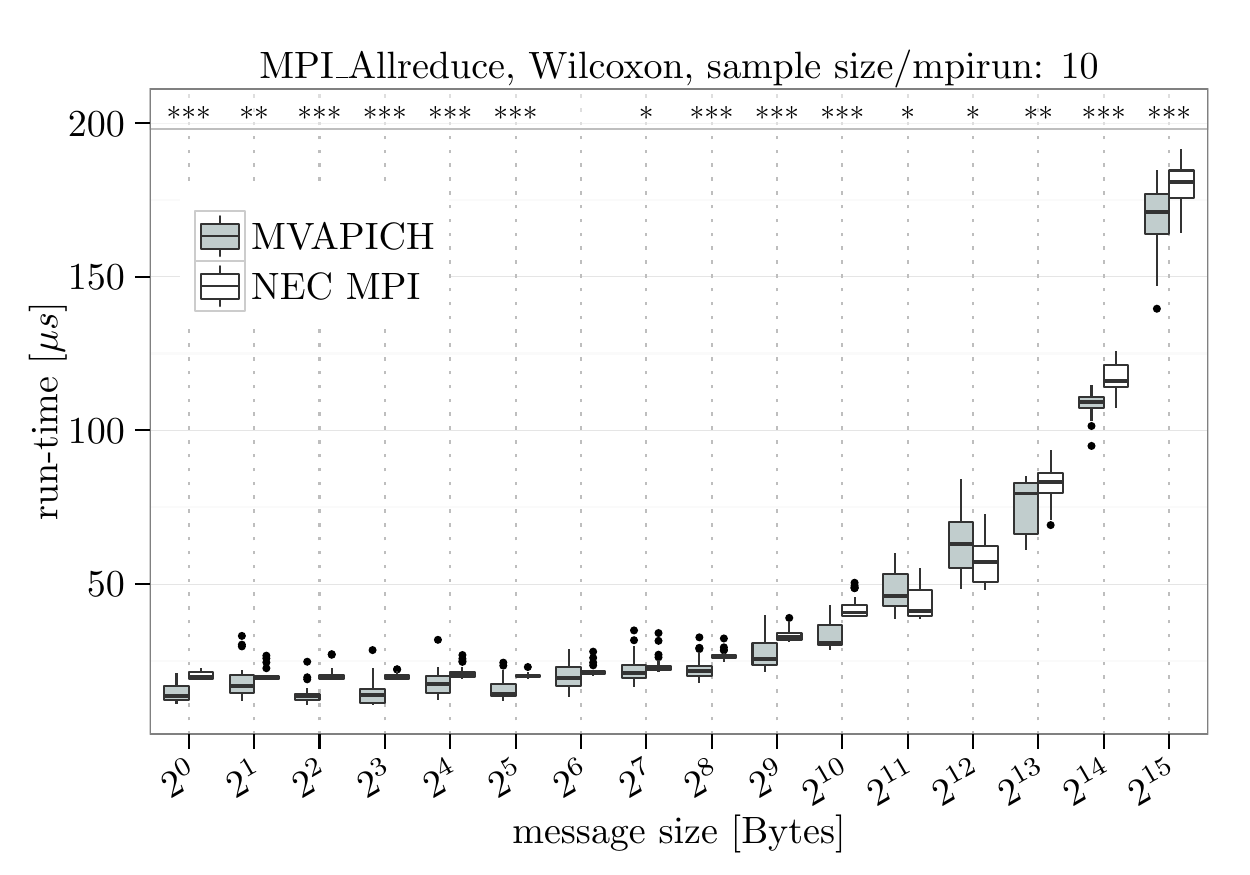}
 \hfill
 \includegraphics[width=.49\linewidth]{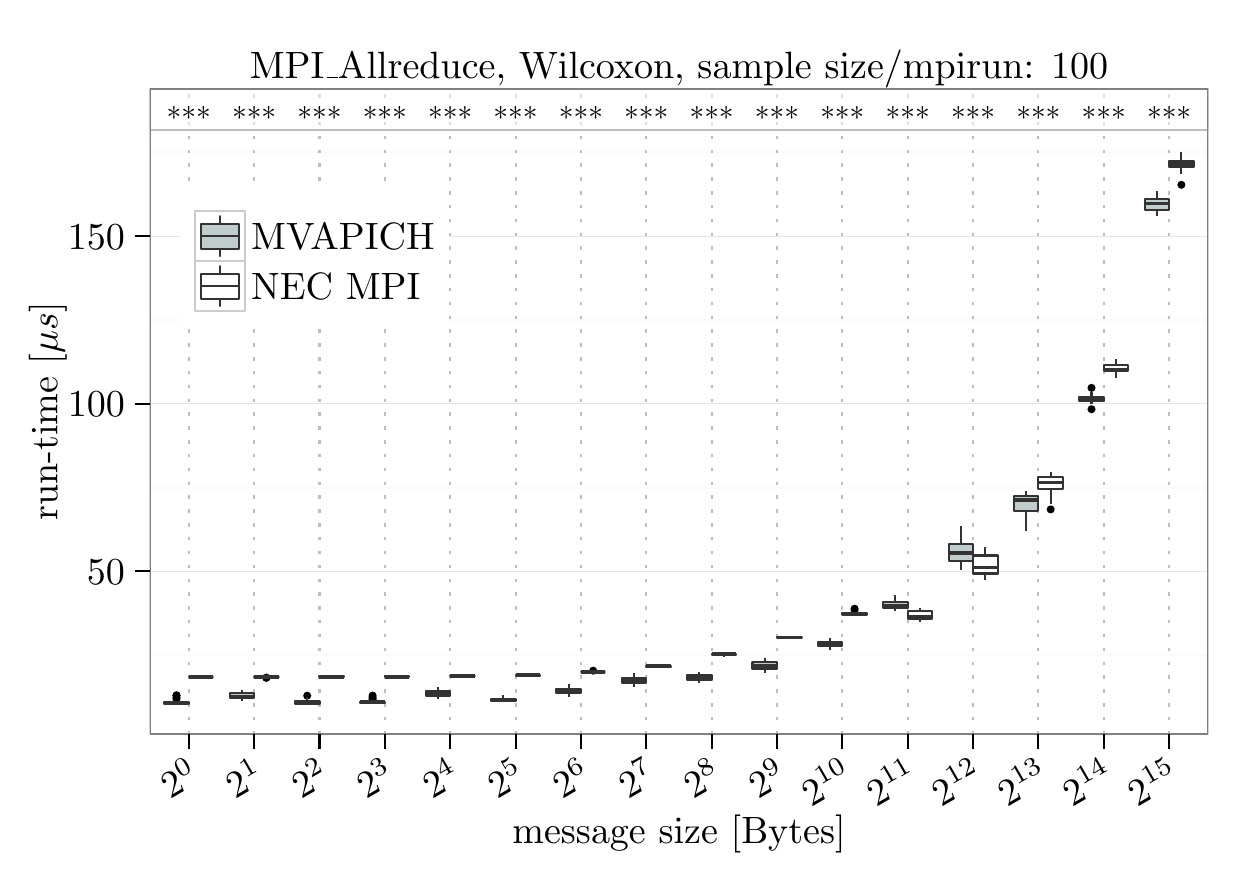}
 \caption{\label{fig:test_res_nrep10}Comparison of the \runtime
   distributions obtained when measuring \mpiallreduce with a sample
   size of \num{10} (left) and \num{100} (right) per message size.
   The statistical significance was computed using the
   \testwilcox (\num{16x1} processes, \num{30} calls to \mpirun,
   \synchca synchronization with window sizes adapted to each message
   size, \mvapichtwoone vs. \necmpitwoeleven, \machone, \expdesc:
   \append~\ref{sec:exp:stat_tests}).}
   
\end{figure*}

Now, the situation is as follows: we run our benchmark on two MPI
implementations $A$ and $B$, perform the data analysis according to
\Sec~\ref{sec:data_analysis}, which gives us a distribution of
averages for each measurement point.
The question then becomes: how can we compare the measured results in
a statistically sound way? We could reduce all the values from the
distribution to a single value using the minimum, the maximum, or the
average, and then compare two MPI implementations based on this single
value. However, our goal is to provide evidence that a measured
performance difference has a high probability of being reproducible
and that it is not merely a result of chance.

\begin{figure}[t]
  \centering
  \includegraphics[width=.7\linewidth]{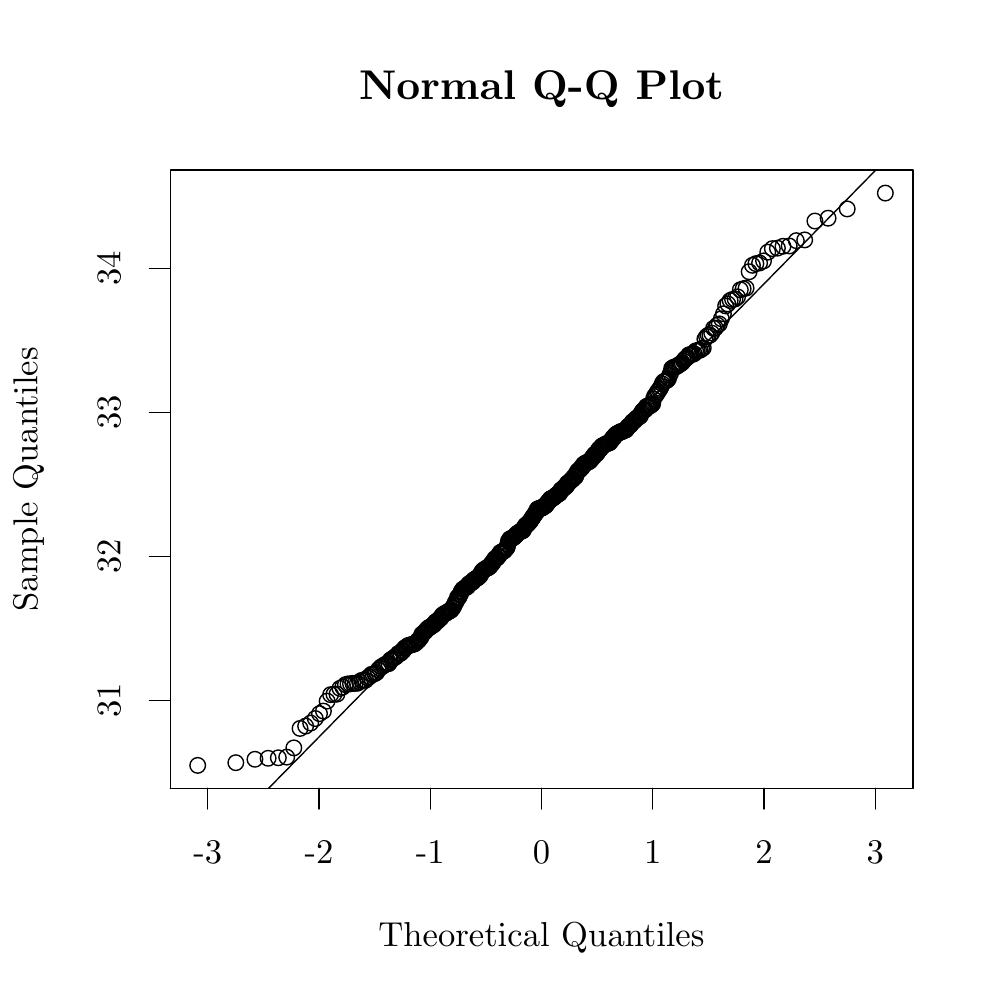}
  \caption{\label{fig:mpiruns_dist_allreduce_qqplot} Q-Q plot of mean
    \runtimes obtained from different calls to \mpirun
    (\cf~\fig~\ref{fig:mpiruns_dist_allreduce}).}
\end{figure}

Since we have two averages (the mean and the median) for each
measurement group, we have several options for selecting a statistical
test. If we use the computed median values as basis for our hypothesis
test, we could employ the nonparametric Wilcoxon–Mann–Whitney test
(Wilcoxon sum-of-ranks, in the remainder: \testwilcox) for comparing
alternatives~\cite{hollander1999nonparametric}. The advantage of the
\testwilcox (besides being nonparametric) is that it makes no
assumption on the underlying distribution; in particular, it ``does
not require the assumption of
normality''~\cite{ross2010introductory}. We could also employ the
\testwilcox on the distribution of means, but in this case the
\testttest for two independent samples is also a promising
candidate. The \testttest assumes that the underlying population is
normally distributed and that the variances of both populations are
equal~\cite{Sheskin:2007}. We first have to make sure that our sample
means computed for each \mpirun are normally distributed.  If the
underlying distributions obtained from each \mpirun are similarly
shaped, then it is possible that also their means are normally
distributed.  For example, the Q-Q plot of the mean \runtimes
(\fig~\ref{fig:mpiruns_dist_allreduce_qqplot}) suggests that the
distribution of means, which was presented in
\fig~\ref{fig:mpiruns_dist_allreduce} of
\Sec~\ref{sec:influence_mpirun}, is normally shaped. In addition, the
\kolmtest and the \shaptest test do not reject the null hypothesis,
such that we can assume normality for the distribution of
means. However, if the means follow a normal distribution, we also
need to verify that the variances are equal. 
If the
homogeneity of variance assumption is violated, several adaptations to
the \testttest have been proposed (\cf \cite[p.~458]{Sheskin:2007},
\cite{hogg2006probability}). One adaptation is the so-called
\testwelch that can be applied when two samples have unequal
variances.

In the remainder of our analysis, we use the \testwilcox
  exclusively.  The reason is that the rigorous verification of the
  distributions of means (over \mpiruns) showed that the mean
  \runtimes (obtained from \mpiruns) are often normally distributed,
  but unfortunately not all of them. Applying a \testttest in cases in
  which the means are not normally distributed will not give us the
  desired statistical confidence, and the test results would be
  misleading (since the assumptions of the test are violated).

We now demonstrate how to apply the \testwilcox to our data and
discuss why the test helps us to provide a fair comparison of MPI
implementations.  \fig~\ref{fig:test_res_nrep10} shows our statistical
comparison method applied to \runtimes measured for \mpiallreduce with
both \NEC and \MVAPICH.  Let us focus first on the graph on the left
of this figure, where we compare the distributions of means recorded
for different message sizes. Each distribution contains \num{30}
elements, which are the median \runtimes measured in each of the 30
calls to \mpirun.  We apply the \testwilcox on the two distributions
of medians for each message size. The test does not only report
whether the null hypothesis (both population means are equal) is
rejected or not, but it also provides a \emph{p-value}. To obtain a
graphical representation of the p-value and therefore the statistical
significance, we represent the p-value by a sequence of asterisks.
One asterisk ($\ast$) represents a p-value of $p \le 0.05$, two
asterisks denote $p \le 0.01$, and three asterisks denote
$p \le 0.001$. It also means that if asterisks are absent in a
specific case, the null hypothesis could not be rejected, and thus,
the statistical test does not provide sufficient evidence which
implementation is better. We used a significance level of \num{0.05}
(\SI{5}{\percent}) for all experiments.

When we look at the left graph of \fig~\ref{fig:test_res_nrep10}, in
which we applied the \testwilcox, we see that using a hypothesis test
can indeed help to separate cases, for which a decision can hardly be
made only by looking at the distributions. For example, the
differences between the distributions for $2^5$ and $2^6$\,\Bytes seem
to be negligible. However, the \testwilcox reveals that there is
evidence that the sample medians are different in the case of
$2^5$\,\Bytes, but not in the case of
$2^6$\,\Bytes. 

The graph on the right of \fig~\ref{fig:test_res_nrep10} presents the
results when applying the \testwilcox with a sample size of \num{100}
per \mpirun. 
It is not surprising that the variances of the distributions of the
averages decrease, and thus, a larger sample size helps the hypothesis
test to separate averages with a higher significance.

\begin{figure}[t]
 \centering
 \includegraphics[width=\linewidth]{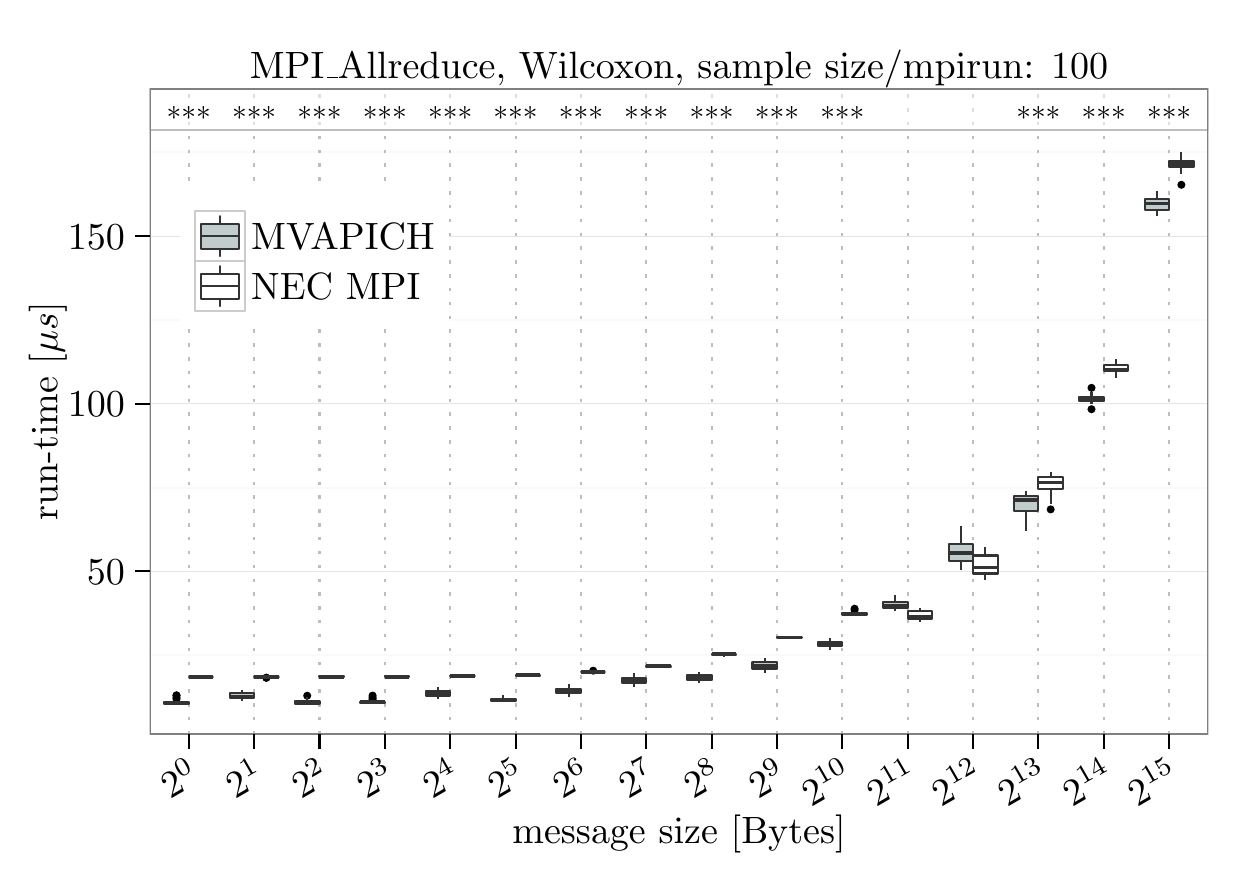}
 \caption{\label{fig:test_res_nrep_less} Comparison of the \runtimes
   of \mpiallreduce while applying the \testwilcox with ``less'' as
   alternative hypothesis  
   (\num{16x1} processes, \num{30} calls to \mpirun, \synchca 
   synchronization with window sizes adapted to each message size, 
   \mvapichtwoone vs. \necmpitwoeleven,
   \machone, \expdesc: \append~\ref{sec:exp:stat_tests}).}
\end{figure}

The graphs in \fig~\ref{fig:test_res_nrep10} compare the \runtime
distributions of two MPI implementations and show the statistical
results when testing whether the population averages are equal. Yet,
in a practical scenario one might rather ask a question like: Is MPI
library $X$ faster than library $Y$ for MPI function $F$? To answer
such a question, we change the alternative hypothesis of the test to
``less'' 
(null hypothesis: $H_0:\, \mu_A = \mu_B$,
alternative hypothesis: $H_a:\, \mu_A < \mu_B$, where $\mu$ denotes
the average).  \fig~\ref{fig:test_res_nrep_less} presents the results
of the same experiments as shown in the bottom right corner of
\fig~\ref{fig:test_res_nrep10}, but now we check whether the \runtime
of \mpiallreduce is smaller with \MVAPICH than with \NEC.

We see that for the two cases $2^{11}$ and $2^{12}$\,Bytes the null
hypothesis could not be rejected, and thus, in these cases the
\runtime of \mpiallreduce using \MVAPICH is not smaller than when
using \NEC. We note that this result does not immediately imply that
\NEC is faster than \MVAPICH in these cases. To verify this, the test 
should use the alternative hypothesis ``greater''.

\subsection{Evaluating the Outcome Reproducibility}

Until now, we have investigated the factors that potentially influence
the benchmarking of MPI functions and have shown how statistical
hypothesis tests help us to fairly compare the performance of two MPI
libraries. One of our initial goals was to develop a benchmarking
method that leads to a reproducible experimental outcome (see
\tab~\ref{tab:moti_intel_mpi}).

\begin{figure*}[t]
\centering
 \subfigure[\intelbench 4.0.2]%
{\includegraphics[width=0.32\linewidth]{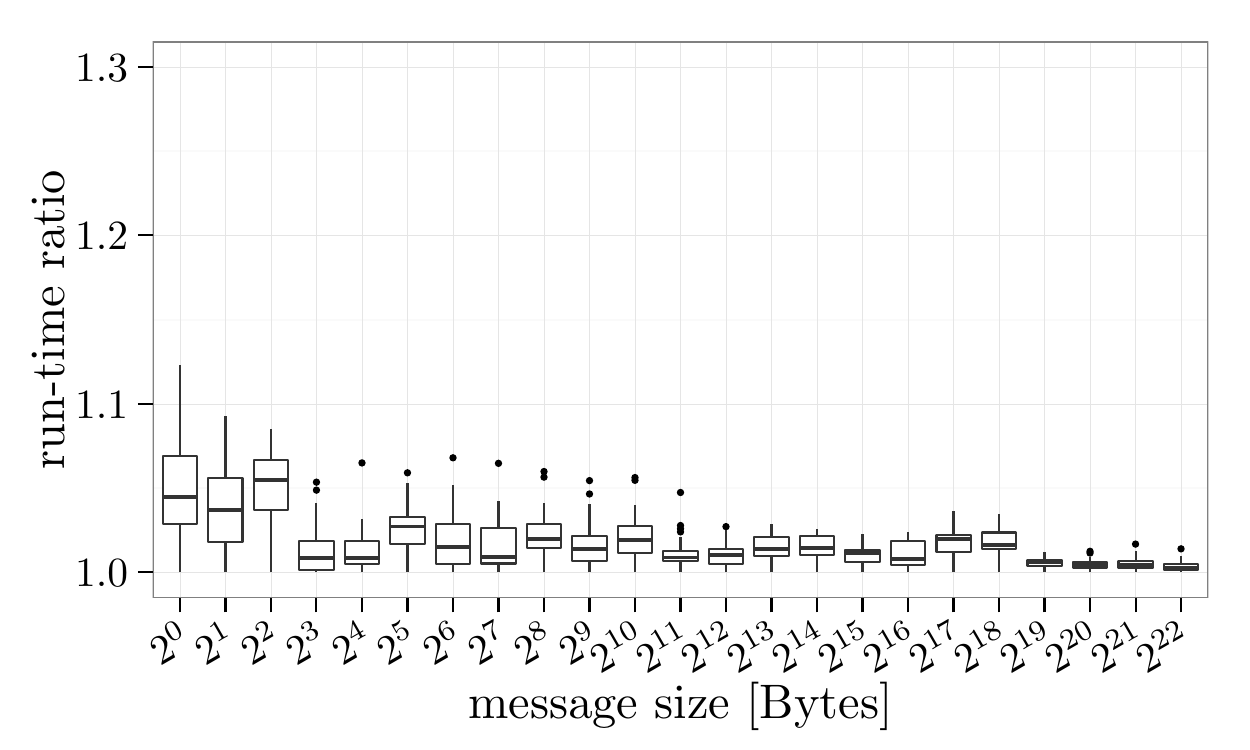}
\label{fig:runtimeratio-a}}%
\hfill
\subfigure[\skampi 5]%
{\includegraphics[width=0.32\linewidth]{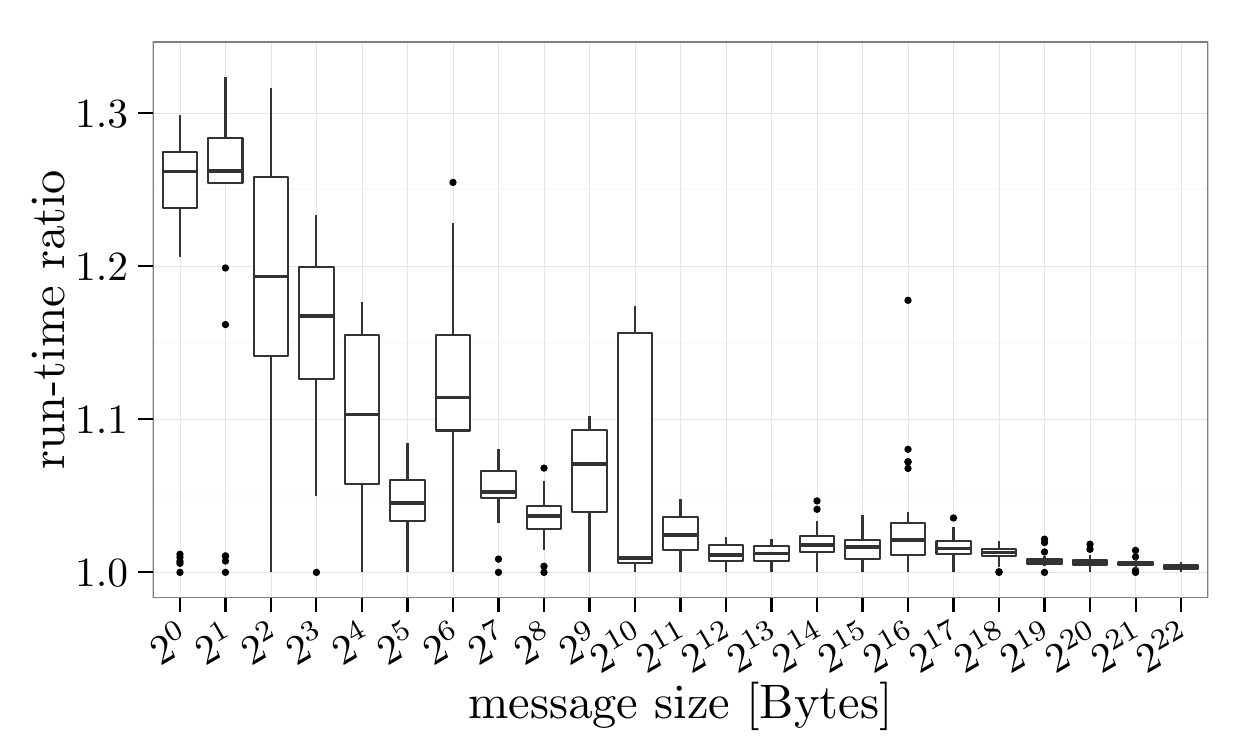}
\label{fig:runtimeratio-b}}%
\hfill
\subfigure[Our method]%
{\includegraphics[width=0.32\linewidth]{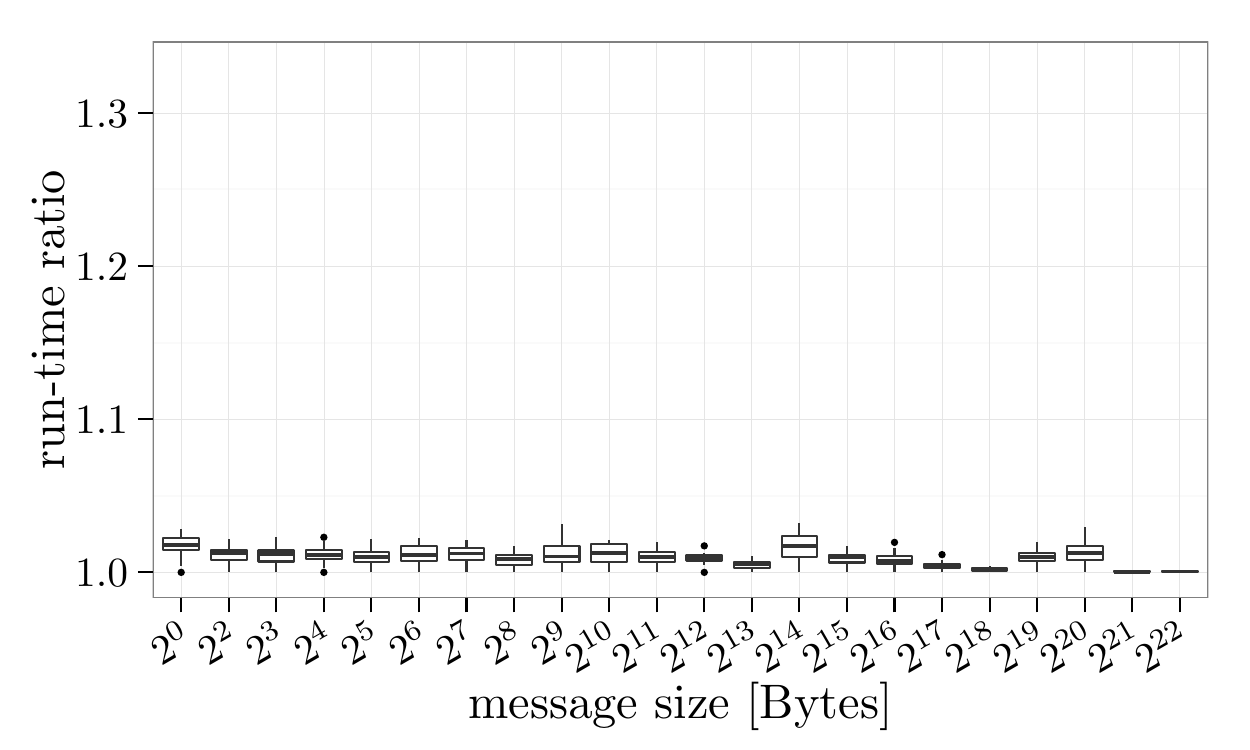}
\label{fig:runtimeratio-c}}%
\caption{\label{fig:runtimeratio}Distribution of normalized \runtimes
  reported by the \intelbench (left), \skampi (center), and
  our method (right) when testing the performance of \mpibcast for
  various message sizes (\num{16x1} processes,  
  \synchca synchronization with window sizes adapted to each message
  size for \fig~\ref{fig:runtimeratio-c}, \mvapichtwoone, \machone, \expdesc: \append~\ref{sec:exp:reprod}).}
\end{figure*}

To examine the reproducibility of our benchmarking method, we
conducted the following experiment: We ran our benchmarking method
(\cf \alg~\ref{alg:repeat_procedure}) for $\ntrial$=\num{30}~times.
Each of the $\ntrial$ runs gave us one distribution of \runtimes per
message size, which contains $\nmpiruns$=\num{30} values. Since we
obtain a distribution of distributions, we collapse the inner
distribution into a single value. To do so, we compute the mean of the
$\nmpiruns$=\num{30}~values measured for one message size in each of
the \ntrial distributions.  Then, we normalize the \runtime values by
computing the ratio of each mean to the minimum mean. We obtain a
distribution of $\ntrial$=\num{30} normalized \runtime values for our
benchmarking method, presented in \fig~\ref{fig:runtimeratio-c}. We
can observe that the maximum relative difference between the
\num{30}~runs is very small (less than~\SI{5}{\percent} for
$2^{14}$\,\Bytes).

As a comparison, we also conducted $\ntrial$=\num{30}~runs of the
\intelbench~4.0.2 and \skampi~5. We used the standard
configuration of the two benchmark suites (in particular, we used the
default values of the number of repetitions for each message size).
We compute the \emph{normalized \runtime} of each measurement for a
specific message size as follows:
$t^{norm}_{\msize,i} = t_{\msize,i} / t_\msize^{*}$\,, for all
$i, 1\le i \le \ntrial=30$, where
$t_\msize^{*} = \min_{1 \le i \le \ntrial} \left( t_{\msize,i} \right)$.  We can see
in \fig~\ref{fig:runtimeratio-a} and \fig~\ref{fig:runtimeratio-b}
that the normalized \runtimes of \intelbench and \skampi
exhibit a significantly larger variance for smaller message sizes than
our benchmarking approach.  The higher variance can be explained by
the influence of system noise on experiments with small message sizes.
In such cases, an MPI benchmark needs to record a sufficiently large
number of repetitions across several calls to \mpirun.  Unfortunately,
the \intelbench and \skampi simply do not implement such
reproducibility policies.

Overall, we can state that our benchmarking approach notably improves
the reproducibility of the performance results compared to the
\intelbench and \skampi. The price for a better reproducibility,
however, is a longer \runtime of the overall benchmark, caused by the
need to take into account the clock drift between processes and to
record a larger number of measurements per message size.

\section{Related Work}
\label{sec:relwork}

The statistically rigorous analysis of experimental data has been the
focus of numerous studies over the last years, driven by the
need for establishing a fair comparison of algorithms across different
computing systems.

Vitek and Kalibera contend that ``[i]mportant results in systems
research should be repeatable, they should be reproduced, and their
evaluation should be carried with adequate rigor''. They show that a
correct experimental design paired with the right statistical tests is
the cornerstone for reproducible experimental
results~\cite{Vitek:2012}. The authors stress the fact that knowing
and understanding the controllable and uncontrollable factors of the
experiment is crucial for obtaining sound experimental results.

The state of performance evaluation in Java benchmarking was
investigated by Georges~\etal~\cite{GeorgesBE07}. They examined the
performance of different garbage collectors for the Java Virtual
Machine (JVM). The paper demonstrates that the answer to the question
of which garbage collector is faster changes completely depending on
the performance values investigated (\eg, mean, median, fastest,
\etcet). The authors show how to conduct a statistically rigorous
analysis of JVM micro-benchmarks. In particular, they explain the need
for considering confidence intervals of the mean and show that the
Analysis of Variance (ANOVA) can be used to compare more than two
alternatives in a sound manner.

Mytkowicz~\etal dedicated an entire article to the problem of
measurement bias in micro-benchmarks~\cite{MytkowiczDHS09}.  The
authors examine the \runtime measurements of several SPEC~CPU2006
benchmarks, when each benchmark is either compiled with the
optimization flag \texttt{-O2} or \texttt{-O3}. In theory, the
programs compiled with \texttt{-O3} should run faster than the ones
compiled with \texttt{-O2}. However, the authors discovered that the
resulting performance not only depends on obvious factors such as the
compilation flags or the input size, but also on less obvious factors,
such as the link order of object files or the size of the UNIX
environment. A possible solution is to apply a randomized experimental
setup. Please refer to the books of Box~\etal~\cite{box2005statistics}
and Montgomery~\cite{Montgomery:2006} for more details on randomizing
experiments.

Touati~\etal developed a statistical protocol called Speedup-Test that
can be used to determine whether the speedup obtained when modifying
an experimental factor, such as the compilation flag (\texttt{-O3}),
is significant~\cite{Touati:2012uo}. The article presents two tests,
one to compare the mean and one to compare the median execution times
of two sets of observations. For a statistically sound analysis, they
base both Speedup-Test protocols on well-known tests, such as the
Student’s \testttest to compare means or the \kolmtest test to
check whether two samples have a common underlying distribution.

Chen~\etal proposed the Hierarchical Performance Testing (HPT)
framework to compare the performance of computer systems using a set
of benchmarks~\cite{ChenCGTWH12}. The authors first contend that it is
generally unknown how large the sample size needs to be, such that the
central limit theorem holds. They show that for some distributions a
sample size ``[i]n the order of 160 to 240'' is required to apply
statistical tests that require normally distributed
data~\cite{ChenCGTWH12}. Since such a high number of experiments seems
infeasible for them, they propose a nonparametric framework to compare
the performance improvement of computer systems. The HPT framework
employs the nonparametric Wilcoxon Rank-Sum Test to compare the
performance score of a single benchmark and the Wilcoxon Signed-Rank
Test to compare the scores over all benchmarks.

Gil~\etal presented a study on micro-benchmarking on the JVM, in which
they show that the mean execution times over several JVM invocations
may significantly differ~\cite{GilLS11}. The described effect is very
similar to the work presented here, as our micro-benchmark also needs
to start an environment (the MPI environment using \mpirun), which can
affect the mean \runtime.

\section{Conclusions}
\label{sec:conclusions}

We have revisited the problem of benchmarking MPI functions. 
Our work was motivated by the need (1)~to fairly compare MPI
implementations
using a sound statistical analysis and (2)~to allow a reproducibility
of the experimental results.

We have experimentally shown that the clock and process
synchronization methods used to benchmark MPI functions have a
tremendous effect on the \runtime. We have also pointed out that the
use of \mpibarrier can potentially skew processes in such a way that
the \runtimes measured are meaningless. To overcome the problem of
synchronizing processes with \mpibarrier, we have investigated the
window-based approach, for which we require globally synchronized
clocks. We have shown that it is essential to consider the clock drift
between processes when seeking accurate MPI timings. For this reason,
the clock synchronization methods used in \netgauge or \skampi---that
only determine clock offsets---will introduce a larger \runtime error
into \runtime measurements of MPI functions unless the experiment is
very short-lived.

We have analyzed experimental factors of MPI experiments, for example,
we have demonstrated that changing the DVFS level or the compiler
flags can alter the outcome of the MPI benchmark. However, our most
important finding is that a call to \mpirun is a factor of the
experiment, \ie, different calls to \mpirun can produce significantly
different means (or medians), even if all other factors and the input
data stay unmodified.

After investigating the implications and consequences of various
synchronization methods and experimental factors, we have proposed a
novel MPI benchmarking method. We have shown how to apply hypothesis
tests such as the \testwilcox 
to increase the fairness and the evidence level when comparing
benchmarking data.  Last, we have demonstrated that our benchmarking
method also improves the reproducibility of results in such a way that
the measured performance values exhibit a much smaller variance across
different experiments compared to other MPI benchmark suites.

\section*{Acknowledgments}

We thank Maciej Drozdowski (Poznan University of Technology), Peter
Filzmoser (TU Wien), Friedrich Leisch and
Bernhard Spangl (University of Natural Resources and Life Sciences,
Vienna), Jesper Larsson Tr\"aff (TU Wien), Alf
Gerisch (TU Darmstadt), and Thomas Geenen (SURFsara) for discussions
and comments. Experiments presented in this paper were carried out
using the Grid'5000 testbed, supported by a scientific interest group
hosted by Inria and including CNRS, RENATER and several Universities
as well as other organizations (see \url{https://www.grid5000.fr}).

\IEEEtriggeratref{10}
\bibliographystyle{IEEEtran}
\bibliography{mpi_stat_arxiv}

\begin{thebibliography}{10}
\providecommand{\url}[1]{#1}
\csname url@samestyle\endcsname
\providecommand{\newblock}{\relax}
\providecommand{\bibinfo}[2]{#2}
\providecommand{\BIBentrySTDinterwordspacing}{\spaceskip=0pt\relax}
\providecommand{\BIBentryALTinterwordstretchfactor}{4}
\providecommand{\BIBentryALTinterwordspacing}{\spaceskip=\fontdimen2\font plus
\BIBentryALTinterwordstretchfactor\fontdimen3\font minus
  \fontdimen4\font\relax}
\providecommand{\BIBforeignlanguage}[2]{{%
\expandafter\ifx\csname l@#1\endcsname\relax
\typeout{** WARNING: IEEEtran.bst: No hyphenation pattern has been}%
\typeout{** loaded for the language `#1'. Using the pattern for}%
\typeout{** the default language instead.}%
\else
\language=\csname l@#1\endcsname
\fi
#2}}
\providecommand{\BIBdecl}{\relax}
\BIBdecl

\bibitem{intel_mpi_bench}
``{Intel(R) MPI Benchmarks},''
  http://software.intel.com/en-us/articles/intel-mpi-benchmarks.

\bibitem{GroveC05}
D.~A. Grove and P.~D. Coddington, ``Communication benchmarking and performance
  modelling of {MPI} programs on cluster computers,'' \emph{The Journal of
  Supercomputing}, vol.~34, no.~2, pp. 201--217, 2005.

\bibitem{LastovetskyRO08a}
A.~L. Lastovetsky, V.~Rychkov, and M.~O'Flynn, ``{MPIBlib}: Benchmarking {MPI}
  communications for parallel computing on homogeneous and heterogeneous
  clusters,'' in \emph{EuroPVM/MPI}, 2008, pp. 227--238.

\bibitem{Traff12}
J.~L. Tr{\"a}ff, ``mpicroscope: Towards an {MPI} benchmark tool for performance
  guideline verification,'' in \emph{EuroMPI}, 2012, pp. 100--109.

\bibitem{GroppL99}
W.~Gropp and E.~L. Lusk, ``Reproducible measurements of {MPI} performance
  characteristics,'' in \emph{EuroPVM/MPI}, 1999, pp. 11--18.

\bibitem{Hoefler:2007fo}
T.~Hoefler, A.~Lumsdaine, and W.~Rehm, ``Implementation and performance
  analysis of non-blocking collective operations for {MPI},'' in
  \emph{Proceedings of the 2007 ACM/IEEE Conference on Supercomputing
  (SC)}.\hskip 1em plus 0.5em minus 0.4em\relax ACM, 2007, pp. 1--10.

\bibitem{osu_benchmarks}
``{{OSU} {MPI} benchmarks},'' http://mvapich.cse.ohio-state.edu/benchmarks/.

\bibitem{phloem_mpi}
``{Phloem MPI Benchmarks},'' \url{https://asc.llnl.gov/sequoia/benchmarks/}.

\bibitem{ReussnerST02}
R.~Reussner, P.~Sanders, and J.~L. Tr{\"a}ff, ``{SKaMPI}: a comprehensive
  benchmark for public benchmarking of {MPI},'' \emph{Scientific Programming},
  vol.~10, no.~1, pp. 55--65, 2002.

\bibitem{Grove:2003to}
\BIBentryALTinterwordspacing
D.~Grove, ``{Performance modelling of message-passing parallel programs},''
  Ph.D. dissertation, University of Adelaide, 2003. [Online]. Available:
  \url{http://hdl.handle.net/2440/37915}
\BIBentrySTDinterwordspacing

\bibitem{hoefler-pmeo08}
T.~Hoefler, T.~Schneider, and A.~Lumsdaine, ``Accurately measuring collective
  operations at massive scale,'' in \emph{Proceedings of the 22nd IEEE
  International Parallel \& Distributed Processing Symposium, PMEO'08
  Workshop}, 04 2008.

\bibitem{Hoefler:2010vr}
------, ``Accurately measuring overhead, communication time and progression of
  blocking and nonblocking collective operations at massive scale,''
  \emph{International Journal of Parallel, Emergent and Distributed Systems},
  vol.~25, no.~4, pp. 241--258, 2010.

\bibitem{Kshemkalyani:2008}
A.~D. Kshemkalyani and M.~Singhal, \emph{Distributed Computing: Principles,
  Algorithms, and Systems}, 1st~ed.\hskip 1em plus 0.5em minus 0.4em\relax New
  York, NY, USA: Cambridge University Press, 2008.

\bibitem{skampi_collectives}
T.~Worsch, R.~Reussner, and W.~Augustin, ``On benchmarking collective {MPI}
  operations,'' in \emph{EuroPVM/MPI}, ser. LNCS, no. 2474, 2002, p. 271–279.

\bibitem{amd-arch-manual}
{AMD}, ``{AMD64 Architecture Programmer's Manual Volume 2: System Programming,
  rev. 3.23},''
  \url{http://amd-dev.wpengine.netdna-cdn.com/wordpress/media/2012/10/24593_APM_v21.pdf},
  2015, accessed: 17/02/2015.

\bibitem{intel-arch-manual}
{Intel}, ``{Intel 64 and IA-32 Architectures Software Developer?s Manual,
  Volume 2B: Instruction Set Reference, N-Z, Order Number: 253667-053US},''
  \url{http://www.intel.com/content/www/us/en/architecture-and-technology/64-ia-32-architectures-software-developer-vol-2b-manual.html},
  2015, accessed: 17/02/2015.

\bibitem{hogg2006probability}
R.~Hogg and E.~Tanis, \emph{Probability and Statistical Inference}.\hskip 1em
  plus 0.5em minus 0.4em\relax Prentice Hall, 2006.

\bibitem{Cristian:1989vn}
F.~Cristian, ``Probabilistic clock synchronization,'' \emph{Distributed
  Computing}, vol.~3, no.~3, pp. 146--158, 1989.

\bibitem{Jones:2012dn}
T.~Jones and G.~A. Koenig, ``{Clock synchronization in high-end computing
  environments: a strategy for minimizing clock variance at runtime},''
  \emph{Concurrency and Computation: Practice and Experience}, vol.~25, no.~6,
  pp. 881--897, Jul. 2012.

\bibitem{Taubenfeld:2006}
G.~Taubenfeld, \emph{Synchronization Algorithms and Concurrent
  Programming}.\hskip 1em plus 0.5em minus 0.4em\relax Upper Saddle River, NJ,
  USA: Prentice-Hall, Inc., 2006.

\bibitem{boudec2010performance}
J.-Y. Le~Boudec, \emph{Performance Evaluation of Computer and Communication
  Systems}, ser. Computer and communication sciences.\hskip 1em plus 0.5em
  minus 0.4em\relax EFPL Press, 2010.

\bibitem{liljameasuring}
D.~Lilja, \emph{Measuring Computer Performance}.\hskip 1em plus 0.5em minus
  0.4em\relax Cambridge University Press, 2005.

\bibitem{ross2010introductory}
S.~Ross, \emph{Introductory Statistics}.\hskip 1em plus 0.5em minus 0.4em\relax
  Elsevier Science, 2010.

\bibitem{ChenCGTWH12}
T.~Chen, Y.~Chen, Q.~Guo, O.~Temam, Y.~Wu, and W.~Hu, ``Statistical performance
  comparisons of computers,'' in \emph{HPCA}.\hskip 1em plus 0.5em minus
  0.4em\relax IEEE, 2012, pp. 399--410.

\bibitem{HoeflerB15}
\BIBentryALTinterwordspacing
T.~Hoefler and R.~Belli, ``Scientific benchmarking of parallel computing
  systems: twelve ways to tell the masses when reporting performance results,''
  in \emph{Proceedings of the International Conference for High Performance
  Computing, Networking, Storage and Analysis, {SC} 2015}, 2015, pp.
  73:1--73:12. [Online]. Available:
  \url{http://doi.acm.org/10.1145/2807591.2807644}
\BIBentrySTDinterwordspacing

\bibitem{Sheskin:2007}
D.~J. Sheskin, \emph{Handbook of Parametric and Nonparametric Statistical
  Procedures}, 4th~ed.\hskip 1em plus 0.5em minus 0.4em\relax Chapman \&
  Hall/CRC, 2007.

\bibitem{Montgomery:2006}
D.~C. Montgomery, \emph{Design and Analysis of Experiments}.\hskip 1em plus
  0.5em minus 0.4em\relax John Wiley \& Sons, 2006.

\bibitem{hollander1999nonparametric}
M.~Hollander and D.~Wolfe, \emph{Nonparametric Statistical Methods}, ser. Wiley
  Series in Probability and Statistics.\hskip 1em plus 0.5em minus 0.4em\relax
  Wiley, 1999.

\bibitem{Vitek:2012}
\BIBentryALTinterwordspacing
J.~Vitek and T.~Kalibera, ``R3: Repeatability, reproducibility and rigor,''
  \emph{SIGPLAN Not.}, vol.~47, no.~4a, pp. 30--36, Mar. 2012. [Online].
  Available: \url{http://doi.acm.org/10.1145/2442776.2442781}
\BIBentrySTDinterwordspacing

\bibitem{GeorgesBE07}
A.~Georges, D.~Buytaert, and L.~Eeckhout, ``Statistically rigorous {Java}
  performance evaluation,'' in \emph{OOPSLA}, 2007, pp. 57--76.

\bibitem{MytkowiczDHS09}
T.~Mytkowicz, A.~Diwan, M.~Hauswirth, and P.~F. Sweeney, ``Producing wrong data
  without doing anything obviously wrong!'' in \emph{ASPLOS}, 2009, pp.
  265--276.

\bibitem{box2005statistics}
G.~Box, J.~Hunter, and W.~Hunter, \emph{Statistics for Experimenters: Design,
  Innovation, and Discovery}.\hskip 1em plus 0.5em minus 0.4em\relax
  Wiley-Interscience, 2005.

\bibitem{Touati:2012uo}
S.~A.~A. Touati, J.~Worms, and S.~Briais, ``{The Speedup-Test: A Statistical
  Methodology for Program Speedup Analysis and Computation},''
  \emph{Concurrency and Computation: Practice and Experience}, vol.~25, no.~10,
  pp. 1410--1426, 2013.

\bibitem{GilLS11}
J.~Gil, K.~Lenz, and Y.~Shimron, ``A microbenchmark case study and lessons
  learned,'' in \emph{SPLASH Workshops}, 2011, pp. 297--308.

\end{thebibliography}

\clearpage

\appendices

\section{List of Variables}

\begin{tabular}{p{3cm} p{3cm}p{8cm}}
variable & data type & description \\
\midrule
\np & uint & number of processes \\
\msize & uint & message size \\
\nrep & uint & number of repetitions (in one \mpirun) \\
\nmpiruns & uint & number of calls to \mpirun \\
\winsize & double & window size \\
\startwin & double & start time in window-based synchronization \\
\func & address & MPI function \\
\mlist & list<uint> & list of message sizes \\
\funclist & list<address> & list of MPI calls \\
\explist & list<exp> & list of experiments (\textit{exp} is a $3$-tuple $(\msize, \func, \nrep)$) \\

\rootproc & uint & rank of root process \\
\myrank & uint & rank of current process \\
\exectime & double & \runtime \\
\stime, \etime & double &  timestamps \\
\proclist & list<uint> & list of processes \\
\exectimes & array<double> & list of \runtimes \\
\stimes, \etimes & array<double> & list of timestamps \\

\textit{myoffset} & double & clock offset of the current process \\
\diff & double & clock offset \\
\difflist & list<double> & list of clock offsets \\
\rtt & double & round trip time (\rtt) \\
\rttlist & list<double> & list of \rtt{}s \\

\nfitpoints & uint & number of points to fit linear model (\synchca, \syncjk) \\
\nexchanges & uint & number of ping-pong messages exchanged to record one \emph{fitpoint}  (\synchca, \syncjk) \\
\npingpongs & uint & number of ping-pong messages for \rtt estimation \\
\textit{lm} & tuple (double, double) & linear model of the clock drift defined by a tuple (\textit{slope}, \textit{intercept}) \\
\modellist & list<address> & list of linear models \\
\tremote & uint & current time on remote process \\
\textit{tlocal} & uint & local time of current process \\
 \textit{tglobal} & uint & global (normalized) time of current process \\
\textit{r}, \server, \client, \pref & uint & process ranks \\
\end{tabular}

\clearpage

\section{Pseudo-Codes of Clock Synchronization Methods}

\subsection{The \skampi Benchmark}

\begin{algorithm}[ht] 
  \caption{\label{alg:skampi_compute_offset}
  Clock offset between two processes.}
  \begin{algorithmic}[1] 
  \Function{SKaMPI\_PingPong}{$p1, p2$}
      \State $td\_min = -\infty$ 
      \State $td\_max = \infty$ 
      \State $\npingpongs= 100 $
      \For{$i$ in $0$ to $\npingpongs - 1$}
          \If {$\myrank == \textit{p1}$}
              \State $\textit{s\_last} = $ \Call{Get\_Time}{}
              \State  \Call{MPI\_Send}{\textit{s\_last}, $1$,  \mpidouble, \textit{p2}}
              \State  \Call{MPI\_Recv}{\textit{t\_last}, $1$,  \mpidouble, \textit{p2}}
              \State $\textit{s\_now} = $ \Call{Get\_Time}{}
	 
	      \State $\textit{td\_min} = $ \Call{max}{\textit{td\_min}, $\textit{t\_last} - \textit{s\_now}$}
              \State $\textit{td\_max} =$ \Call{min}{\textit{td\_max}, $\textit{t\_last} - \textit{s\_last}$}                
          \ElsIf {$\myrank == \textit{p2}$}
              \State  \Call{MPI\_Recv}{\textit{s\_last}, $1$,  \mpidouble, \textit{p1}}
              \State $\textit{t\_last} = $ \Call{Get\_Time}{}
              \State  \Call{MPI\_Send}{\textit{t\_last}, $1$,  \mpidouble, \textit{p1}}
              \State $\textit{t\_now} = $ \Call{Get\_Time}{}
	 
	      \State $\textit{td\_min} = $ \Call{max}{\textit{td\_min}, $\textit{s\_last} - \textit{t\_now}$}
              \State $\textit{td\_max} = $ \Call{min}{\textit{td\_max}, $\textit{s\_last} - \textit{t\_last}$}  
          \EndIf
      \EndFor
      \State $\diff = (\textit{td\_min} + \textit{td\_max})/2$
      \State \textbf{return} \diff
 \EndFunction
 \end{algorithmic}
\end{algorithm}

\begin{algorithm}[ht]  
  \caption{\label{alg:skampi_time_sync}
Clock offsets measurement relative to the root and synchronization window initialization.}
  \begin{algorithmic}[1] 
  \Statex \difflist\ - list of clock offsets of the current process relative to each of 
  \Statex \hspace*{14pt} the others 
  \Statex \startwin\ - timestamp of the first synchronization window
  \Procedure{Compute\_And\_Set\_Clock\_Offsets}{} \label{lst:line:skampi_sequential}
      \For{$i$ in $0$ to $\np - 1$} 
          \State $\difflist[i] = 0$
      \EndFor      
      \For{$i$ in $0$ to $\np - 1$} 
          \State \Call{MPI\_Barrier}{}
          \If {$\myrank == \rootproc$}
              \State $\difflist[i] = $ \Call{SKaMPI\_PingPong}{\rootproc, $i$}
          \ElsIf {$\myrank == i$}
              \State $\difflist[\rootproc] = $ \Call{SKaMPI\_PingPong}{\rootproc, $i$} 
          \EndIf          
      \EndFor
      \If {$\myrank == \rootproc$}
          \State \tmpdifflist = \difflist
      \EndIf
      \State  \Call{MPI\_Bcast}{\tmpdifflist, \np, \mpidouble, \rootproc}     

      \For{$i$ in $1$ to $\np - 1$ }
              \State $\difflist[i] = \tmpdifflist[i] + \difflist[\rootproc]$ 
      \EndFor
  \EndProcedure
  \Statex

  \Procedure{Initialize\_First\_Sync\_Window}{}
      \If {$\myrank == \rootproc$}
          \State  $\startwin= $ \Call{Get\_Time}{}
      \EndIf
      \State  \Call{MPI\_Bcast}{\startwin, 1, \mpidouble, \rootproc}  
  \EndProcedure
  \Statex

  \Function{Start\_Sync}{\winsize, \textit{counter}}
      \State $\textit{sync\_error} = 0$
      \State $\nextwin = \startwin - \difflist[\rootproc] + (\textit{counter}+1) \cdot \winsize$
      \State $\textit{time} = $  \Call{Get\_Time}{}
      \If {$\textit{time} > \nextwin$} 
          \State $\textit{sync\_error} = \texttt{STARTED\_LATE}$
      \EndIf
      
      \While {$\textit{time} < \nextwin$}
          \State $\textit{time} = $ \Call{Get\_Time}{}
      \EndWhile
      \State \textbf{return} \textit{sync\_error}
  \EndFunction
  \Statex
  
  \Function{Stop\_Sync}{\winsize}
      \State $\textit{sync\_error} = 0$
      \State $\textit{time} = $ \Call{Get\_Time}{}
      \If {$\textit{time}  - \nextwin > \winsize$}
          \State $\textit{sync\_error} = \texttt{TOOK\_TOO\_LONG}$
      \EndIf
      \State \textbf{return} \textit{sync\_error}
  \EndFunction
 \end{algorithmic}
\end{algorithm}

\begin{algorithm}[ht]  
  \caption{Timing procedure (\skampi and \netgauge).}
  \begin{algorithmic}[1]   
      \Procedure{Measure}{\func, \msize}
          \State \stime = \Call{Get\_Time}{}
          \State \Call{\func}{\msize}
          \State \etime = \Call{Get\_Time}{}
          \State $\exectime = \etime - \stime$
          \State \textbf{return} \exectime
      \EndProcedure
      \Statex
 \end{algorithmic} 
\end{algorithm}

\begin{algorithm}[ht] 
  \caption{\label{alg:skampi_bench} The \skampi benchmark.}
  \begin{algorithmic}[1] 
  
  \Procedure{Benchmark}{\funclist, \mlist, \textit{max\_rep}, \textit{min\_rep}}
      \Statex \mycomment{\funclist\ - MPI functions to benchmark}
      \Statex \mycomment{\mlist\ - list of message sizes}
      \Statex \mycomment{max\_rep/min\_rep - max/min number of measurements for each message size}
      \Statex \texttt{max\_std\_err} - max standard error of measurements
      
      \Statex
      \State \Call{Compute\_And\_Set\_Clock\_Offsets}{}
      \State \Call{Initialize\_First\_Sync\_Window}{} 
      \For {\msize in \mlist $\And$ \func in \funclist}
          \State $\winsize = 0 $
          \While {$\textit{stop} \not= \texttt{TRUE} $}
              \State set \nrep
          
              \For {$i$ in $0$ to $\nrep - 1$}
                  \State $\localerrorlist[i] = $ \Call{Start\_Sync}{\winsize, \textit{i}}
      	          \State $\localexectimes[i] = $ \Call{Measure}{\func, \msize}
                  \State $\localerrorlist[i] = \localerrorlist[i] + $\Call{Stop\_Sync}{\winsize}
              \EndFor
              \State compute \textit{total\_time} 
              \State \Call{MPI\_Gather}{\localexectimes, \nrep, \mpidouble, 
              \Statex  \hspace*{76pt} \exectimes, \nrep, \mpidouble, \rootproc}
              \State \Call{MPI\_Allreduce}{\localerrorlist, \nrep, \mpidouble, 
              \Statex  \hspace*{88pt} \errorlist, \nrep, \mpidouble, \mpimax}
              
              \If {$\myrank == \rootproc$}
                  \State \exectimesfinal.add($\{\, \exectimes[i] \ \forall\ i<\nrep\ \textbf{s.t.}\ \errorlist[i]==0 \}$)
 
                  \If {$\textit{n\_errors} \ge \nrep/4$}
          	         \State $\winsize =  $\Call{max}{$2 \cdot \winsize$,  $\textit{total\_time}/(\nrep +1) \cdot 1.5 $}
                  \EndIf
                  \State $\textit{max\_consec\_errors} =$ \Call{max}{$(j-i)\  \textbf{s.t.}$
                  \Statex \hspace*{120pt} $\textit{\errorlist[k]}>0, i \le k \le j$ }
                  \If {$ \textit{max\_consec\_errors} > \nrep/2 $ }
                      \State \nrep =  \Call{max}{$\nrep/2 $,  $4$}
                 \EndIf
              
                  \State $\textit{std\_error} = $ \Call{Compute\_Std\_Error}{\exectimesfinal}
                  \State $\textit{stop}= (\Call{len}{\exectimesfinal} \ge \textit{max\_rep}) \  \Or $
                  \Statex  \hspace*{38pt} $(\Call{len}{\exectimesfinal} \ge \textit{min\_rep} \And \textit{std\_error} \le \texttt{max\_std\_error})  $
	      \EndIf
	      
	      \State  \Call{MPI\_Bcast}{\winsize, 1, \mpidouble, \rootproc}  
	      \State  \Call{MPI\_Bcast}{\nrep, 1, \mpidouble, \rootproc}  
	      \State  \Call{MPI\_Bcast}{\textit{stop}, 1, \mpidouble, \rootproc}  
	      \State \Call{Initialize\_First\_Sync\_Window}{} 
          \EndWhile
          \State print \exectimesfinal
    \EndFor
              
  \EndProcedure
  \end{algorithmic}  
\end{algorithm}

\FloatBarrier

\clearpage

\subsection{\netgauge/\nbcbench Synchronization}

\begin{algorithm}[!ht]
\caption{\label{alg:netgauge_time_sync}Clock offsets measurement 
relative to the root and synchronization window initialization.}
\begin{algorithmic}[1] 
    \Statex $r$ - current process rank
    \Statex \np\ - number of processes
    \Statex \texttt{maxpower} $= 2^{\lfloor log_2{\np} \rfloor} $
    \Statex \difflist\ - list of clock offsets of the current process relative to each of 
    \Statex \hspace*{14pt} the others 
    \Statex \textit{myoffset} - clock offset of the current process relative to \rootproc
    \Statex \startwin\ - next window start time, updated after each sync
    \Statex

   \Procedure{Sync\_Clocks\_Pow2}{}
       \Statex \mycomment {compute clock offsets of processes with ranks between $0$ and $\texttt{maxpower} -1$}  
       \State $\currentround = 1$
       
       \If {$r \ge \texttt{maxpower}$} \textbf{return}
       \EndIf
       \While {$2^{\currentround} \le \texttt{maxpower}$}
           \If {$r \bmod 2^{\currentround} == 0$} \mycomment{client}
               \State $\server = r + 2^{\currentround-1}$ 
               \State $\difflist[\server]=$ \Call{Compute\_Offset}{$r$, \server}
               
               \Statex \mycomment {receive time differences collected by the server}
               \State \Call{MPI\_Recv}{\recvbuf, $2^{\currentround-1}-1$, \mpidouble, \server}

               \Statex \mycomment {compute final time differences}
               \For{$i$ in $0$ to $(2^{\currentround-1}-2)$}
 	           \State $\difflist[\server+i+1] =  \difflist[\server] + \recvbuf[i]$
 	       \EndFor
          \EndIf
          \If {$r \bmod 2^{\currentround} == 2^{\currentround-1}$} \mycomment{server}
              \State $\client = r - 2^{\currentround-1}$  
              \State $\diff = $ \Call{Compute\_Offset}{\client, $r$}      
              
              \Statex \mycomment {send all the time differences to the client}
              \State \Call{MPI\_Send}{$\difflist[r+1]$, $2^{\currentround-1}-1$, \mpidouble, \client}
          \EndIf
          \State $\currentround = \currentround + 1$
      \EndWhile

      \Statex \mycomment{send final time differences from \rootproc to all processes}
      \State \Call{MPI\_Scatter}{\difflist, $1$, \mpidouble,
      \Statex \hspace*{55pt} \textit{myoffset}, $1$, \mpidouble, \rootproc} 
    \EndProcedure
    \Statex

    \Procedure{Sync\_Clocks\_Remaining}{}
        \Statex \mycomment {compute clock offsets of processes with ranks between \texttt{maxpower} and $\np - 1$}      
        \If {$\texttt{maxpower} == \np$} \textbf{return}
       \EndIf
        \If {$r < \np - \texttt{maxpower}$}
            \State $\server = r+\texttt{maxpower}$
            \State $\diff = \Call{Compute\_Offset}{$r$, \server}$
            \State $\diff = \diff + \textit{myoffset}$
            \State \Call{MPI\_Send}{\diff, $1$, \mpidouble, \server}
        \ElsIf  {$r \ge \texttt{maxpower}$} 
            \State $\client = r - \texttt{maxpower}$
            \State $\diff =$ \Call{Compute\_Offset}{\client, $r$}
            \State  \Call{MPI\_Recv}{\textit{myoffset}, $1$, \mpidouble, \client}            
        \EndIf
    \EndProcedure
    \Statex
  
    \Procedure{Initialize\_First\_Sync\_Window}{}
        \State $\texttt{N} = 10$
        \State \textit{local\_bcasttime} = run-time of \texttt{N} executions of \mpibcast
        \State \Call{MPI\_Reduce}{\textit{local\_bcasttime}, \textit{bcasttime}, $1$, \mpidouble, \rootproc}
        
        \State $\startwin = \Call{Get\_Time}{} + \textit{bcasttime}$
        \State \Call{MPI\_Bcast}{\startwin, $1$, \mpidouble, \rootproc}

        \State $\startwin = \startwin - \textit{myoffset}$\ \mycomment{adjust next start time to local clock}
    \EndProcedure
 \end{algorithmic} 
\end{algorithm}

\begin{algorithm}[ht]  
  \caption{\label{alg:netgauge_compute_offset}
  Clock offset between two processes.}
  \begin{algorithmic}[1] 
\Procedure{Compute\_Offset}{\client, \server} 
\State $\npingpongs = 100$
\State $\rtt = 0$

\While {$\rtt  \le $ min(last \npingpongs)}
     \If {$\myrank == \client$}
          \State $\stime =$ \Call{Get\_Time}{}
          \State  \Call{MPI\_Send}{\stime, $1$, \mpidouble, \server}
          \State  \Call{MPI\_Recv}{\tremote, $1$, \mpidouble, \server}
          \State $\etime =$ \Call{Get\_Time}{}
	 \State $\rtt = \etime - \stime$
          \State $\diff = \stime + \rtt/2 - \tremote$   
     \EndIf

     \If {$\myrank == \server$}
          \State  \Call{MPI\_Recv}{\stime, $1$, \mpidouble, \client}    
          \State $\tremote =$ \Call{Get\_Time}{}
          \State \Call{MPI\_Send}{\tremote, $1$, \mpidouble, \client}
     \EndIf
\EndWhile

\State \textbf{return} \diff
\EndProcedure
 \end{algorithmic}
\end{algorithm}

\begin{algorithm}[ht] 
  \caption{\label{alg:netgauge_sync}Synchronization function.}
  \begin{algorithmic}[1] 
  
  \Procedure{Sync}{\winsize}
  \State \textit{time} = \Call{Get\_Time}{}
  \If {$\textit{time} > \startwin$} \mycomment{sync started too late}
      \State $\textit{err} = \textit{time} - \startwin$
  \Else 
      \While {$\textit{time} < \startwin$}
           \State $\textit{time} =$ \Call{Get\_Time}{} 
      \EndWhile
  \EndIf

  \State $\startwin = \startwin + \winsize$
  \State \textbf{return} \textit{err}
  \EndProcedure

 \end{algorithmic} 
\end{algorithm}

\FloatBarrier

\begin{algorithm}[ht]  
  \caption{\label{alg:nbc_bench} \nbcbench measurement procedure.}
  \begin{algorithmic}[1] 
      \Procedure{Benchmark}{\mlist, \func, \nrep}
          \Statex \mycomment{\mlist\ - list of message sizes}
          \Statex \mycomment{\func\ - MPI function to benchmark}
          \Statex \mycomment{\nrep\ - number of measurements for each message size}
          \State \Call{Sync\_Clocks\_Pow2}{}
          \State \Call{Sync\_Clocks\_Remaining}{}
  
          \For{\msize in \mlist} 
              \For {i in $0$ to \warmup} 
              \State \Call{Measure}{\func, \msize}
              \EndFor
              \For {i in $0$ to $(\nrep/10 - 1)$} 
              \State \textit{runtimes}[i] = \Call{Measure}{\func, \msize}
              \EndFor
	     \State $\textit{local\_est\_runtime} = $ \Call{Min}{\textit{runtimes}}
             \State \Call{MPI\_Allreduce}{\textit{local\_est\_runtime}, $1$, \mpidouble,
             \Statex \hspace*{78pt} \textit{estimated\_runtime}, $1$, \mpidouble, \mpimax}
             \If{ $\textit{estimated\_runtime} \cdot 5 \cdot \nrep > 10\ \textit{seconds}$}
                \State $\nrep = \Call{max}{\nrep/2, 4}$
             \EndIf  
             \State \Call{MPI\_Bcast}{\nrep, $1$, \mpidouble, \rootproc}
	     \Statex \mycomment{main measurement loop}
             \State \Call{Initialize\_First\_Sync\_Window}{}
             \For{$i$ in $0$ to $\nrep - 1$}
                 \State $\localerrorlist[i] = \Call{Sync}{\winsize}$
      	         \State $\localexectimes[i] = $ \Call{Measure}{\func, \msize}
             \EndFor
             \State \Call{MPI\_Allreduce}{\localerrorlist, \nrep, \mpidouble,
             \Statex \hspace*{78pt} \errorlist, \nrep, \mpidouble, \mpimax}
             \State $\localexectimes = \{\localexectimes[i]\ \forall\ i<\nrep\ \textbf{s.t.}\ \errorlist[i]==0 \}$
           
             \If{ $\textit{n\_errors} > 0.25 \cdot \nrep \  \Or \  \Call{len}{\localexectimes} <4$}
                 \State $\winsize = \winsize \cdot 1.5$
                 \State \textsc{repeat} \text{measurement for current} \msize 
             \EndIf
           
             \State \Call{MPI\_Gather}{\localexectimes, \nrep, \mpidouble, 
             \Statex \hspace*{65pt} \exectimes, \nrep, \mpidouble, \rootproc}
             \State print \exectimes
       \EndFor
  \EndProcedure
  \end{algorithmic} 
\end{algorithm}

\FloatBarrier
\clearpage

\subsection{Clock Synchronization Algorithm of Jones \& Koenig (\syncjk)}

\begin{algorithm}[ht]
  \caption{\label{alg:jk_clock_drift} Linear Model of the clock drift.} 
  \begin{algorithmic}[1]     
  
    \Statex \np\ - number of processes
    \Statex $r$ - current process rank ($0$ to $\np - 1$)
    \Statex \lm\ - linear model of the current process (defined by a \textit{slope} and an
    \Statex \hspace*{10pt}  \textit{intercept}) to adjust the local clock to the reference time of \rootproc
    \Statex \startwin\ - next window start time, updated after each synchronization
    \Statex
  
    \Function{Learn\_Model}{\nfitpoints, \nexchanges, 
                                                    \rtt, \rootproc} 
        \State $\textit{slope} = 0, \textit{intercept} = 0$
        \If {$\myrank == \rootproc$}
            \For{\textit{idx} in $0$ to $\nfitpoints - 1$ }
                \For{$r$ in $0$ to $(\np-1)$ $\And \ r \not= \rootproc$}
                    \For{$i$ in $0$ to $\nexchanges - 1$ }
                        \State  \Call{MPI\_Recv}{\textit{tdummy}, $1$, \mpidouble, \textit{r}}    
                        \State $\tremote =$ \Call{Get\_Time}{}
                        \State \Call{MPI\_Send}{\tremote, $1$, \mpidouble, \textit{r}}
                    \EndFor
                \EndFor
            \EndFor
        \EndIf
        
        \If {$\myrank \not= \rootproc $}
            \For{\textit{idx} in $0$ to $\nfitpoints - 1$ }
                \For{$i$ in $0$ to $\nexchanges - 1$}
                    \State  \Call{MPI\_Send}{$\textit{tdummy}, 1, \mpidouble, \rootproc$}
                    \State  \Call{MPI\_Recv}{$\tremote, 1, \mpidouble, \rootproc$}
                    \State $\textit{local\_times}[i] =$ \Call{Get\_Time}{}
	            \State $\difflist[i] = \textit{local\_times}[i] - \tremote - \rtt/2$
	        \EndFor
	        \State $\difflist = $ \Call{Sort}{\difflist}
	        \State $\textit{yfit}[\textit{idx}] = $ \Call{Compute\_Median}{\difflist}
	        \State $\textit{idx\_median} = i \ \textbf{s.t.}\  0 \le i < \nexchanges \ \And\ $
	        \Statex \hspace*{90pt} $ \difflist[i] == \textit{yfit}[\textit{idx}]$
	        \State $\textit{xfit}[\textit{idx}] = \textit{local\_times}[\textit{median\_idx}] $	        
           \EndFor
           \State $(\textit{slope}, \textit{intercept}) = $ \Call{Linear\_Fit}{\textit{xfit}, \textit{yfit}, \nfitpoints}
        \EndIf

    \State \textbf{return} \Call{New\_LM}{\textit{slope}, \textit{intercept}}
    \EndFunction    
    \Statex
    
    \Procedure{Sync\_Clocks}{}
        \State \Call{\warmup}{}
        \For{$i$ in $1$ to $\np - 1$}
            \State $\rttlist[i] = $ \Call{Compute\_Rtt}{$\rootproc,i$}          
        \EndFor
        \State  \Call{MPI\_Scatter}{\rttlist, $1$, \mpidouble, \rtt, $1$, \mpidouble, \rootproc}     
        \State  $\lm = $\Call{Learn\_Model}{\nfitpoints, \nexchanges, \rtt, \rootproc}
        \State \Call{MPI\_Barrier}{}
        \Statex

        \State $\startwin = \Call{Get\_Normalized\_Time}{\Call{Get\_Time}{}}  + \winsize$
        \State  \Call{MPI\_Bcast}{\startwin, $1$, \mpidouble}   

  \EndProcedure
  \end{algorithmic}
\end{algorithm}

\begin{algorithm}[ht]
  \caption{\label{alg:hca_normalize_time} Local time normalization to reference clock (\syncjk and \synchca).}
  \begin{algorithmic}[1] 
  
    \Function{Get\_Normalized\_Time}{\textit{local\_time}}
        \State \textbf{return} $\textit{local\_time} - (\textit{local\_time} \cdot \lm.\textit{slope} + \lm.\textit{intercept})$
    \EndFunction
    
\end{algorithmic}
\end{algorithm}

\begin{algorithm}[ht]
  \caption{\label{alg:hca_mean_rtt} Measurement of the RTT between two nodes (\syncjk and \synchca).}  
  \begin{algorithmic}[1] 
 \Function{Compute\_Rtt}{$\textit{p1}, \textit{p2}$} 
        \State $\textit{mean\_rtt} = 0$
        \State \Call{Warmup\_Rounds}{}\ \mycomment{send dummy ping-pong messages}

        \If {$\myrank == \textit{p1}$}
            \For{$i$ in $0$ to $\npingpongs - 1$}
                \State  \Call{MPI\_Recv}{\textit{tdummy}, $1$, \mpidouble, \textit{p2}}    
                \State $\tremote =$ \Call{Get\_Time}{}
                \State \Call{MPI\_Send}{\tremote, $1$, \mpidouble, \textit{p2}}
            \EndFor
        
        \ElsIf {$\myrank == \textit{p2}$}
            \For{$i$ in $0$ to $\npingpongs - 1$ }
              \State $\stime =$ \Call{Get\_Time}{}
              \State  \Call{MPI\_Send}{\stime, $1$, \mpidouble, \textit{p1}}
              \State  \Call{MPI\_Recv}{\tremote, $1$, \mpidouble, \textit{p1}}
              \State $\etime =$ \Call{Get\_Time}{}
	      \State $\rttlist[i] = \etime - \stime$
           \EndFor
            \State $\rttlist = $ \Call{Remove\_Outliers}{\rttlist}
            \State $\textit{mean\_rtt} = $ \Call{Mean}{\rttlist}
        \EndIf
        \State \textbf{return} \textit{mean\_rtt}
    \EndFunction  
\end{algorithmic}
\end{algorithm}

\clearpage

\section{Overview of MPI Experiments}

This section details the experimental setup and the pseudo-code of
each of the experiments presented in this paper. Unless otherwise
stated, the following configuration is common to all the obtained
results.\\[1ex]
  \begin{scriptsize}
  \begin{tabularx}{\columnwidth}{lX}
    \toprule
    Parameter & Values  \\
    \midrule 
      parallel machine & \machone \\
      compiler   &   \gcc~4.4.7  \\
      compiler flags   &   \texttt{-O2}  \\
      DVFS       &   CPU frequency fixed to the highest available frequency level  \\
      cache       &  No cache control    \\
      process pinning   & Core pinning (\texttt{-bind-to rr} / \texttt{-pin\_mode consec-rev})  \\
      timing mechanism & \rdtscp \\
    \bottomrule
  \end{tabularx}
  \end{scriptsize}

\subsection{Experiment: Clock Drift}
\label{sec:exp_clock_drift}

  \begin{scriptsize}
  \begin{tabularx}{\columnwidth}{llX}
    \toprule
    Parameter & Values & Details  \\
    \midrule 
    \multicolumn{3}{l}{\textbf{Experiment configuration} - \fig~\ref{fig:clock_skew_jupiter}}\\
    \midrule 
      \np       &  \num{16 x 1}  & Number of nodes and processes per node  \\
      \nrep       &  \num{100} & Number of measured ping-pongs   \\
      \warmup       & \num{10}   &  Number of warmup rounds performed before measurement \\
    \bottomrule
  \end{tabularx}
  \end{scriptsize}

\begin{algorithm}[ht]
  \caption{\label{exp:clock_drift}Experiment: Clock drift.}  
  \begin{algorithmic}[1] 
    \State \Call{Warmup\_Rounds}{}\ \mycomment{dummy ping-pong messages between \rootproc and $r$}
    \If {\myrank == \rootproc}
    \For{$r$ in $0$ to $\np - 1$}
    \If {$r \ne \rootproc$}
    \For{\textit{rep} in $0$ to $\nrep - 1$}
    \State $\localstimes[\textit{rep}] = \Call{Get\_Time}{}$
    \State \Call{MPI\_Send}{$\localstimes[\textit{rep}]$, $1$, \mpidouble, $r$}
    \State \Call{MPI\_Recv}{$\remotetimes[\textit{rep}]$, $1$, \mpidouble, $r$}
    \State $\localetimes[\textit{rep}] = \Call{Get\_Time}{}$
    \State \Call{Nanosleep}{$0.5\ s$}
    \EndFor
 
    \For{\textit{rep} in $0$ to $\nrep - 1$}
    \State \textbf{print} $\textit{r}, \localstimes[\textit{rep}], \localetimes[\textit{rep}], \remotetimes[\textit{rep}]$
    \EndFor
    \EndIf
    \EndFor
    \ElsIf {$\myrank \ne \rootproc$}
    \For{\textit{rep} in $0$ to $\nrep - 1$}
    \State \Call{MPI\_Recv}{\textit{tdummy}, $1$, \mpidouble, \rootproc}
    \State $\textit{local\_time} = \Call{Get\_Time}$
    \State \Call{MPI\_Send}{\textit{local\_time}, $1$, \mpidouble, \rootproc}
    \EndFor
    \EndIf
  \end{algorithmic}
\end{algorithm}

\subsection{Experiment: Frequency Calibration}
\label{sec:exp_freq_calibration}

  \begin{scriptsize}
  \begin{tabularx}{\columnwidth}{llX}
    \toprule
    Parameter & Values & Details  \\
    \midrule 
    \multicolumn{3}{l}{\textbf{Experiment configuration} - \fig~\ref{fig:core_freq_diff}}\\
    \midrule 
      \np       &  \num{16 x 1}  & Number of nodes and processes per node  \\
      \nrep       &  \num{100} & Number of repetitions  \\
      \hrtcal       &  - & Frequency estimation function from \netgauge 2.4.6 \\
    \bottomrule
  \end{tabularx}
  \end{scriptsize}

\begin{algorithm}[ht]
  \caption{\label{exp:freq_calibration} Experiment: Frequency calibration.}  
    \begin{algorithmic}[1] 
      \For{$i$ in $0$ to $\nrep -1$}
      \State \Call{\hrtcal}{\textit{freq}}
      \State $\freqlist[i] = \textit{freq}$
      \EndFor
      \State \Call{MPI\_Gather}{\freqlist, \nrep, \mpiuint, 
      \Statex \hspace*{43pt} \textit{all\_freqs}, \nrep, \mpiuint, \rootproc}
      \If{\myrank == \rootproc}
          \For{$r$ in $0$ to $\np - 1$}
              \For{\textit{rep} in $0$ to $\nrep - 1$}
                   \State \textbf{print} $r$, $rep$, $\textit{all\_freqs}[r \cdot \nrep + \textit{rep}]$
              \EndFor
          \EndFor
      \EndIf
    \end{algorithmic}
\end{algorithm}

\subsection{Experiment: Clock Offset after Synchronization}
\label{sec:exp:clock_drift_in_time}

\newcolumntype{b}{X}
\newcolumntype{s}{{\hsize=.5\hsize}X}

  \begin{scriptsize}
  \begin{tabularx}{\columnwidth}{p{1.8cm}p{2.2cm}X}
    \toprule
   Parameter & Values & Details  \\
      \midrule 
    \multicolumn{3}{l}{\textbf{Experiment configuration} - \fig~\ref{fig:netgauge_drift}}\\
    \midrule 
      \np        & \num{16x1}   &  Number of processes \\
      \nmpiruns & \num{10} & Number of experiments \\
      \nrep       &  \num{10} & Number of \texttt{PingPong} operations performed in each 
      									step to measure the clock offset \\
      \texttt{sleep\_time} & \SI{1}{\second} & Waiting time between clock offset measurements \\
      \textit{nsteps}       &  \num{10} & Number of clock offset measurements  \\
      MPI  & \mvapichtwoone & MPI implementation \\
    \bottomrule
  \end{tabularx}
  \end{scriptsize}

\begin{algorithm}[ht]
  \caption{\label{exp:clock_drift_in_time}Experiment: Clock offset after sync.}  
  \begin{algorithmic}[1] 
    \State compute \rttlist - the list of \rtts between \rootproc and process $r$
    \State \Call{Init\_Sync\_Module}{}
    \For{\textit{step} in $1$ to \textit{nsteps}}
        \If {\myrank == \rootproc}
            \For{$r$ in $0$ to $\np - 1$}
            \If {$r \ne \rootproc$}
                \For{\textit{j} in 0 to $\textit{nrounds} - 1$}
                    \State \Call{MPI\_Send}{\stime, $1$, \mpidouble, $r$}
                    \State \Call{MPI\_Recv}{\tremote, $1$, \mpidouble, $r$}
                    \If {\textit{sync\_type} == \synchca}     
                       \State $\textit{tlocal} =  \Call{Get\_Adjusted\_Time}{} - \rttlist[r]/2$
                    \Else
           	      \State $\textit{tlocal} = \Call{Get\_Time}{} - \rttlist[r]/2$
	            \EndIf
                    \State $\globaltimes[r][\textit{j}] = \tremote$
                \EndFor
            \EndIf
            \EndFor
            \State \Call{Sleep}{\texttt{sleep\_time}}
        \ElsIf  {$\myrank \ne \rootproc$}
            \State \Call{MPI\_Recv}{\textit{tdummy}, $1$, \mpidouble, $0$}   
            \If {\textit{sync\_type} == \synchca}     
                  \State $\textit{tlocal} = $ \Call{Get\_Adjusted\_Time}{} 
            \Else
           	 \State $\textit{tlocal} = $ \Call{Get\_Time}{}
	    \EndIf
            \State $\textit{tglobal} = $ \Call{Get\_Normalized\_Time}{\textit{tlocal}}
            \State \Call{MPI\_Send}{\textit{tglobal}, $1$, \mpidouble, $0$}  
        \EndIf
    \EndFor
    \If {\myrank == \rootproc}
        \For{$r$ in $0$ to $\np -1$}
         \For{\textit{j} in 0 to $\textit{nrounds} - 1$}
            \State \textbf{print} $r, \reftimes[\textit{j}], \globaltimes[\textit{j}]$
        \EndFor
        \EndFor
    \EndIf
  \end{algorithmic}
\end{algorithm}

\subsection{Experiment: \Runtime Drift in \netgauge and \skampi}
\label{sec:exp:clock_drift_ng_skampi}

This experiment is based on \alg~\ref{alg:repeat_procedure}, in which
the synchronization method has been set to either \netgauge, \skampi or
\mpibarrier.\\[1ex]
  \begin{scriptsize}
  \begin{tabularx}{\columnwidth}{p{1.8cm}p{1.5cm}X}
    \toprule
    Parameter & Values & Details  \\
     \midrule 
    \multicolumn{3}{l}{\textbf{Experiment configuration} - \fig~\ref{fig:runtime_increase_sync}}\\
     \midrule 
      \np        & \num{32x16}   &  Number of processes \\
      \nmpiruns & \num{1} & Number of experiments \\
      \nrep       &  \num{4000} & Number of measurements \\
      \func	& \mpibcast & Benchmarked function \\
      \msize & \SI{8192}{\Bytes} & message size \\
      \winsize & \SI{300}{\micro\second} & Window size \\
      MPI  & \mvapichtwoone & MPI implementation \\
    \bottomrule
  \end{tabularx}
  \end{scriptsize}

\clearpage
\subsection{Experiment: Clock Offset after Synchronization}
\label{sec:exp:clock_sync_offset}

This experiment relies on \alg~\ref{exp:clock_drift_in_time}, where
only one measurement round is performed after synchronization, instead
of multiple \textit{nsteps}.\\[1ex]
  \begin{scriptsize}
  \begin{tabularx}{\columnwidth}{p{1.8cm}p{1.5cm}X}
    \toprule
    Parameter & Values & Details  \\
     \midrule 
    \multicolumn{3}{l}{\textbf{Algorithm parameters}} \\
    \midrule
      \textsc{sleep\_time} & \SI{0}{\second} & No waiting time between synchronization and measurements \\
      \textit{nsteps}       &  \num{1} & Number of clock offset estimations  \\
      \textit{nrounds}       &  \num{10} & Number of \texttt{PingPong} operations performed in each
      						 step to measure clock offset \\
      \nfitpoints (\synchca /\syncjk sync.) & \num{1000} & Number of fitpoints used to fit linear models\\
      \nexchanges (\synchca/\syncjk sync.) & \num{100} &  Number of ping-pong messages 
      						exchanged to obtain the difference between local 
						and reference times corresponding to a single fit point \\
      MPI  & \mvapichtwoone & MPI implementation \\
    \midrule
    \multicolumn{3}{l}{\textbf{Experiment configuration} - \fig~\ref{fig:clock_drift_ppn1}}\\
     \midrule 
      \np        & [$2$--$36$]$\times$\num{1} &  Number of processes \\
      \nmpiruns & \num{10} & Number of experiments \\
    \midrule
    \multicolumn{3}{l}{\textbf{Experiment configuration} - \fig~\ref{fig:clock_drift_ppn16}}\\
     \midrule 
      \np        & [$2$--$36$]$\times$\num{16}   &  Number of processes \\
      \nmpiruns & \num{10} & Number of experiments \\
    \bottomrule
  \end{tabularx}
  \end{scriptsize}

\subsection{Experiment: Comparison of Synchronization Methods w.r.t. the Clock Drift }
\label{sec:exp:clock_drift_all}

This experiment is based on \alg~\ref{exp:clock_drift_in_time}, in which
the synchronization method has been set to either \synchca, \syncjk,
\skampi or \netgauge.\\[1ex]
  \begin{scriptsize}
  \begin{tabularx}{\columnwidth}{p{1.8cm}p{1.5cm}X}
    \toprule
    Parameter & Values & Details  \\
     \midrule 
    \multicolumn{3}{l}{\textbf{Algorithm parameters}} \\
    \midrule
      \textsc{sleep\_time} & \SI{1}{\second} & Waiting time between clock offset measurements \\
      \textit{nsteps}       & \num{20} & Number of clock offset estimations  \\
      \textit{nrounds}         &  \num{10} & Number of \texttt{PingPong} operations performed in each 
      									step to measure clock offset \\
      \nfitpoints (\synchca/\syncjk sync.) & \num{1000} & Number of fitpoints used to fit linear models\\
      \nexchanges (\synchca/\syncjk sync.) & \num{100} &  Number of ping-pong messages 
      						exchanged to obtain the difference between local 
						and reference times corresponding to a single fit point \\    
    \midrule
    \multicolumn{3}{l}{\textbf{Experiment configuration} - \fig~\ref{fig:clock_skew_in_time_ppn16}}\\
     \midrule 
      \np        & \num{16x16}   &  Number of processes \\
      \nmpiruns & \num{10} & Number of experiments \\
    MPI  & \mvapichtwoone & MPI implementation \\
    \midrule
    \multicolumn{3}{l}{\textbf{Experiment configuration} - \fig~\ref{fig:clock_skew_in_time_ppn1}}\\
     \midrule 
      \np        & \num{16x1}  &  Number of processes \\
      \nmpiruns & \num{10} & Number of experiments \\
      MPI  & \mvapichtwoone & MPI implementation \\
    \midrule
    \multicolumn{3}{l}{\textbf{Experiment configuration} - \fig~\ref{fig:g5k_clock_drift_120nodes}}\\
     \midrule 
      \np        & \num{15x8}  &  Number of processes \\
      \nmpiruns & \num{10} & Number of experiments \\
      MPI  & \mvapichonenine & MPI implementation \\
      machine & \multicolumn{2}{l}{\machthree}  \\
    \midrule
    \multicolumn{3}{l}{\textbf{Experiment configuration} - \fig~\ref{fig:surfsara_clock_drift_16nodes}}\\
     \midrule 
      \np        & \num{16x1}  &  Number of processes \\
      \nmpiruns & \num{1} & Number of experiments \\
      machine & \multicolumn{2}{l}{\machfour}  \\
      MPI  & \intelmpi & MPI implementation \\
    \bottomrule
  \end{tabularx}
  \end{scriptsize}

\subsection{Experiment: \Runtime of \mpibarrier}
\label{sec:exp:mpi_barrier_time}

  \begin{scriptsize}
  \begin{tabularx}{\columnwidth}{llX}
    \toprule
    Parameter & Values & Details  \\
     \midrule 
    \multicolumn{3}{l}{\textbf{Algorithm parameters}} \\
     \midrule 
      \nmpiruns      &  \num{30} & Number of \textsc{mpirun}s the experiment was repeated  \\
      \nrep            &  \num{100000} & Number of \mpibarrier calls  \\
      \warmup      & \num{10}   &  Number of warmup rounds performed before measurement \\
    \midrule 
    \multicolumn{3}{l}{\textbf{Experiment configuration} - \fig~\ref{fig:sync_efficiency_nodes512}}\\
     \midrule 
      \np        & \num{32x16}  &  Number of processes \\
      MPI  & \mvapichtwoone & MPI implementation \\
    \midrule 
    \multicolumn{3}{l}{\textbf{Experiment configuration} - \fig~\ref{fig:jupiter_sync_efficiency_nodes16}}\\
    \midrule 
      \np        & \num{16x1}  &  Number of processes \\
      MPI  & \mvapichtwoone & MPI implementation \\  
     \midrule
    \multicolumn{3}{l}{\textbf{Experiment configuration} - \fig~\ref{fig:g5k_sync_efficiency_nodes120}}\\
     \midrule 
      \np        & \num{15x8}  &  Number of processes \\
     MPI  & \mvapichonenine & MPI implementation \\
      machine & \multicolumn{2}{l}{\machthree}  \\
    \bottomrule
  \end{tabularx}
  \end{scriptsize}

\begin{algorithm}[ht]
  \caption{\label{exp:mpi_barrier}Experiment: \Runtime of MPI\_Barrier}  
    \begin{algorithmic}[1]
      \State \warmup of \mpibarrier
      \State $\stime =$ \Call{Get\_Time}{}
      \For {$i$ in $0$ to $\nrep - 1$}
      \State \Call{MPI\_Barrier}{}
      \EndFor
      \State $\etime =$ \Call{Get\_Time}{}
      \State $\localexectime = (\etime - \stime) / \nrep$
      \State \Call{MPI\_Reduce}{\localexectime, \exectime, 1, \mpidouble, \mpimax, \rootproc}
      \If{\myrank == \rootproc}
      \State print barrier time \exectime
      \EndIf
    \end{algorithmic}
\end{algorithm}

\vspace*{-15pt}  
  
\subsection{Experiment: Comparison of Synchronization Methods - Clock Offset vs. Synchronization Time }
\label{sec:exp:sync_duration}

  \begin{scriptsize}
  \begin{tabularx}{\columnwidth}{p{1.8cm}p{1.5cm}X}
    \toprule
    Parameter & Values & Details  \\
     \midrule 
    \multicolumn{3}{l}{\textbf{Algorithm parameters (\algs~\ref{exp:clock_drift_in_time} and~\ref{exp:sync_duration})}} \\
     \midrule 
      \textsc{sleep\_time} & \SI{1}{\second} & Waiting time between clock offset measurements \\
      \textit{nsteps}       &  \num{20} & Number of clock offset estimations  \\
      \textit{nrounds}       &  \num{10} & Number of \texttt{PingPong} operations performed in each 
      									step to measure clock offset \\
      \nfitpoints (\synchca/\syncjk sync.) & \numlist{10;100;200;300;500;700;1000} & Number of fitpoints used to fit linear models\\
      \nexchanges (\synchca/\syncjk sync.) & [$10$--$100$] &  Number of ping-pong messages 
      						exchanged to obtain the difference between local 
						and reference times corresponding to a single fit point \\    
    \midrule
    \multicolumn{3}{l}{\textbf{Experiment configuration} - \fig~\ref{fig:sync_efficiency_nodes512}}\\
     \midrule 
      \np        & \num{32x16}   &  Number of processes \\
      \nmpiruns & \num{10} & Number of experiments \\
            MPI  & \mvapichtwoone & MPI implementation \\
    \midrule
    \multicolumn{3}{l}{\textbf{Experiment configuration} - \fig~\ref{fig:jupiter_sync_efficiency_nodes16}}\\
     \midrule 
      \np        & \num{16x1}  &  Number of processes \\
      \nmpiruns & \num{10} & Number of experiments \\
            MPI  & \mvapichtwoone & MPI implementation \\
    \midrule
    \multicolumn{3}{l}{\textbf{Experiment configuration} - \fig~\ref{fig:g5k_sync_efficiency_nodes120}}\\
     \midrule 
      \np        & \num{15x8}  &  Number of processes \\
      \nmpiruns & \num{10} & Number of experiments \\
            MPI  & \mvapichonenine & MPI implementation \\
            machine & \multicolumn{2}{l}{\machthree}  \\
    \bottomrule
  \end{tabularx}
  \end{scriptsize}

\begin{algorithm}[!ht]
  \caption{\label{exp:sync_duration} Experiment: Sync. duration.}
  \begin{small}
  \begin{algorithmic}[1] 
    \State $\stime = $ \Call{Get\_Time}{}
    \State \Call{Init\_Sync\_Module}{} \mycomment{compute clock drifts, linear models}
    \State $\etime = $ \Call{Get\_Time}{}
    \State $\textit{sync\_time\_local} =  \etime - \stime$
    \State \Call{MPI\_Reduce}{\textit{sync\_time\_local}, \textit{sync\_time}, $1$, \mpidouble, \mpimax, \rootproc}        
    \State \textbf{print} \textit{sync\_time}
  \end{algorithmic}
\end{small}
\end{algorithm}

\FloatBarrier

\newpage
\subsection{Experiment: Impact of the Timing Mechanism -- Local Times vs. Global Times}
\label{sec:exp:barrier_local_vs_global_times}

This experiment is based on \alg~\ref{alg:repeat_procedure}, in which the 
synchronization method has been set to either \synchca or \mpibarrier.\\[1ex]
  \begin{scriptsize}
  \begin{tabularx}{\columnwidth}{p{1.8cm}p{2.2cm}X}
    \toprule
    Parameter & Values & Details  \\
     \midrule 
    \multicolumn{3}{l}{\textbf{Experiment configuration} - \fig~\ref{fig:vsc3_barrier_vs_hca_Allreduce}}\\
     \midrule 
      \np        & \num{16x1}   &  Number of processes \\
      \nmpiruns & \num{1} & Number of experiments \\
      \nrep & \num{4000}  & Number of measurements per experiment \\
      \func	& \mpiallreduce & Benchmarked function \\
      \msize & \SI{32}{\kibi\byte} & message size \\
      \winsize & \SI{1}{\milli\second} & Window size \\
      \nfitpoints (\synchca sync.) & \num{1000} & Number of fitpoints used to fit linear models\\
      \nexchanges (\synchca sync.) & \num{100} &  Number of ping-pong messages 
      						exchanged to obtain the difference between local 
						and reference times corresponding to a single fit point \\    
      MPI  & \mvapichtwoa & MPI implementation \\
      machine & \multicolumn{2}{l}{\machfive}  \\
    \bottomrule
  \end{tabularx}
  \end{scriptsize}

\subsection{Experiment: \mpibarrier Exit Times}
\label{sec:exp:barrier_exit_times}

This experiment is based on \alg~\ref{alg:repeat_procedure}, in which
the synchronization method has been set to \synchca.\\[1ex]
  \begin{scriptsize}
  \begin{tabularx}{\columnwidth}{p{1.8cm}p{2.2cm}X}
    \toprule
    Parameter & Values & Details  \\
     \midrule 
    \multicolumn{3}{l}{\textbf{Experiment configuration} - \fig~\ref{fig:vsc3_barrier_processes}}\\
     \midrule 
      \np        & \num{16x1}   &  Number of processes \\
      \nmpiruns & \num{1} & Number of experiments \\
      \nrep & \num{1000}  & Number of measurements per experiment \\
      \func	& \mpibarrier & Benchmarked function \\
      \winsize & \SI{100}{\micro\second} & Window size \\
      \nfitpoints (\synchca sync.) & \num{1000} & Number of fitpoints used to fit linear models\\
      \nexchanges (\synchca sync.) & \num{100} &  Number of ping-pong messages 
      						exchanged to obtain the difference between local 
						and reference times corresponding to a single fit point \\    
      MPI  & \mvapichtwoa, \intelmpifive & MPI implementation \\
      machine & \multicolumn{2}{l}{\machfive}  \\
    \bottomrule
  \end{tabularx}
  \end{scriptsize}

\subsection{Experiment: Barrier Implementation Impact}
\label{sec:exp:barrier_implem_impact}

This experiment is based on \alg~\ref{alg:repeat_procedure}, in which
the synchronization method has been set to either \mpibarrier or a
dissemination barrier implemented into the benchmark.\\[1ex]
  \begin{scriptsize}
  \begin{tabularx}{\columnwidth}{p{1.8cm}p{2.2cm}X}
    \toprule
    Parameter & Values & Details  \\
     \midrule 
    \multicolumn{3}{l}{\textbf{Experiment configuration} - \fig~\ref{fig:vsc3_intel_vs_mvapich_barriers}}\\
     \midrule 
      \np        & \num{16x1}   &  Number of processes \\
      \nmpiruns & \num{10} & Number of experiments \\
      \nrep & \num{1000}  & Number of measurements per experiment \\
      \func	& \mpibcast & Benchmarked function \\
      \msize & \SIrange[range-phrase = --, range-units=single, exponent-base=2]{d6}{d15}{\Bytes}  & Message size \\
      MPI  & \mvapichtwoa, \intelmpifive & MPI implementation \\
       machine & \multicolumn{2}{l}{\machfive}  \\
    \bottomrule
  \end{tabularx}
  \end{scriptsize}

\newpage
\subsection{Experiment: The Influence of \mpirun}
\label{sec:exp:infl_mpirun}

This experiment is based on \alg~\ref{alg:repeat_procedure}, in which
the synchronization method has been set to \mpibarrier.\\[1ex]
  \begin{scriptsize}
  \begin{tabularx}{\columnwidth}{p{1.8cm}p{1.8cm}X}
    \toprule
    Parameter & Values & Details  \\
     \midrule 
    \multicolumn{3}{l}{\textbf{Algorithm parameters}} \\
     \midrule 
      \nmpiruns & \num{30} & Number of experiments \\
      \nrep & \num{1000}  & Number of measurements per experiment \\
     \midrule 
     \multicolumn{3}{l}{\textbf{Experiment configuration} - \fig~\ref{fig:mpirun_distinct} (a)}\\
     \midrule 
      \np        & \num{16x1}   &  Number of processes \\
      \msize & \SI{8}{\kibi\byte}  & Message size \\
      \func	& \mpibcast & Benchmarked function \\
      MPI  & \necmpitwoeight & MPI implementation \\
     \midrule 
     \multicolumn{3}{l}{\textbf{Experiment configuration} - \fig~\ref{fig:mpirun_distinct} (b)}\\
     \midrule 
      \np        & \num{32x1}   &  Number of processes \\
      \msize & \SI{2}{\kibi\byte}  & Message size \\
      \func	& \mpibcast & Benchmarked function \\
      MPI  & \intelmpi & MPI implementation \\
             machine & \multicolumn{2}{l}{\machtwo}  \\
     \midrule 
     \multicolumn{3}{l}{\textbf{Experiment configuration} - \fig~\ref{fig:mpirun_distinct} (c)}\\
     \midrule 
      \np        & \num{16x1}   &  Number of processes \\
      \msize & \SI{8}{\kibi\byte}  & Message size \\
      \func	& \mpibcast & Benchmarked function \\
      MPI  & \mvapichonenine & MPI implementation \\
             machine & \multicolumn{2}{l}{\machthree}  \\
    \bottomrule
  \end{tabularx}
  \end{scriptsize}

\subsection{Experiment: Distribution of \Runtimes with Window-based Synchronization}
\label{sec:exp:runtime_dist_allreduce}

This experiment is based on \alg~\ref{alg:repeat_procedure}, in which the 
synchronization method has been set to \syncjk.\\[1ex]
  \begin{scriptsize}
  \begin{tabularx}{\columnwidth}{p{1.8cm}p{1.8cm}X}
    \toprule
    Parameter & Values & Details  \\
     \midrule 
    \multicolumn{3}{l}{\textbf{Experiment configuration} - \fig~\ref{fig:mpiruns_dist_allreduce}}\\
     \midrule 
      \np        & \num{16x1}   &  Number of processes \\
      \nmpiruns & \num{500} & Number of experiments \\
      \nrep & \num{1000}  & Number of measurements per experiment \\
      \msize & \SI{1000}{\Bytes} & Message size \\
      \func	& \mpiallreduce & Benchmarked function \\
      \winsize & \SI{1}{\milli\second} & Window size \\
      \nfitpoints & \num{1000} & Number of fitpoints used to fit linear models\\
      \nexchanges & \num{20} &  Number of ping-pong messages 
      						exchanged to obtain the difference between local 
						and reference times corresponding to a single fit point \\    
      MPI  & \mvapichtwoone & MPI implementation \\
    \bottomrule
  \end{tabularx}
  \end{scriptsize}

\newpage
\subsection{Experiment: \Runtime Histograms}
\label{sec:exp:runtime_hist}

This experiment is based on \alg~\ref{alg:repeat_procedure}, in which
the synchronization method has been set to either \mpibarrier or \synchca.\\[1ex]
  \begin{scriptsize}
  \begin{tabularx}{\columnwidth}{p{1.8cm}p{2.2cm}X}
    \toprule
    Parameter & Values & Details  \\
     \midrule 
    \multicolumn{3}{l}{\textbf{Experiment configuration} - \fig~\ref{fig:distribution_examples}, \fig~\ref{fig:system_noise2}}\\
     \midrule 
      \np        & \num{16x1}   &  Number of processes \\
      \nmpiruns & \num{10} & Number of experiments \\
      \nrep & \num{1000}  & Number of measurements per experiment \\
      \msize & \SI{10000}{\Bytes}, \SI{1000}{\Bytes} & Message size \\
      \func	& \mpiscan, \mpiallreduce & Benchmarked function \\
      MPI  & \necmpitwoeight & MPI implementation \\
     \midrule 
    \multicolumn{3}{l}{\textbf{Experiment configuration} - \fig~\ref{fig:g5k_histograms_hca}}\\
     \midrule 
      \np        & \num{16x8}   &  Number of processes \\
      \nmpiruns & \num{20} & Number of experiments \\
      \nrep & \num{4000}  & Number of measurements per experiment \\
      \msize & \SI{1}{\kibi\byte}, \SI{8}{\kibi\byte} & Message size \\
      \func	& \mpibcast, \mpiallreduce & Benchmarked function \\
      \winsize & \SI{500}{\micro\second} & Window size \\
      \nfitpoints (\synchca sync.) & \num{1000} & Number of fitpoints used to fit linear models\\
      \nexchanges (\synchca sync.) & \num{100} &  Number of ping-pong messages 
      						exchanged to obtain the difference between local 
						and reference times corresponding to a single fit point \\    
      MPI  & \mvapichonenine & MPI implementation \\
       machine & \multicolumn{2}{l}{\machthree}  \\
     \midrule 
    \multicolumn{3}{l}{\textbf{Experiment configuration} - \fig~\ref{fig:surfsara_histograms_hca}}\\
     \midrule 
      \np        & \num{16x1}   &  Number of processes \\
      \nmpiruns & \num{3} & Number of experiments \\
      \nrep & \num{4000}  & Number of measurements per experiment \\
      \msize & \SIrange[range-phrase = --, range-units=single, exponent-base=2]{d6}{d15}{\Bytes} & Message size \\
      \func	& \mpibcast, \mpiallreduce, \mpiscan & Benchmarked function \\
      \winsize & \SI{500}{\micro\second} & Window size \\
      \nfitpoints (\synchca sync.) & \num{1000} & Number of fitpoints used to fit linear models\\
      \nexchanges (\synchca sync.) & \num{100} &  Number of ping-pong messages 
      						exchanged to obtain the difference between local 
						and reference times corresponding to a single fit point \\    
      MPI  & \intelmpi & MPI implementation \\
      machine & \multicolumn{2}{l}{\machfour}  \\
     \midrule 
    \multicolumn{3}{l}{\textbf{Experiment configuration} - \fig~\ref{fig:vsc3_histograms_hca}}\\
     \midrule 
      \np        & \num{16x1}   &  Number of processes \\
      \nmpiruns & \num{10} & Number of experiments \\
      \nrep & \num{1000}  & Number of measurements per experiment \\
      \msize & \SIrange[range-phrase = --, range-units=single, exponent-base=2]{d6}{d15}{\Bytes}  & Message size \\
      \func	& \mpibcast, \mpiallreduce & Benchmarked function \\
      \winsize & \SI{1}{\milli\second} & Window size \\
      \nfitpoints (\synchca sync.) & \num{1000} & Number of fitpoints used to fit linear models\\
      \nexchanges (\synchca sync.) & \num{100} &  Number of ping-pong messages 
      						exchanged to obtain the difference between local 
						and reference times corresponding to a single fit point \\    
      MPI  & \mvapichtwoa & MPI implementation \\      
       machine & \multicolumn{2}{l}{\machfive}  \\
    \bottomrule
  \end{tabularx}
  \end{scriptsize}

\newpage
\subsection{Experiment: \Runtime Drift --  \syncjk  vs. \mpibarrier Synchronization}
\label{sec:exp:runtime_drift_jk}

This experiment is based on \alg~\ref{alg:repeat_procedure}, in which
the synchronization method has been set to either \syncjk or
\mpibarrier.\\[1ex]
  \begin{scriptsize}
  \begin{tabularx}{\columnwidth}{p{1.8cm}p{1.8cm}X}
    \toprule
    Parameter & Values & Details  \\
     \midrule 
    \multicolumn{3}{l}{\textbf{Experiment configuration} - \fig~\ref{fig:jones_skew_allreduce}}\\
     \midrule 
      \np        & \num{16x1}   &  Number of processes \\
      \nmpiruns & \num{1} & Number of experiments \\
      \nrep & \num{500000}  & Number of measurements per experiment \\
      \msize & \SI{1000}{\Bytes} & Message size \\
      \func	& \mpiallreduce & Benchmarked function \\
      \winsize & \SI{1}{\milli\second} & Window size \\
      \nfitpoints (\syncjk) & \num{1000} & Number of fitpoints used to fit linear models\\
      \nexchanges (\syncjk) & \num{20} &  Number of ping-pong messages 
      						exchanged to obtain the difference between local 
						and reference times corresponding to a single fit point \\    
      MPI  & \mvapichtwoone & MPI implementation \\

    \bottomrule
  \end{tabularx}
  \end{scriptsize}

\subsection{Experiment: Impact of Window Size on \Runtime and Number of Invalid Results}
\label{sec:exp:runtime_infl_winsize}

This experiment is based on \alg~\ref{alg:repeat_procedure}, which has
been repeated for \synchca, \skampi and \netgauge.\\[1ex]
  \begin{scriptsize}
  \begin{tabularx}{\columnwidth}{p{1.8cm}p{1.8cm}X}
    \toprule
    Parameter & Values & Details  \\
     \midrule 
    \multicolumn{3}{l}{\textbf{Experiment configuration} - \fig~\ref{fig:win_errors_alltoall}, \fig~\ref{fig:win_effect_scan}}\\
     \midrule 
      \np        & \num{16x1}   &  Number of processes \\
      \nmpiruns & \num{10} & Number of experiments \\
      \nrep & \num{1000}  & Number of measurements per experiment \\
      \msize & \SI{8}{\kibi\byte} & Message size \\
      \func	& \mpialltoall, \mpiscan & Benchmarked function \\
      \winsize & [$150$--\num{10000}]\,\si{\micro\second} & Window size \\
      \nfitpoints (\synchca) & \num{1000} & Number of fitpoints used to fit linear models\\
      \nexchanges (\synchca) & \num{100} &  Number of ping-pong messages 
      						exchanged to obtain the difference between local 
						and reference times corresponding to a single fit point \\    
      MPI  & \mvapichtwoone & MPI implementation \\
    \bottomrule
  \end{tabularx}
  \end{scriptsize}
  
\newpage
\subsection{Experiment: \Runtime Drift Comparison}
\label{sec:exp:runtime_drift_all}

This experiment is based on \alg~\ref{alg:repeat_procedure}, which has
been repeated for each synchronization method.\\[1ex]
  \begin{scriptsize}
  \begin{tabularx}{\columnwidth}{p{1.8cm}p{1.8cm}X}
    \toprule
    Parameter & Values & Details  \\
     \midrule 
    \multicolumn{3}{l}{\textbf{Algorithm parameters} }\\
     \midrule 
      \nrep & \num{4000}  & Number of measurements per experiment \\
      \winsize & \SI{500}{\micro\second} & Window size \\
      \nfitpoints (\syncjk, \synchca) & \num{1000} & Number of fitpoints used to fit linear models\\
      \nexchanges (\syncjk, \synchca) & \num{100} &  Number of ping-pong messages 
      						exchanged to obtain the difference between local 
						and reference times corresponding to a single fit point \\    
           \midrule 
     \multicolumn{3}{l}{\textbf{Experiment configuration} - \fig~\ref{fig:all_runtime_drift_allreduce}}\\
     \midrule 
      \np        & \num{32x16}   &  Number of processes \\
      \nmpiruns & \num{10} & Number of experiments \\
      \msize & \SI{8}{\kibi\byte} & Message size \\
      \func	& \mpiallreduce & Benchmarked function \\
      MPI  & \mvapichtwoone & MPI implementation \\
     \midrule 
     \multicolumn{3}{l}{\textbf{Experiment configuration} - \fig~\ref{fig:g5k_measurement_drift_nodes120}}\\
     \midrule 
      \np        & \num{15x8}   &  Number of processes \\
      \nmpiruns & \num{10} & Number of experiments \\
      \msize & \SI{32}{\kibi\byte}  & Message size \\
      \func	& \mpibcast & Benchmarked function \\
      MPI  & \mvapichonenine & MPI implementation \\
             machine & \multicolumn{2}{l}{\machthree}  \\
     \midrule 
     \multicolumn{3}{l}{\textbf{Experiment configuration} - \fig~\ref{fig:surfsara_sync_efficiency_nodes16}}\\
     \midrule 
      \np        & \num{16x1}   &  Number of processes \\
      \msize & \SI{8}{\kibi\byte}  & Message size \\
      \nmpiruns & \num{3} & Number of experiments \\
      \func	& \mpiscan & Benchmarked function \\
      MPI  & \intelmpi & MPI implementation \\
             machine & \multicolumn{2}{l}{\machfour}  \\
     \midrule 
     \multicolumn{3}{l}{\textbf{Experiment configuration} - \fig~\ref{fig:vsc3-drift-sync-methods}}\\
     \midrule 
      \np        & \num{16x1}   &  Number of processes \\
      \msize & \SI{1}{\kibi\byte}  & Message size \\
      \nmpiruns & \num{10} & Number of experiments \\
      \func	& \mpiallreduce & Benchmarked function \\
      MPI  & \mvapichtwoa & MPI implementation \\
             machine & \multicolumn{2}{l}{\machfive}  \\
    \bottomrule
  \end{tabularx}
  \end{scriptsize}

\subsection{Experiment: Pinning Effect}
\label{sec:exp:pinning}

This experiment is based on \alg~\ref{alg:repeat_procedure} and was
conducted using the \synchca synchronization.\\[1ex]
  \begin{scriptsize}
  \begin{tabularx}{\columnwidth}{p{1.8cm}p{1.8cm}X}
    \toprule
    Parameter & Values & Details  \\
     \midrule 
    \multicolumn{3}{l}{\textbf{Experiment configuration} - \fig~\ref{fig:pinning_hist_nodes16_ppn16}}\\
     \midrule 
      \np        & \num{16x16}   &  Number of processes \\
      \nmpiruns & \num{10} & Number of experiments \\
      \nrep & \num{1000}  & Number of measurements per experiment \\
      \msize & [\num{1000}--\num{10000}]\,\Bytes & Message size \\
      \func	& \mpiallreduce & Benchmarked function \\
      \winsize & \SI{1}{\milli\second} & Window size \\
      \nfitpoints & \num{1000} & Number of fitpoints used to fit linear models\\
      \nexchanges & \num{100} &  Number of ping-pong messages 
      						exchanged to obtain the difference between local 
						and reference times corresponding to a single fit point \\    
      MPI  & \necmpitwoeleven & MPI implementation \\
      clock & \mpiwtime & clock source for time measurements \\
    \bottomrule
  \end{tabularx}
  \end{scriptsize}

\newpage

\subsection{Experiment: Compiler Effect}
\label{sec:exp:compiler}

This experiment is based on \alg~\ref{alg:repeat_procedure} and 
was conducted using the \synchca synchronization.\\[1ex]
  \begin{scriptsize}
  \begin{tabularx}{\columnwidth}{p{1.8cm}p{1.8cm}X}
    \toprule
    Parameter & Values & Details  \\
     \midrule 
    \multicolumn{3}{l}{\textbf{Experiment configuration} - \fig~\ref{fig:compiler_nodes16_ppn1}}\\
     \midrule 
      \np        & \num{16x1}   &  Number of processes \\
      \nmpiruns & \num{30} & Number of experiments \\
      \nrep & \num{1000}  & Number of measurements per experiment \\
      \msize & \SIrange[range-phrase = --, range-units=single, exponent-base=2]{d6}{d13}{\Bytes} & Message size \\
      \func	& \mpiallreduce & Benchmarked function \\
      \winsize & \SI{1}{\milli\second} & Window size \\
      \nfitpoints & \num{1000} & Number of fitpoints used to fit linear models\\
      \nexchanges & \num{100} &  Number of ping-pong messages 
      						exchanged to obtain the difference between local 
						and reference times corresponding to a single fit point \\    
      MPI  & \mvapichtwoone & MPI implementation \\
    \bottomrule
  \end{tabularx}
  \end{scriptsize}

\subsection{Experiment: DVFS Effect}
\label{sec:exp:dvfs}

This experiment is based on \alg~\ref{alg:repeat_procedure} and was
conducted using the \synchca synchronization.\\[1ex]
  \begin{scriptsize}
  \begin{tabularx}{\columnwidth}{p{1.8cm}p{1.8cm}X}
    \toprule
    Parameter & Values & Details  \\
     \midrule 
    \multicolumn{3}{l}{\textbf{Experiment configuration} - \fig~\ref{fig:dvfs_nodes16_ppn1}}\\
     \midrule 
      \np        & \num{16x1}   &  Number of processes \\
      \nmpiruns & \num{30} & Number of experiments \\
      \nrep & \num{1000}  & Number of measurements per experiment \\
      \msize & \SIrange[range-phrase = --, range-units=single, exponent-base=2]{d6}{d13}{\Bytes} & Message size \\
      \func	& \mpiallreduce & Benchmarked function \\
      \winsize & \SI{1}{\milli\second} & Window size \\
      \nfitpoints & \num{1000} & Number of fitpoints used to fit linear models\\
      \nexchanges & \num{100} &  Number of ping-pong messages 
      						exchanged to obtain the difference between local 
						and reference times corresponding to a single fit point \\    
      MPI  & \mvapichtwoone,  \necmpitwoeleven & MPI implementation \\
    \bottomrule
  \end{tabularx}
  \end{scriptsize}

\subsection{Experiment: Caching Effect}
\label{sec:exp:caching}

This experiment is based on \alg~\ref{alg:repeat_procedure} and 
was conducted using the \synchca synchronization.\\[1ex]
  \begin{scriptsize}
  \begin{tabularx}{\columnwidth}{p{1.8cm}p{1.8cm}X}
    \toprule
    Parameter & Values & Details  \\
     \midrule 
    \multicolumn{3}{l}{\textbf{Experiment configuration} - \fig~\ref{fig:factor_caching_allreduce}}\\
     \midrule 
      \np        & \num{16x1}   &  Number of processes \\
      \nmpiruns & \num{10} & Number of experiments \\
      \nrep & \num{1000}  & Number of measurements per experiment \\
      \msize & \SIrange[range-phrase = --, range-units=single, exponent-base=2]{d6}{d13}{\Bytes} & Message size \\
      \func	& \mpiallreduce & Benchmarked function \\
      \winsize & \SI{1}{\milli\second} & Window size \\
      \nfitpoints & \num{1000} & Number of fitpoints used to fit linear models\\
      \nexchanges & \num{100} &  Number of ping-pong messages 
      						exchanged to obtain the difference between local 
						and reference times corresponding to a single fit point \\    
      MPI  & \mvapichtwoone & MPI implementation \\
    \bottomrule
  \end{tabularx}
  \end{scriptsize}

\newpage

\subsection{Experiment: Statistical Testing}
\label{sec:exp:stat_tests}

This experiment is based on \alg~\ref{alg:repeat_procedure} and 
was conducted using the \synchca synchronization.\\[1ex]
  \begin{scriptsize}
  \begin{tabularx}{\columnwidth}{p{1.8cm}p{2cm}X}
    \toprule
    Parameter & Values & Details  \\
     \midrule 
    \multicolumn{3}{l}{\textbf{Experiment configuration} - \fig~\ref{fig:reg_stats_mean}, 
    				 \fig~\ref{fig:test_res_nrep10}, \fig~\ref{fig:test_res_nrep_less}}\\
     \midrule 
      \np        & \num{16x1}   &  Number of processes \\
      \nmpiruns & \num{30} & Number of experiments \\
      \nrep & \num{1000}  & Number of measurements per experiment \\
      \msize & \SIrange[range-phrase = --, range-units=single, exponent-base=2]{1}{d15}{\Bytes} & Message size \\
      \func	& \mpiallreduce & Benchmarked function \\
      \winsize & adapted for each message size & Window size \\
      \nfitpoints & $1000$ & Number of fitpoints used to fit linear models\\
      \nexchanges & $100$ &  Number of ping-pong messages 
      						exchanged to obtain the difference between local 
						and reference times corresponding to a single fit point \\    
      MPI  & \mvapichtwoone,  \necmpitwoeleven & MPI implementation \\
    \bottomrule
  \end{tabularx}
  \end{scriptsize}

\subsection{Experiment: Reproducibility Evaluation}
\label{sec:exp:reprod}

This experiment relies on the \intelbench~4.0.2, \skampi~5
and \alg~\ref{alg:repeat_procedure} (using the \synchca
synchronization method) to measure the \runtime of \mpibcast.\\[1ex]
  \begin{scriptsize}
  \begin{tabularx}{\columnwidth}{p{1.8cm}p{2cm}X}
    \toprule
    Parameter & Values & Details  \\
     \midrule 
     \multicolumn{3}{l}{\textbf{Algorithm parameters} }\\
     \midrule 
      Experiment runs & \num{30} & Number of times the full experiment was repeated \\
      \np        & \num{16x1}   &  Number of processes \\
      \msize & \SIrange[range-phrase = --, range-units=single, exponent-base=2]{1}{d15}{\Bytes} & Message size \\
      \func	& \mpibcast & Benchmarked function \\
      MPI  & \mvapichtwoone & MPI implementation \\
    \midrule
    \multicolumn{3}{l}{\textbf{Experiment configuration} - \fig~\ref{fig:runtimeratio-c}}\\
     \midrule 
      \nmpiruns & \num{30} & Number of \mpiruns \\
      \nrep & \num{1000}  & Number of measurements per experiment \\
      \winsize & adapted for each message size & Window size \\
      \nfitpoints & \num{1000} & Number of fitpoints used to fit linear models\\
      \nexchanges & \num{100} &  Number of ping-pong messages 
      						exchanged to obtain the difference between local 
						and reference times corresponding to a single fit point \\    
    \bottomrule
  \end{tabularx}
  \end{scriptsize}

\subsection{Experiment: \Runtime Comparison}
\label{sec:exp:runtime_barrier_vs_hca}

This experiment is based on \alg~\ref{alg:repeat_procedure}, which has
been repeated for each synchronization method.\\[1ex]
  \begin{scriptsize}
  \begin{tabularx}{\columnwidth}{p{1.8cm}p{1.8cm}X}
    \toprule
    Parameter & Values & Details  \\
     \midrule 
     \multicolumn{3}{l}{\textbf{Experiment configuration} - \fig~\ref{fig:winsync_vs_barrier}}\\
     \midrule 
      \np        & \num{32x16}   &  Number of processes \\
      \nmpiruns & \num{10} & Number of experiments \\
      \msize & \SIrange[range-phrase = --, range-units=single, exponent-base=2]{1}{d15}{\Bytes}  & Message size \\
      \func	& \mpiallreduce & Benchmarked function \\
      \nrep & \num{1000}  & Number of measurements per experiment \\
      \winsize & \SI{150}{\micro\second} & Window size \\
      \nfitpoints (\syncjk, \synchca) & \num{1000} & Number of fitpoints used to fit linear models\\
      \nexchanges (\syncjk, \synchca) & \num{100} &  Number of ping-pong messages 
      						exchanged to obtain the difference between local 
						and reference times corresponding to a single fit point \\    
      MPI  & \mvapichtwoone & MPI implementation \\
    \bottomrule
  \end{tabularx}
  \end{scriptsize}

\cleardoublepage

\section{Thematic Summary of Measurements}

\subsection{Investigating the Round Trip Time (RTT)}
\vspace{-12pt}
\begin{figure}[!ht]
  \centering
  \includegraphics[width=\linewidth]{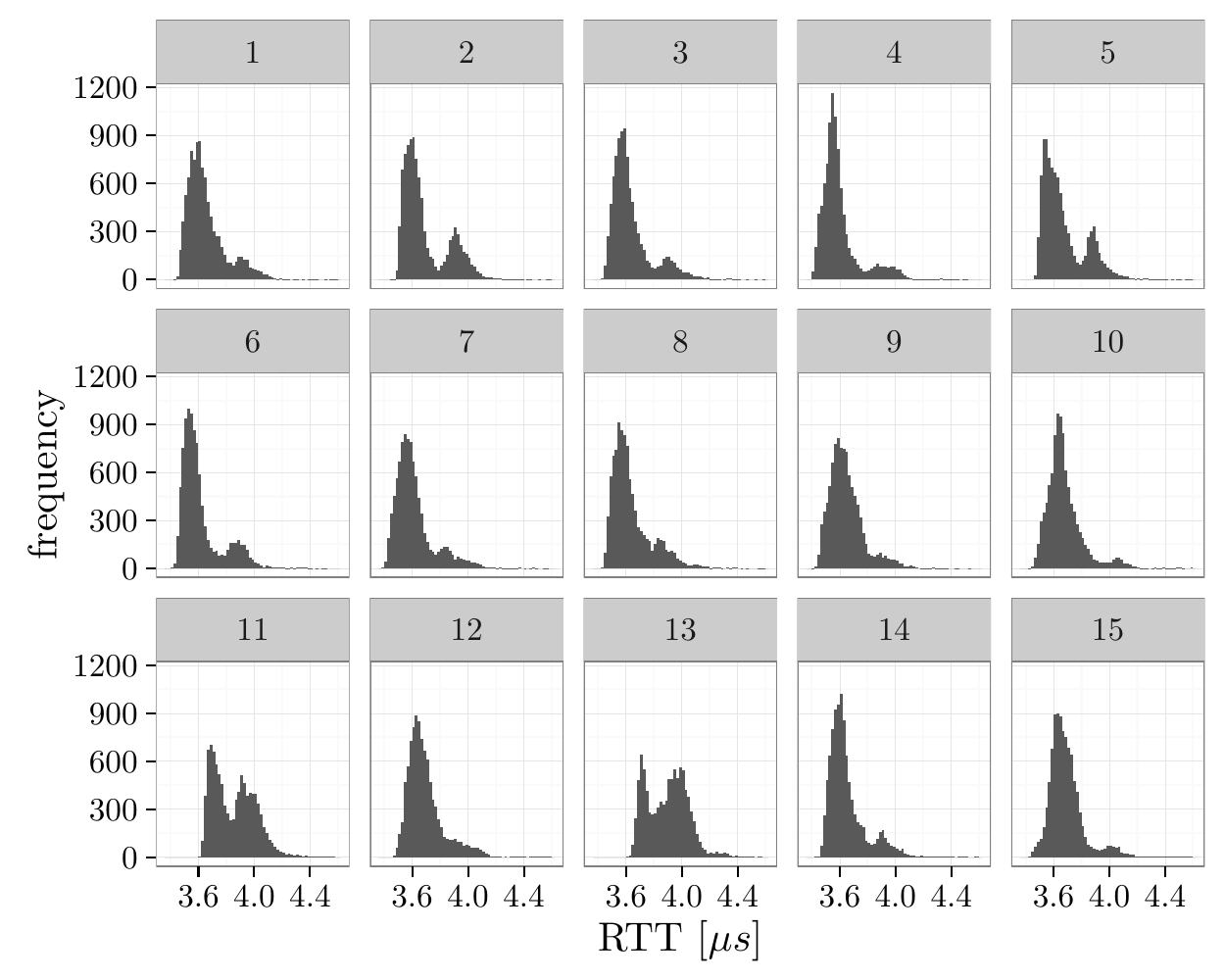}
  \caption{\label{fig:rtt_dist}Histogram of RTTs when sending \num{1000}
    ping-pong messages with $1$~double (\SI{8}{\Bytes}) payload between
    reference host ($0$) and $15$ other hosts ($1$--$15$), (\mvapichtwoone,
    \machone).}
\end{figure}

\begin{figure}[!ht]
  \centering
  \includegraphics[width=.9\linewidth]{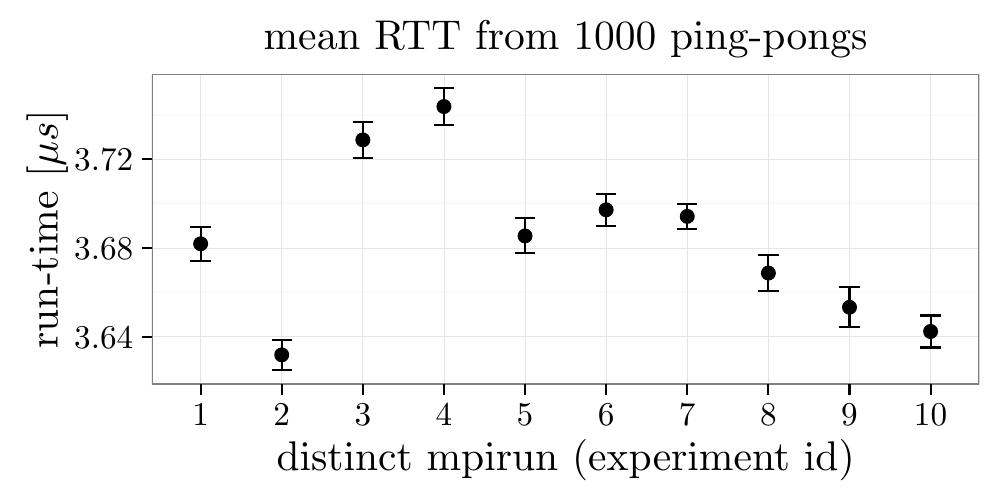}
  \caption{\label{fig:mean_rtt}Mean RTT (after outlier removal)
    computed from \num{1000} ping-pongs between one pair of nodes
    (\mvapichtwoone, \machone).}
\end{figure}

\FloatBarrier

\subsection{Investigating the Number of Invalid Measurement for Window-based Process Synchronization}

\begin{figure}[!ht]
  \centering
  \includegraphics[width=\linewidth]{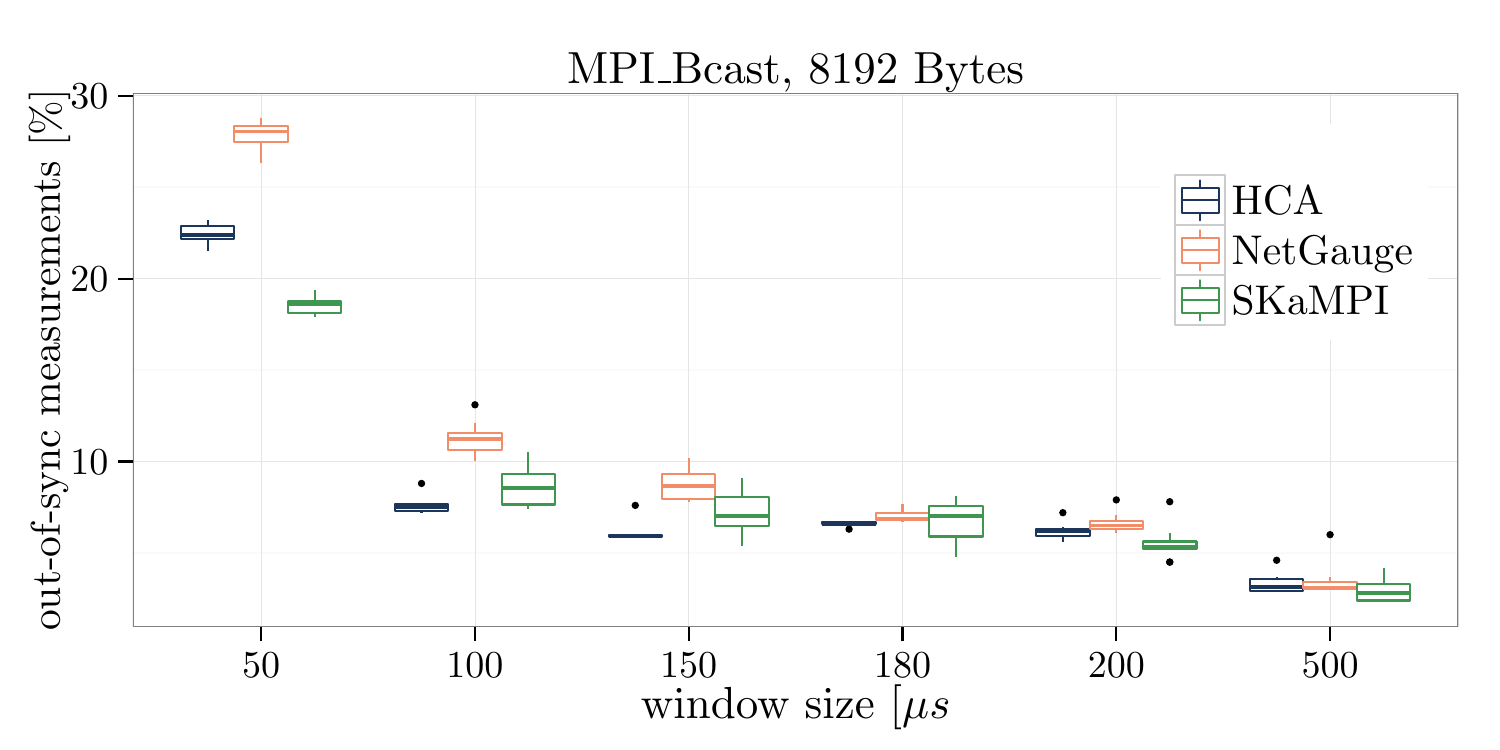}
  \caption{\label{fig:win_errors_bcast}Percentage of incorrect
    measurements for MPI\_Bcast, (\SI{8}{\kibi\byte}, \num{16x1}
    processes, \num{1000} measurements, \num{10} calls to \mpirun,
    \mvapichtwoone, \machone).}
\end{figure}

\FloatBarrier

\newpage
\subsection{Investigating the Clock Drift after Synchronization}
\subsubsection{\machone}
\vspace{-10pt}
\begin{figure}[!ht]
  \centering
  \includegraphics[width=\linewidth]{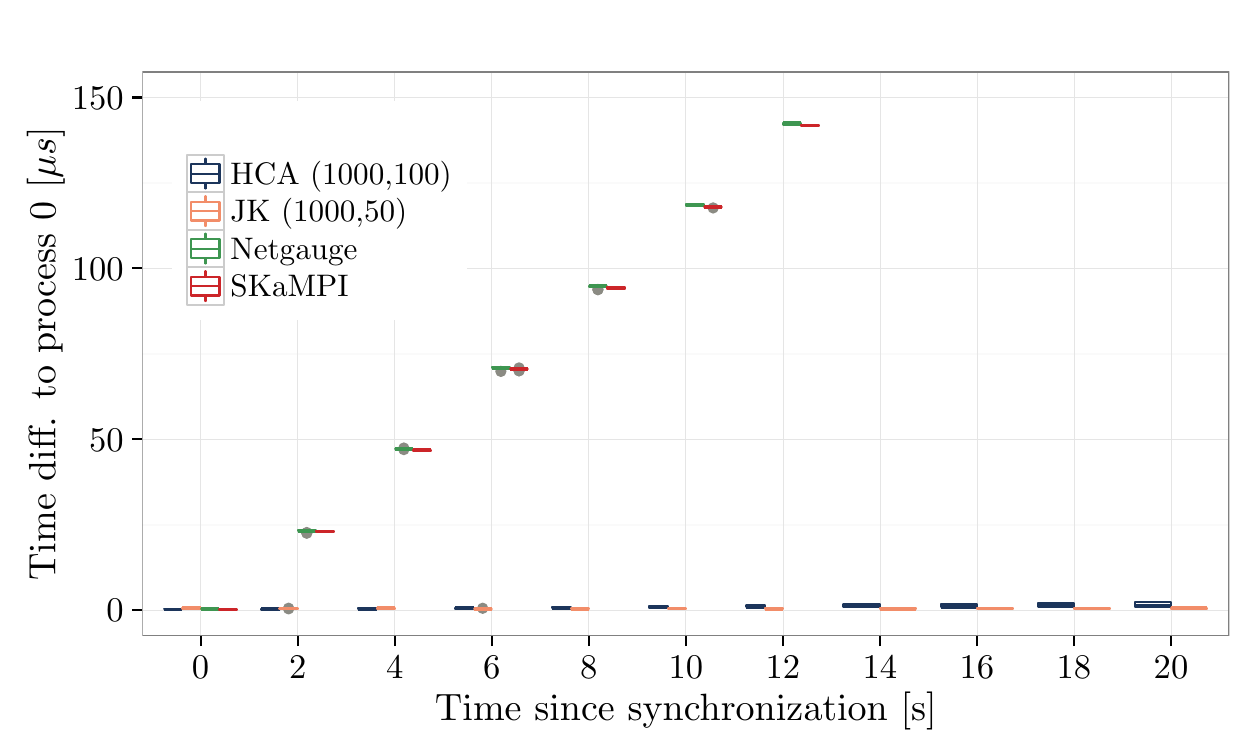}
  \caption{\label{fig:clock_skew_in_time_ppn1} Clock drift for 
  \num{16x1}~processes after \num{0},~\num{1},~\num{2},~$\ldots$,~\SI{20}{\second} 
  (distribution of maximum offsets over \num{10}~calls to \mpirun, 
  \mvapichtwoone, \machone, \expdesc: \append~\ref{sec:exp:clock_drift_all}).}
\end{figure}

\FloatBarrier

\subsubsection{\machthree}
\vspace{-12pt}
\begin{figure}[!ht]
  \centering
  \includegraphics[width=\linewidth]{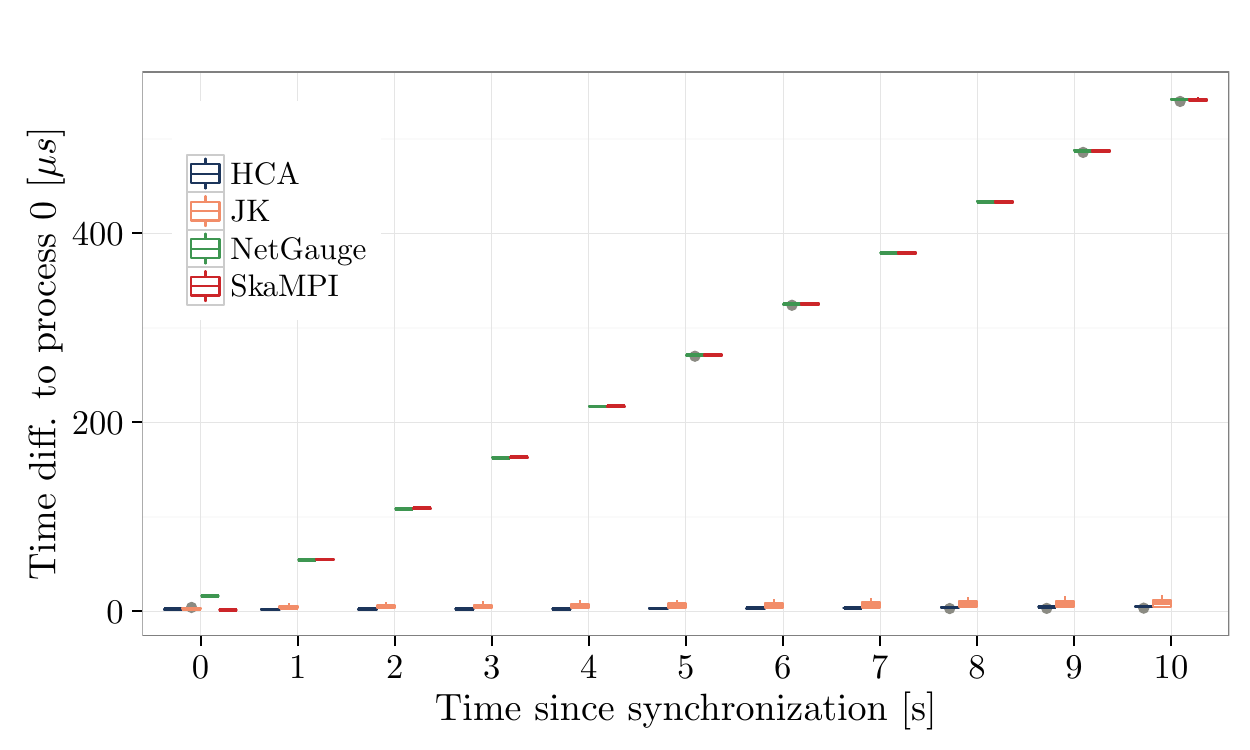}
  \caption{\label{fig:g5k_clock_drift_120nodes} Clock drift for 
  \num{120} (\num{15x8})~processes after \num{0},~\num{1},~\num{2},~$\ldots$,~\SI{10}{\second} 
  (distribution of maximum offsets over \num{10}~calls to \mpirun, 
  \mvapichonenine, \machthree, \expdesc: \append~\ref{sec:exp:clock_drift_all}).}
\end{figure}

\FloatBarrier

\subsubsection{\machfour}
\vspace{-10pt}
\begin{figure}[!ht]
  \centering
  \includegraphics[width=.9\linewidth]{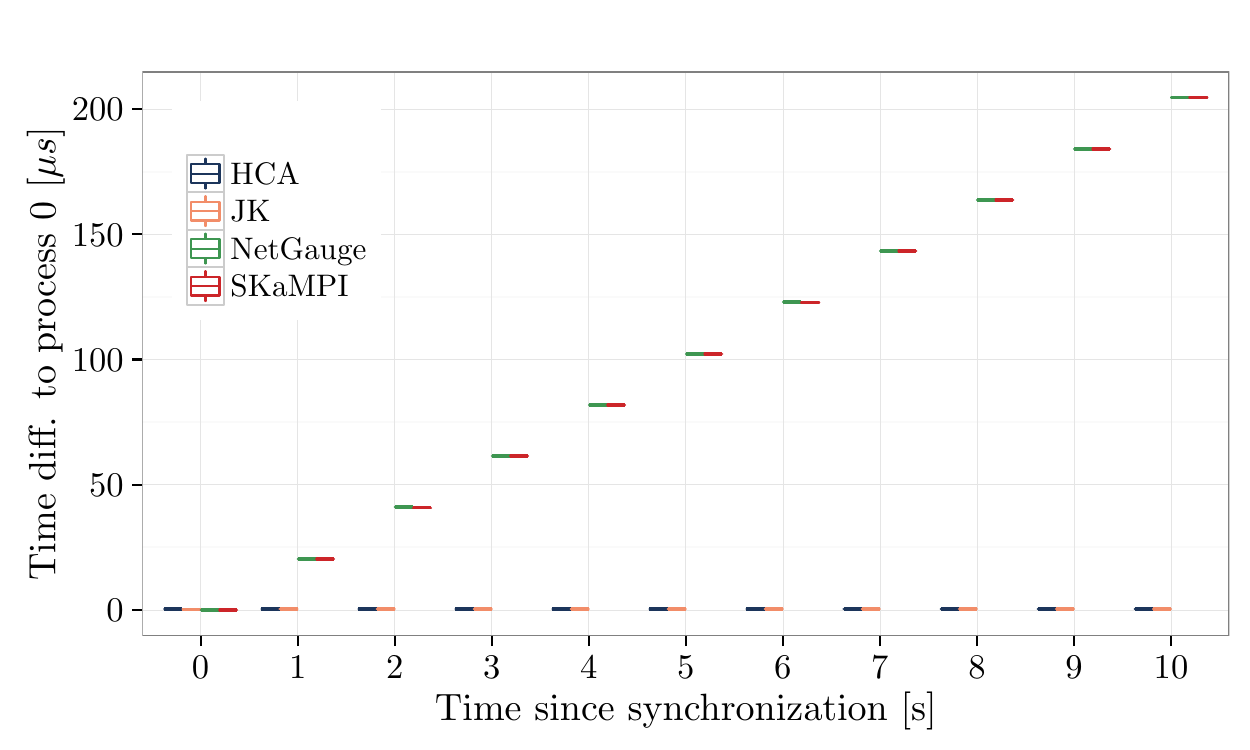}
  \caption{\label{fig:surfsara_clock_drift_16nodes} Clock drift for 
  \num{16x1}~processes after \num{0},~\num{1},~\num{2},~$\ldots$,~\SI{10}{\second} 
  (\num{1}~call to \mpirun, 
  \intelmpi, \machfour, \expdesc: \append~\ref{sec:exp:clock_drift_all}).}
\end{figure}

\FloatBarrier

\newpage
\subsection{Investigating the Synchronization Efficiency}

\subsubsection{\machone}
\begin{figure}[!ht]
  \centering
  \includegraphics[width=\linewidth]{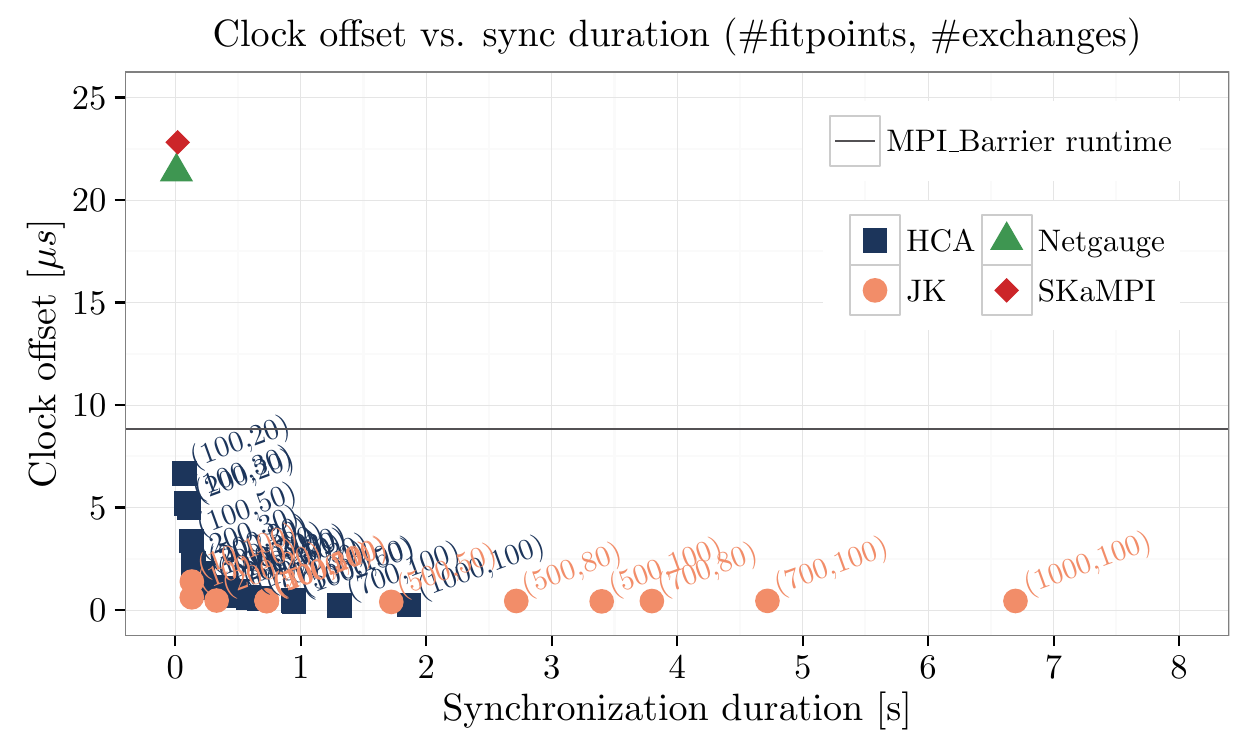}
  \caption{\label{fig:jupiter_sync_efficiency_nodes16}Median clock offset
    (after \SI{2}{\second}) vs. synchronization phase duration for
    \num{16x1}~processes (\num{10}~calls to \mpirun{}, 
    \mvapichtwoone, \machone, \expdesc: \append~\ref{sec:exp:sync_duration}).}
\end{figure}

\FloatBarrier

\subsubsection{\vspace{-10pt}\machthree}
\begin{figure}[!ht]
  \centering
  \includegraphics[width=\linewidth]{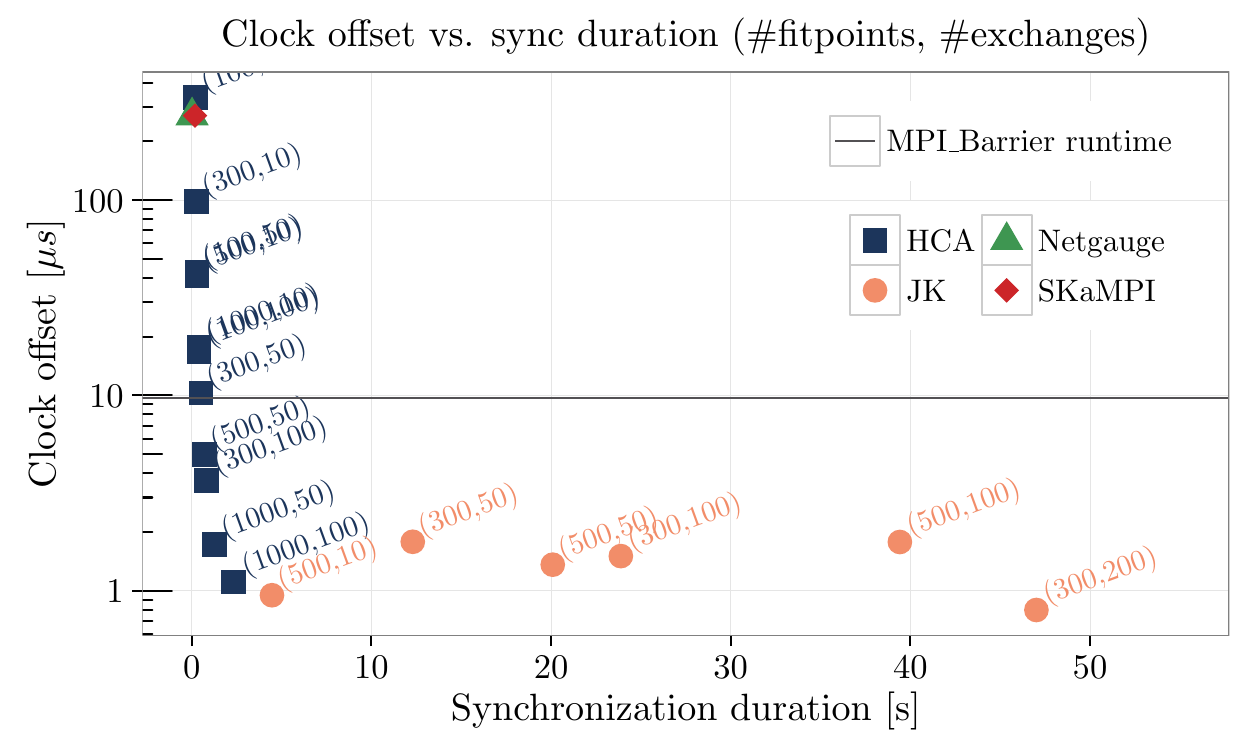}
  \caption{\label{fig:g5k_sync_efficiency_nodes120}Median clock offset
    (after \SI{5}{\second}) vs. synchronization phase duration for
    \num{120}~processes (\num{15x8}) (\num{10}~calls to \mpirun{}, 
    \mvapichonenine, \machthree, \expdesc: \append~\ref{sec:exp:sync_duration}).}
\end{figure}

\FloatBarrier

\vspace{-12pt}
\myneedspace
\subsection{Measuring Drift over Time}

\subsubsection{\machthree}
\vspace{-12pt}
\begin{figure}[!ht]
  \centering
  \includegraphics[width=\linewidth]{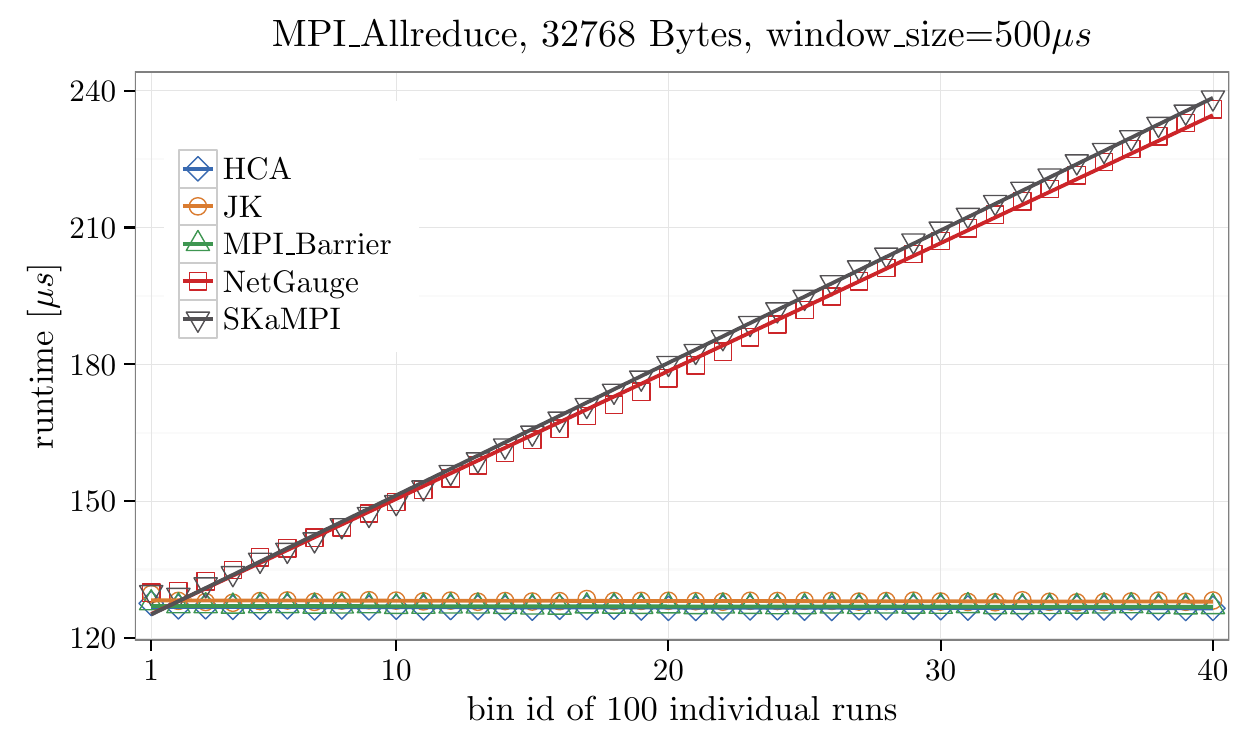}
  \caption{\label{fig:g5k_measurement_drift_nodes120}Drifting \runtimes
    of \mpiallreduce (median of \num{10}~experiments,
    \SI{32}{\kibi\byte}, \num{15x8}~processes, \num{4000}~runs, bin
    size \num{100}, window size: \SI{500}{\micro\second}, 
    \mvapichonenine, \machthree, \expdesc:
    \append~\ref{sec:exp:runtime_drift_all}).}
\end{figure}

\FloatBarrier

\newpage
\subsubsection{\machfour}
\vspace{-12pt}

\begin{figure}[!ht]
  \centering
  \includegraphics[width=\linewidth]{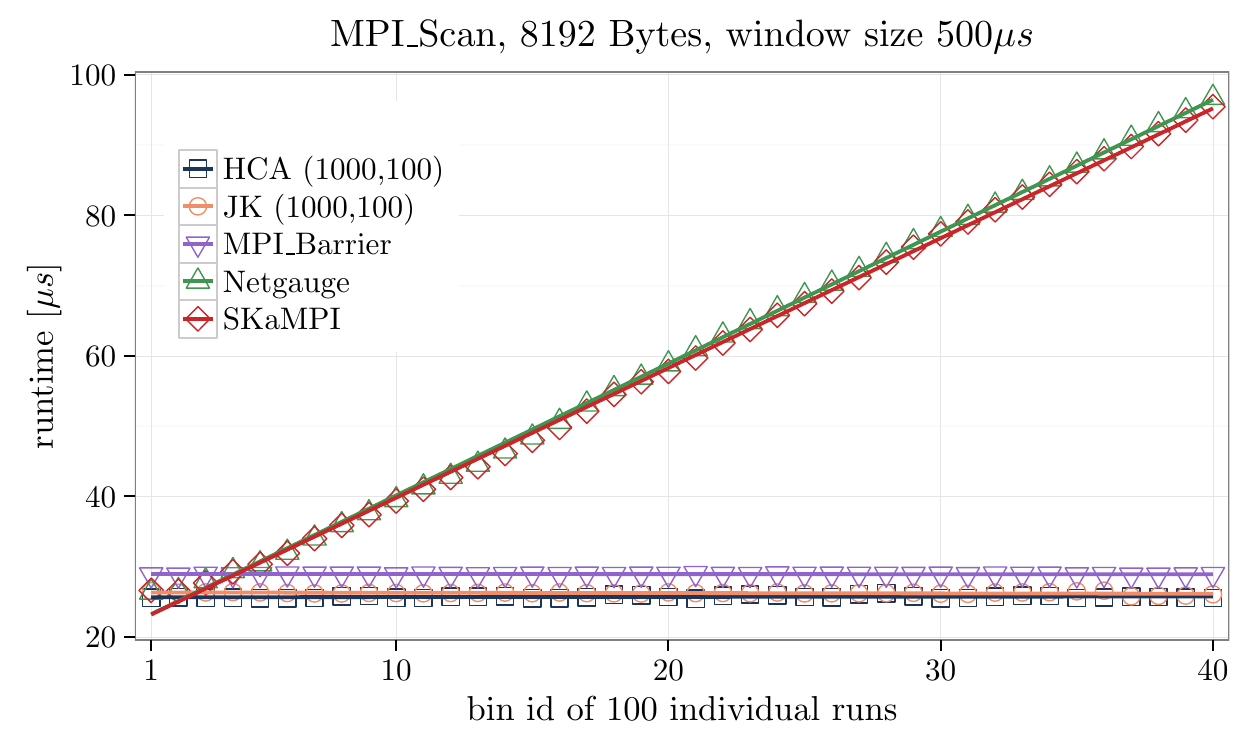}
  \caption{\label{fig:surfsara_sync_efficiency_nodes16}Drifting \runtimes
    of \mpiscan (median of \num{3}~experiments,
    \SI{8}{\kibi\byte}, \num{16x1}~processes, \num{4000}~runs, bin
    size \num{100}, window size: \SI{500}{\micro\second}, 
    \intelmpi, \machfour, \expdesc:
    \append~\ref{sec:exp:runtime_drift_all}).}
\end{figure}

\FloatBarrier

\myneedspace
\subsubsection{\machfive}
\vspace{-12pt}

\begin{figure}[!ht]
  \centering
  \includegraphics[width=0.9\linewidth]{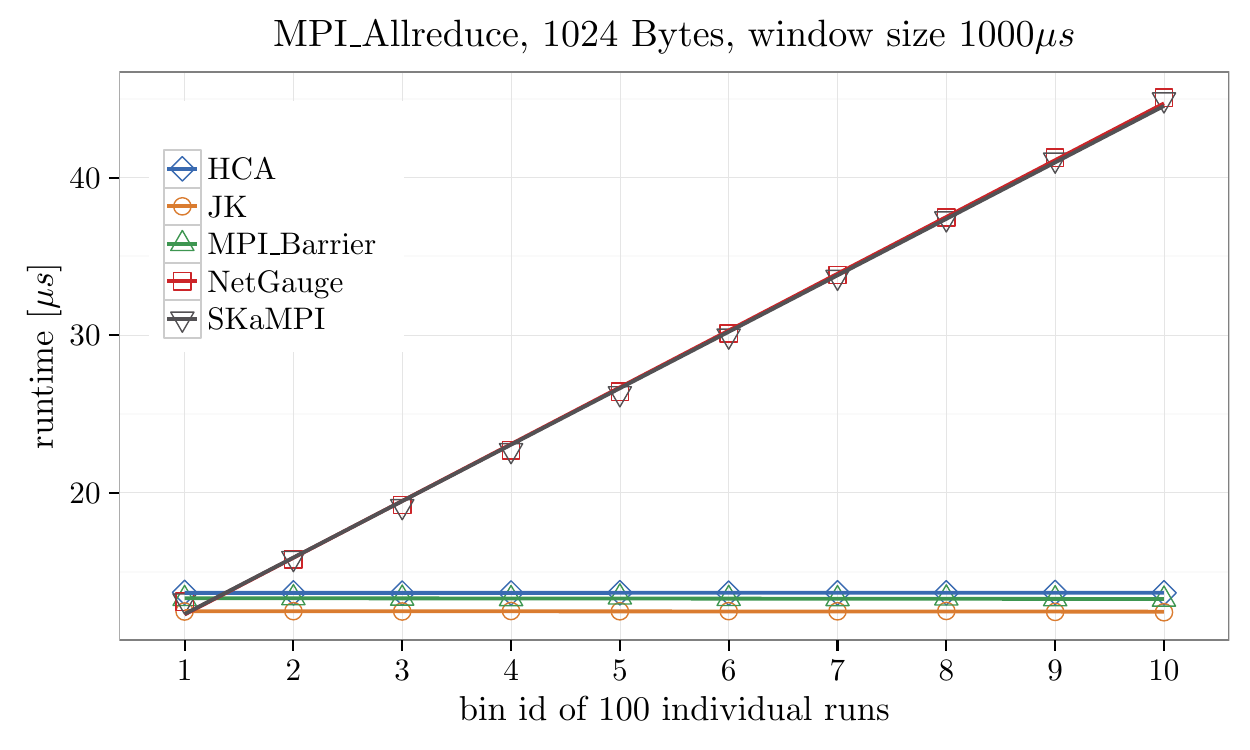}
  \caption{\label{fig:vsc3-drift-sync-methods} Drifting \runtimes
    of \mpiallreduce (median of \num{10}~experiments,
    \SI{1}{\kibi\byte}, \num{16x1}~processes, \num{4000}~runs, bin
    size \num{100}, window size: \SI{1}{\milli\second}, 
    \mvapichtwoa, \machfive, \expdesc:
    \append~\ref{sec:exp:runtime_drift_all}).}
\end{figure}

\FloatBarrier

\newpage
\subsection{Distribution of Measured \Runtimes}
\label{app:meas:distributions}
\subsubsection{\machthree}
\vspace{-12pt}

\FloatBarrier

\myneedspace
\subsubsection{\machfour}

\FloatBarrier

\newpage
\subsubsection{\machfive}

\FloatBarrier

\newpage
\subsection{Comparing \Runtimes measured using \mpibarrier or Window-based Synchronization}

\begin{figure}[!ht]
  \centering
  \includegraphics[width=\linewidth]{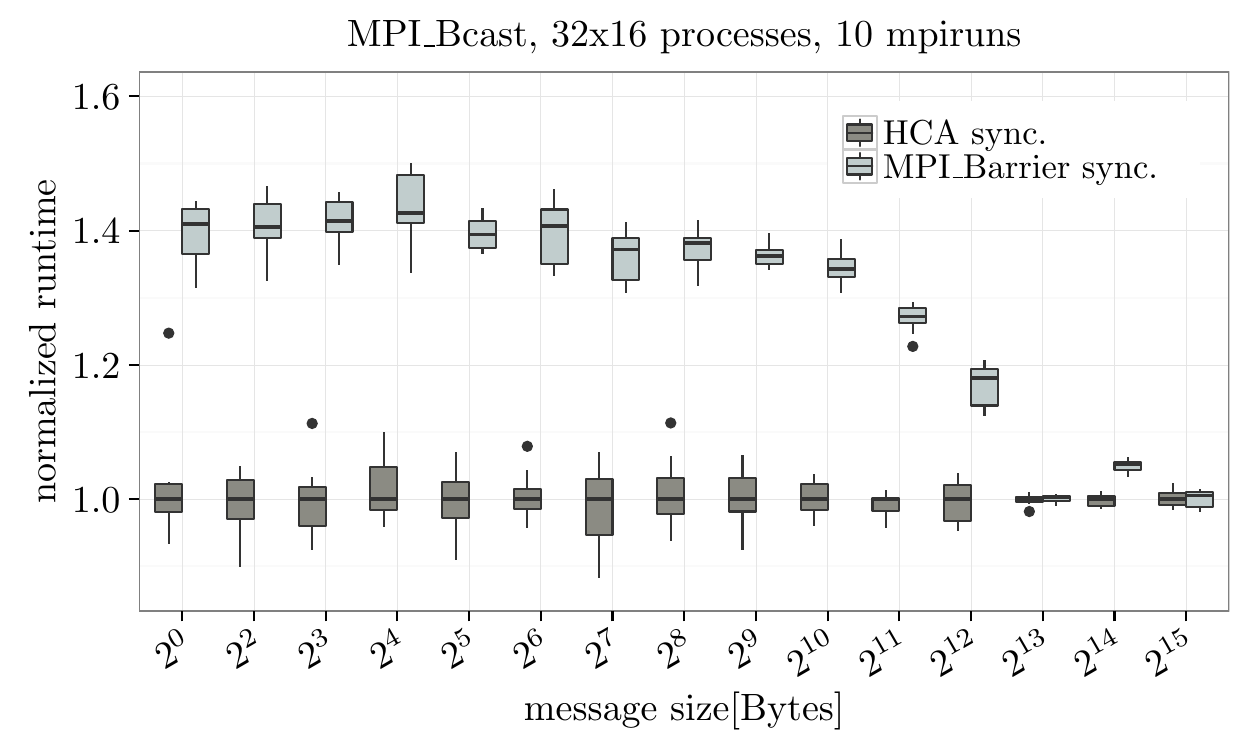}
  \caption{\label{fig:winsync_vs_barrier}Distribution of normalized
    median \runtimes reported for \mpibcast with various message
    sizes, for the window-based synchronization method \synchca
    (window size: \SI{150}{\micro\second}) and \mpibarrier,
    (\num{10} calls to \mpirun, \num{32x16}~processes, 
    \num{1000} runs, \mvapichtwoone, \machone, 
    \expdesc: \append~\ref{sec:exp:runtime_barrier_vs_hca}).}
\end{figure}

\end{document}